\documentclass[a4paper,11pt]{article}
\pdfoutput=1
\usepackage{jheppub}
\usepackage{amsmath,amssymb,latexsym}
\usepackage{bm}
\usepackage{hepunits}
\usepackage[svgnames]{xcolor}
\usepackage{comment}

\usepackage{graphicx}
\usepackage{float}
\usepackage{subfigure}

\usepackage{mathtools}
\usepackage{braket}
\usepackage{upgreek}

\usepackage{natbib}
\usepackage{graphicx}

\usepackage{shuffle} 
\usepackage{slashed} 

\usepackage{multirow}

\usepackage{longtable}
\usepackage{lipsum}

\usepackage{xspace}
\newcommand*{\eg}{e.g.,\@\xspace}
\newcommand*{\ie}{i.e.,\@\xspace}

\newcommand*{\cm}{c.m.\@\xspace}

\newcommand{\boldirrep}{\mathbf}
\newcommand{\irrepbase}[1]{\ensuremath{\boldirrep{#1}}}

\newlength{\irrepwidth}
\newlength{\irrepbarthickness}
\newlength{\irrepbarheight}
\newcommand{\irrepbarbase}[1]{%
    \settoheight{\irrepbarheight}{\irrepbase{#1}}%
    \settowidth{\irrepwidth}{\irrepbase{#1}}%
    \makebox[0pt][l]{\irrepbase{#1}}%
    \rule[1.2\irrepbarheight]{\irrepwidth}{\irrepbarthickness}%
}

\def\primes#1#2{\count0=#1 \loop \ifnum\count0>0 \advance\count0 by -1 #2\repeat}
\newcommand{\irrep}[2][0]{\ensuremath{\irrepbase{#2}^{\primes{#1}{\prime}}}}
\newcommand{\irrepbar}[2][0]{\ensuremath{\irrepbarbase{#2}^{\primes{#1}{\prime}}}}

\newcommand{\harpoon}{\overset{\rightharpoonup}}

\newcommand\ident{{\cal I}}
\newcommand\Ione{\ident_1}
\newcommand\Itwo{\ident_2}
\def\msbar{\overline{\mathrm{MS}}}
\def\kira{{\sc Kira}}
\def\qgraf{{\sc Qgraf}}
\def\feynarts{{\sc FeynArts}}

\usepackage{subfigure,tikz}
\usetikzlibrary{positioning,arrows}
\usetikzlibrary{decorations.pathmorphing}
\usetikzlibrary{decorations.markings}
\usetikzlibrary{shapes.geometric}
\tikzset{
    vector/.style={decorate, decoration={snake}, draw},
    provector/.style={decorate, decoration={snake,amplitude=2.5pt}, draw},
    antivector/.style={decorate, decoration={snake,amplitude=-2.5pt}, draw},
    fermion/.style={draw=red, line width=0.4mm,
      postaction={decorate},decoration={markings,mark=at position .55
        with {\arrow[draw=red]{>}}}},
    fermionbar/.style={draw=red, postaction={decorate},
                       decoration={markings,mark=at position .55 with {\arrow[draw=red]{<}}}},
    fermionnoarrow/.style={draw=black},
    top/.style={draw=black, line width=1mm,
      postaction={decorate},decoration={markings,mark=at position .55
        with {\arrow[draw=black]{>}}}},
    wilson/.style={draw=black, line width=1mm,
      postaction={decorate},decoration={markings}},
    gluon/.style={decorate, draw=red,decoration={coil,amplitude=5pt, segment length=7pt}},
    photon/.style={decorate, draw=red,decoration={snake,amplitude=5pt, segment length=9pt}},
    ghost/.style={dotted,draw=red,
      postaction={decorate},decoration={markings,mark=at position .55
        with {\arrow[draw=red]{>}}}},
    ghostbar/.style={dotted,draw=red,
      postaction={decorate},decoration={markings,mark=at position .55
        with {\arrow[draw=red]{<}}}},
    scalar/.style={dashed,draw=black,
      postaction={decorate},decoration={markings,mark=at position .55
        with {\arrow[draw=black]{>}}}},
    scalarbar/.style={dashed,draw=black,
      postaction={decorate},decoration={markings,mark=at position .55
        with {\arrow[draw=black]{<}}}},
    scalarnoarrow/.style={dashed,draw=black},
    electron/.style={draw=black,
      postaction={decorate},decoration={markings,mark=at position .55
        with {\arrow[draw=black]{>}}}},
    bigvector/.style={decorate, decoration={snake,amplitude=4pt}, draw},
}

\allowdisplaybreaks

\title{Analytic NNLO transverse-momentum-dependent soft function for heavy quark pair hadroproduction at threshold}

\author[a]{Hua-Sheng Shao,}
\author[a]{Guoxing Wang}
\affiliation[a]{Laboratoire de Physique Th\'eorique et Hautes Energies (LPTHE), UMR 7589, Sorbonne Universit\'e et CNRS, 4 place Jussieu, 75252 Paris Cedex 05, France}
\emailAdd{huasheng.shao@lpthe.jussieu.fr}
\emailAdd{wangguoxing2015@pku.edu.cn}

\abstract{The transverse-momentum-dependent (TMD) soft function for non-relativistic heavy quark pair production at hadron colliders is analytically computed at next-to-next-to-leading order (NNLO) in the strong coupling expansion. We present the details of our computational approach and analyze the general two-loop structure of the soft function. The final result, which takes a particularly simple form, provides the last missing ingredient for a complete NNLO calculation of color-octet $S$-wave quarkonium hadroproduction--including charmonium, bottomonium, and toponium--using the $q_T$-slicing formalism. It also enables next-to-next-to-next-to-leading-logarithmic (N$^3$LL) resummation at small transverse momentum for the same process.}

\begin{document}

\maketitle


\section{Introduction}
Heavy quark pair production is one of the most fundamental and extensively studied processes at high-energy colliders, such as the CERN Large Hadron Collider (LHC). Given that the heavy quark mass $m_Q$ is much larger than the intrinsic Quantum Chromodynamics (QCD) scale $\Lambda_{\rm QCD}$, the production cross section can be reliably computed in perturbative QCD. This is possible because of asymptotic freedom, which renders the strong coupling $\alpha_s(\mu_R)$ small at high energies~\cite{Gross:1973id,Politzer:1973fx}. Here, the renormalization scale $\mu_R$ is typically chosen to be of the order of the heavy quark mass $m_Q$.

Among all known particles, the top quark--the heaviest elementary particle discovered to date--plays a unique role. In the Standard Model (SM) of particle physics, the top quark couples more strongly to the Higgs boson than any other fermion, with the Yukawa coupling strength $y_t=\sqrt{2}m_t/v$ that is approximately unity. Here, $v\simeq 246$ GeV denotes the vacuum expectation value of the Higgs field. This strong coupling suggests that the top quark may play a crucial role in revealing the mechanism of electroweak symmetry breaking. In addition, unlike other heavy quarks such as charm and bottom, the top quark has an extremely short lifetime--much shorter than $\Lambda_{\rm QCD}^{-1}$. As a result, it decays before hadronization occurs. Consequently, color-singlet bound states composed of a top-antitop pair, known as \emph{toponium}, do not manifest as resonance peaks, in contrast to their charmonium and bottomonium counterparts. 

Top quarks are predominantly produced in pairs with their antiparticles at high-energy colliders. The production of top-quark pairs near threshold--\ie\ in the non-relativistic limit--is of particular interest. It has long been proposed to determine the top-quark pole mass and width with high precision via beam energy scans at future lepton colliders operating near the top-pair production threshold~\cite{Fadin:1987wz,Fadin:1988fn,Strassler:1990nw,Martinez:2002st,Seidel:2013sqa,Horiguchi:2013wra}. The top-quark mass is a key parameter of the SM, significantly affecting theoretical predictions of Higgs boson properties, searches for physics beyond the SM (BSM), and playing a crucial role in the stability of the electroweak vacuum at asymptotically high energies~\cite{Alekhin:2012py}. Near the threshold region, the resummation of Coulombic soft-gluon exchanges becomes particularly important, both for top–antitop bound states below the $2m_t$ threshold and for scattering states just above it. While this effect is relatively rare, it has been extensively studied in the context of both lepton~\cite{Fadin:1987wz,Fadin:1988fn,Strassler:1990nw} and hadron~\cite{Fadin:1990wx,Hagiwara:2008df} colliders. More recently, the inclusion of Coulomb resummation~\cite{Ju:2019mqc,Ju:2020otc} and toponium effects~\cite{Fuks:2021xje} has been shown to be essential for interpreting small deviations observed between experimental measurements at the LHC~\cite{ATLAS:2019hau,CMS:2018adi} and the most advanced theoretical predictions, which include fixed order calculations~\cite{Czakon:2013goa,Czakon:2017wor}, soft-gluon resummation~\cite{Czakon:2018nun}, and next-to-leading order (NLO) matching to parton showers~\cite{Alioli:2010xd,Alwall:2014hca}. This conclusion has been further supported by recent measurements from the ATLAS~\cite{ATLAS:2023fsd} and CMS~\cite{CMS:2024pts,CMS:2025kzt} collaborations in the threshold regime. These developments strongly motivate improved theoretical predictions for top-quark pair-- and, more generally, heavy-quark pair--hadroproduction near the $2m_t$ threshold, as highlighted in several recent studies~\cite{Maltoni:2024tul,Fuks:2024yjj,Garzelli:2024uhe,Nason:2025hix}.

Let us briefly review the precise cross-section calculations for heavy-quark pair hadroproduction in the SM that have been developed over the past fifteen years. The next-to-NLO (NNLO) QCD corrections have been computed in refs.~\cite{Baernreuther:2012ws,Czakon:2012zr,Czakon:2012pz,Czakon:2013goa,Czakon:2015owf,Czakon:2016dgf, Catani:2019iny, Catani:2019hip, Catani:2020tko}, while the NLO electroweak (EW) corrections are presented in refs.~\cite{Hollik:2011ps, Kuhn:2011ri, Bernreuther:2012sx, Pagani:2016caq, Czakon:2017wor,  Gutschow:2018tuk, Frederix:2018nkq, Czakon:2019txp}, along with their combined treatment in ref.~\cite{Czakon:2017lgo}. Off-shell effects in top-quark pair production have been thoroughly studied at NLO in refs.~\cite{Denner:2010jp,Bevilacqua:2010qb,Denner:2012yc,Frederix:2013gra,Heinrich:2013qaa,Denner:2016jyo,Heinrich:2017bqp,Denner:2017kzu}. These fixed-order results can be further improved through various resummation techniques that account for logarithmically enhanced contributions, as well as through parton-shower matching. In particular, soft-gluon threshold resummation has been extensively investigated~\cite{Kidonakis:2014pja,Kidonakis:2023juy,Kidonakis:2010dk,Kidonakis:2009ev,Kidonakis:2014isa,Kidonakis:2019yji,Ahrens:2010zv,Ferroglia:2012ku,Ferroglia:2013awa,Pecjak:2016nee,Czakon:2018nun,Almeida:2008ug,Pecjak:2018lif}; Coulomb-gluon resummation has been addressed in refs.~\cite{Hagiwara:2008df,Kiyo:2008bv,Ju:2020otc,Ju:2019mqc}; combined soft- and Coulomb-gluon resummation in refs.~\cite{Beneke:2009ye,Beneke:2010da,Beneke:2011mq,Cacciari:2011hy,Piclum:2018ndt}; and small transverse-momentum ($q_T$) resummation in refs.~\cite{Zhu:2012ts,Li:2013mia,Catani:2014qha,Catani:2017tuc,Catani:2018mei,Alioli:2021ggd,Ju:2022wia}. Furthermore, particle-level event simulation for $pp\to t\bar{t}+X$ production at NNLO QCD accuracy matched to parton showers has become feasible, as demonstrated in ref.~\cite{Mazzitelli:2021mmm}.

In this work, we focus on the analytical calculation of the transverse-momentum-dependent (TMD) soft function for heavy-quark pair production at the $2m_Q$ threshold, up to second order in $\alpha_s$. The TMD soft function, which encodes the dynamics of soft parton exchanges between colored external particles, is a universal quantity determined solely by the quantum numbers and kinematics of these external states. It plays an essential role in small-$q_T$ resummation calculations and in fixed-order computations based on the $q_T$ slicing method~\cite{Catani:2007vq}. A notable application of the latter is the NNLO QCD computation of heavy-quark pair production cross sections at the LHC, as performed in refs.~\cite{Catani:2019iny,Catani:2019hip,Catani:2020tko,Catani:2020kkl}.

Among the process-independent components in the small-$q_T$ factorization formula, the TMD soft function is by far the most intricate. Many calculations, both numerical and analytical, have been
performed in the literature. The analytic expressions for the azimuthal-angle-averaged and azimuthal-angle-dependent soft functions in back-to-back heavy-quark pair hadroproduction are known at NLO, as given in refs.~\cite{Li:2013mia} and \cite{Catani:2014qha}, respectively. Numerical results for the azimuthal-angle-averaged NNLO soft function for the same process have been presented in refs.~\cite{Angeles-Martinez:2018mqh,Catani:2023tby}. For a generic process involving massive colored particles in the final state, the NLO soft function is available either as one-fold integrals requiring numerical evaluation~\cite{Catani:2021cbl,Ju:2022wia} or in closed analytic form involving multiple polylogarithms~\cite{Shao:2025qgv}.~\footnote{Ref.~\cite{Shao:2025qgv} also provides the fully analytic one-loop TMD soft function to higher orders in the dimensional regulator, which is required in higher-order computations.} Recently, the azimuthal-angle-dependent TMD soft function for the process $e^+e^-\to Q\bar{Q}V$ (with $V$ denoting a generic color-singlet system) was analytically computed at NNLO in ref.~\cite{Liu:2024hfa}.

In the following, we focus on heavy-quark pair hadroproduction at threshold, which is a simpler case than the general kinematic configuration and therefore more amenable to analytic calculation of the NNLO soft function.  The resulting expressions are applicable to the production of $S$-wave quarkonium--such as charmonium, bottomonium, and toponium-- at small transverse momentum.

This paper is organized as follows. In section~\ref{sec:formula}, we introduce the notations and present the factorization formulae in both the threshold and small-$q_T$ limits. Section~\ref{sec:soft_function} discusses the operator definition of the soft function, its general structure, and the methodology for computing the NNLO TMD soft function for the process of interest: heavy-quark pair hadroproduction in the non-relativistic limit. Our conclusions are summarized in section~\ref{sec:conclusion}. A complete set of two-loop integrals is provided in appendix \ref{sec:tabofints}.

\section{Two-step factorization formalism}\label{sec:formula}

We consider the following hadronic process
\begin{equation}\label{eq:procdef}
  N_1(P_1) + N_2(P_2) \to Q(p_3) + \bar{Q}(p_4) + X(k_X)\,,
\end{equation}
where $N_1$ and $N_2$ are the incoming hadrons with momenta $P_1$ and $P_2$, $Q$ and $\bar{Q}$ are the final-state heavy quark and antiquark with momenta $p_3$ and $p_4$, and $X$ represents all unresolved hadronic radiation originating from beam remnants and real emissions. The on-shell conditions are $p_3^2=p_4^2=m_Q^2$ and $P_1^2=P_2^2=0$, where the hadron masses are neglected. At LO, there are two types of partonic subprocesses: quark-antiquark annihilation and gluon fusion, given by
\begin{eqnarray}
  q(p_1) + \bar{q}(p_2) & \to & Q(p_3) + \bar{Q}(p_4)\,, \\
  g(p_1) + g(p_2)      & \to & Q(p_3) + \bar{Q}(p_4)\,.
\end{eqnarray}
In the collinear factorization framework, the momenta of the initial-state partons are $p_1 = \xi_1 P_1$ and $p_2 = \xi_2 P_2$, with $0<\xi_{1},\xi_{2}<1$.

For convenience, we define the following kinematic variables:
\begin{align}\label{eq:kintematicsv}
  S&=(P_1+P_2)^2 \, , \quad s=(p_1+p_2)^2 \, , \quad M^2=(p_3+p_4)^2 \, , \nonumber \\
  t_1&=(p_1-p_3)^2-m_Q^2 \, , \quad u_1=(p_1-p_4)^2-m_Q^2 \, ,  \nonumber \\
  Y & =\frac{1}{2}\log{\left(\frac{\xi_1}{\xi_2}\right)} \, , \quad \beta = \sqrt{1-\frac{4m_Q^2}{M^2}}\, ,
\end{align}
where $Y$~\footnote{$Y$ is defined in the center-of-mass (\cm) frame of the incoming hadrons $N_1$ and $N_2$.} and $M$ denote the rapidity, and invariant mass of the $Q\bar{Q}$ system, respectively. The Mandelstam variables $s$, $t_1$, and $u_1$ satisfy $s + t_1 + u_1 = 0$ by momentum conservation.
The cross section for $Q\bar{Q}$ production is given by
\begin{align}\label{eq:dsigma_1}
\sigma &= \frac{1}{2S} \int{\frac{d^4p_3}{(2\pi)^4} \frac{d^4p_4}{(2\pi)^4} \, (2\pi) \delta^{(+)}(p_3^2-m_Q^2) \, (2\pi) \delta^{(+)}(p_{4}^2-m_Q^2)} \nonumber \\ 
&\times \int\limits_{X}\hspace{-0.55cm}\sum \,(2\pi)^4 \delta^{(4)}(P_1 + P_2 - p_3-p_4 - k_X) \, \big| \mathcal{A}(N_1 + N_2 \to Q + \bar{Q} + X) \big|^2 \,,
\end{align}
where $\delta^{(+)}(p^2-m^2)=\delta(p^2-m^2)\Theta(p^0)$, with $\Theta()$ denoting the Heaviside step function. Since the threshold limit $\beta\to0$ and the small-$q_T$~\footnote{$q_T$ denotes the transverse momentum of the $Q\bar{Q}$ system in the \cm\ frame of the incoming hadrons.} limit $q_T\to 0$ do not generally commute, we adopt a two-step factorization approach: we first take the threshold limit $\beta\to 0$, and subsequently consider the small-$q_T$ limit.~\footnote{For instance, if we instead take the $\beta\to 0$ limit of the NNLO TMD azimuthal-angle-averaged soft function for a heavy-quark pair--\ie\ by swapping the order of the two limits--a difference, including the $\log{\beta}$ and finite terms in $\beta$, arises from the imaginary part of the three-Wilson-line correlators in the real-virtual contributions (cf. figure 18 of ref.~\cite{Angeles-Martinez:2018mqh}).}

In the threshold region $M\to 2m_Q$ (\ie\ $\beta\to 0$), the production of the heavy-quark pair $Q\bar{Q}$ can be factorized into a short-distance cross section, describing the creation of the pair at a hard scale $\mu\sim M$, and a potential function that accounts for the exchanges of Coulomb-type and soft virtual gluons between the heavy quark and antiquark. These contributions from Coulomb and soft gluons can be systematically described within non-relativistic QCD (NRQCD)~\cite{Bodwin:1994jh} and potential NRQCD (pNRQCD)~\cite{Pineda:1997bj, Brambilla:1999xf, Beneke:1999zr, Beneke:1999qg}, up to next-to-leading power (NLP) in the expansion around $\beta\sim 0$. Following refs.~\cite{Kiyo:2008bv, Beneke:2010da,Ju:2019mqc, Ju:2020otc}, the cross section in eq.~\eqref{eq:dsigma_1} can be factorized in pNRQCD as
\begin{align}\label{eq:dsigma_2}
\sigma &= \sum_{C\in\{1,8\}}{\int{\frac{dM^2}{2\pi M^2}\sigma_{[C]}(M) J^{[C]}(E)}}\,,
\end{align}
where $J^{[C]}(E)$ is the long-distance potential function that resums multiple Coulomb-gluon exchanges and depends only on the color representation ($C$) of the $Q\bar{Q}$ system. This dependence arises because the NLO pNRQCD Lagrangian is spin-independent, and thus $J^{[C]}(E)$ does not depend on the total spin of the heavy quark pair. Spin-dependent non-Coulomb potentials and soft contributions enter only beyond NLP in $\beta$ (see eq.(2.13) of ref.~\cite{Beneke:2011mq}). The heavy-quark pair can be in either a color-singlet ($C=1$) or color-octet ($C=8$) configuration. The quantity $\sigma_{[C]}(M)$ denotes the short-distance cross section for producing a $Q\bar{Q}$ pair in color configuration $C$. The potential function $J^{[C]}(E)$ is defined through an operator product expansion (OPE) as~\cite{Beneke:2010da,Ju:2020otc}
\begin{align}
J^{[C]}(E) &= M^{2} \int \frac{d^4p_3}{(2\pi)^4} \frac{d^4p_4}{(2\pi)^4} \, (2\pi) \delta^{(+)}(p_3^2-m_Q^2) \, (2\pi) \delta^{(+)}(p_4^2-m_Q^2)  \, (2\pi)^4 \delta^{(4)}\left(P_{Q\bar{Q}}-p_3-p_4\right)\nonumber
\\
&\times \mathbb{P}_{C,c_4c_3}\cdot \mathbb{P}_{C,c_3^\prime c_4^\prime}\langle 0 | \chi^\dagger \psi | Q_{c^\prime_3}\bar{Q}_{c^\prime_4}\rangle \langle Q_{c_3}\bar{Q}_{c_4} | \psi^\dagger \chi | 0\rangle \, ,\label{eq:JinOPE}
\end{align}
where $\psi$ and $\chi$ are the two-component NRQCD fields describing the heavy-quark and antiquark after integrating out the hard modes, and the indices $c_i$ denote their color components. The color projectors are given by
\begin{equation}
\mathbb{P}_{1,c_4c_3}=\frac{\delta_{c_4,c_3}}{\sqrt{N_c}}\,,\quad
\mathbb{P}_{8,c_4c_3}=\sqrt{2}t^a_{c_4,c_3}\,,\label{eq:colorprojector}
\end{equation}
where $\delta_{c_4,c_3}$ is the Kronecker delta, $t^a$ are the Gell-Mann matrices, and $N_c=3$ in QCD. The normalization of the projectors ensures the completeness relation:
\begin{equation}
\sum_{C\in\{1,8\}}{\mathbb{P}_{C,c_4c_3}\cdot \mathbb{P}_{C,c_3^\prime c_4^\prime}}=\delta_{c_3,c_3^\prime}\delta_{c_4,c_4^\prime}\,.
\end{equation}
The potential function $J^{[C]}(E)$ can also be expressed in terms of the imaginary part of the Green function $G^{[C]}(\vec{r}_1,\vec{r}_2;E)$ of the heavy-quark pair in pNRQCD~\cite{Beneke:2010da}:
\begin{align}
J^{[C]}(E) = 2 {\rm Im} \, G^{[C]}(\vec{0},\vec{0};E) \, .\label{eq:JinG}
\end{align}
This Green function is obtained by solving the Schr\"odinger equation with the corresponding $\left(Q\bar{Q}\right)_C$ potential and admits a spectral representation: 
\begin{equation}
G^{[C]}(\vec{r}_1,\vec{r}_2;E)=-\sum_{E^\prime}{\frac{\Psi^{[C]}_{E^\prime}(\vec{r}_1)\Psi^{[C],\dagger}_{E^\prime}(\vec{r}_2)}{E-E^\prime+i0^+}}\,,
\end{equation}
where $\Psi^{[C]}_{E^\prime}(\vec{r})$ is the coordinate-space wavefunction with binding energy eigenvalue $E^\prime$. While eq.\eqref{eq:JinOPE} applies only to continuum states with positive binding energy ($E>0$), eq.\eqref{eq:JinG} also extends naturally to bound states with $E<0$. For well-separated bound states,~\footnote{Strictly speaking, there are no bound states in the repulsive color-octet potential. However, for $C=8$, the potential function can still be non-zero for $E<0$ due to the finite decay width.} the potential function simplifies to
\begin{equation}
J^{[C]}(E)=2 {\rm Im} \, G^{[C]}(\vec{0},\vec{0};E) = 2\pi \sum_{E^\prime}{\left|\Psi^{[C]}_{E^\prime}(\vec{0})\right|^2\delta(E^\prime-E)}\,,\quad E<0\,.
\end{equation}
In the case of toponium, where the interaction is Coulombic, the Green function can be solved analytically~\cite{Beneke:1999qg, Beneke:1999zr, Pineda:2006ri, Beneke:2011mq} (see, \eg\ eq.(A.1) in ref.~\cite{Beneke:2011mq} or eq.(2.33) in ref.~\cite{Ju:2020otc}). 
The short-distance cross section is given by
\begin{align}\label{eq:SDxs1}
\sigma_{[C]}(M)&=\frac{1}{2S} \int{\frac{d^4P_{Q\bar{Q}}}{(2\pi)^4}(2\pi)\delta^{(+)}\left(P_{Q\bar{Q}}^2-M^2\right)} \nonumber \\
&\times\int\limits_{X}\hspace{-0.55cm}\sum \,(2\pi)^4 \delta^{(4)}(P_1 + P_2 - P_{Q\bar{Q}} - k_X) \, \big| \mathcal{A}(N_1 + N_2 \to (Q\bar{Q})_{C} + X) \big|^2\,.
\end{align}
The amplitude in the integrand should be evaluated in the non-relativistic limit $M\to 2m_Q$.

On the other hand, for $S$-wave charmonium and bottomonium production in NRQCD, one must further project the intermediate $Q\bar{Q}$ pair into spin-singlet ($j=0$) and spin-triplet ($j=1$) states. Consequently, eq.\eqref{eq:dsigma_2} should be modified as follows:
\begin{align}\label{eq:dsigmaonium}
\sigma &= \sum_{C\in\{1,8\}}\sum_{j=0}^{1}{\int{\frac{dM^2}{2\pi M^2}\sigma_{[C,j]}(M) J^{[C,j]}(E)}}\,,
\end{align}
where the potential function is given by
\begin{equation}
J^{[C,j]}(E)=\frac{2\pi}{\left(2j+1\right)N_{[C]}}\sum_{n=1}^{+\infty}{\left\langle \mathcal{O}^{n}_{^{(2j+1)}S_j^{[C]}}\right\rangle\delta(E-E_{n,j,C})}\,,
\end{equation}
where $N_{[1]}=2N_c$ and $N_{[8]}=N_c^2-1$~\cite{AH:2024ueu}. Here, $n$ is the principal quantum number, $E_{n,j,C}$ is the binding energy of the $n$th radial excitation, and $\left\langle \mathcal{O}^{n}_{^{(2j+1)}S_j^{[C]}}\right\rangle$ denotes the long-distance matrix element (LDME), which can be related to the wavefunction at the origin in pNRQCD via 
\begin{equation}
\left\langle \mathcal{O}^{n}_{^{(2j+1)}S_j^{[C]}}\right\rangle=(2j+1)N_{[C]}\left|\Psi^{[C]}_{E_{n,j,C}}(\vec{0})\right|^2\,.
\end{equation}
The Fock state $^{(2j+1)}S_j^{[C]}$ ($j=0,1$) follows the usual spectroscopic notation. In practice, instead of computing $\Psi^{[8]}_{E_{n,j,8}}(\vec{0})$ from the Schr\"odinger equation using the poorly known color-octet potential, the corresponding color-octet LDMEs can either be expressed in terms of the color-singlet wavefunction at the origin and a universal gluonic correlator within pNRQCD~\cite{Brambilla:2022rjd,Brambilla:2022ayc}, or extracted directly from experimental data~\cite{Butenschoen:2011yh,Chao:2012iv,Gong:2012ug,Shao:2014yta,Han:2014kxa,Bodwin:2014gia,Brambilla:2024iqg}.~\footnote{Additionally, the color-octet LDMEs are suppressed by $\mathcal{O}(\beta^4)$ relative to the color-singlet ones. This suppression originates from the long-distance evolution from color-octet Fock states to physical color-singlet quarkonium. Such suppression does not occur for toponium, whose constituents decay weakly and promptly, leaving no time for hadronization.} The corresponding short-distance cross section--analogous to eq.\eqref{eq:SDxs1}--is given by
\begin{align}\label{eq:SDxs2}
\sigma_{[C,j]}(M)&=\frac{1}{2S} \int{\frac{d^4P_{Q\bar{Q}}}{(2\pi)^4}(2\pi)\delta^{(+)}\left(P_{Q\bar{Q}}^2-M^2\right)} \nonumber \\
&\times\int\limits_{X}\hspace{-0.55cm}\sum \,(2\pi)^4 \delta^{(4)}(P_1 + P_2 - P_{Q\bar{Q}} - k_X) \, \big| \mathcal{A}(N_1 + N_2 \to (Q\bar{Q})[^{(2j+1)}S_j^{[C]}] + X) \big|^2\,,
\end{align}
where the notation $(Q\bar{Q})[^{(2j+1)}S_j^{[C]}]$ in the amplitude implies that the heavy-quark pair is projected onto the corresponding Fock state ${}^{(2j+1)}S_j^{[C]}$ as detailed in section 2 of ref.~\cite{AH:2024ueu}.

 Our next step is to write down the factorization formulae for the short-distance cross sections, $\sigma_{[C]}(M)$ and $\sigma_{[C,j]}(M)$, in the small-$q_T$ limit, defined by the hierarchy
 \begin{equation}\label{eq:small_qt_limit}
  s, M^2, |t_1|, |u_1|, m_Q^2 \gg q_T^2 \gg \Lambda^2_{\rm QCD}\,.
\end{equation}
In this regime, only soft or collinear emissions contribute significantly. Before proceeding, we introduce light-cone coordinates in dimensional regularization with $d=4-2\epsilon$. Any $d$-dimensional vector $k^\mu$ can be decomposed as
\begin{eqnarray}
k^\mu&=&\underbrace{n\cdot k}_{\equiv k_-} \frac{\bar{n}^\mu}{2}+\underbrace{\bar{n}\cdot k}_{\equiv k_{+}}\frac{n^\mu}{2}+\underbrace{k_\perp^\mu}_{=g_\perp^{\mu\nu}k_{\nu}}\equiv k_{-}^\mu+k_{+}^\mu+k_\perp^\mu\,,
\end{eqnarray}
where the light-like reference vectors aligned with the beam directions are defined as 
\begin{eqnarray}
n&=&(1,\underbrace{0,\cdots,0}_{d-2},1)=(1,\harpoon{0},1)\,,\\
\bar{n}&=&(1,\underbrace{0,\cdots,0}_{d-2},-1)=(1,\harpoon{0},-1)\,,
\end{eqnarray}
with $k_{\pm}=k^0\pm k^z$, and the transverse metric is defined
by $g_\perp^{\mu\nu}=\mathrm{diag}\left(0,-\harpoon{1},0\right)$. The spatial components of $n^\mu$ and $\bar{n}^\mu$ point along the directions of the incoming momenta $P_1^\mu$ and $P_2^\mu$, respectively. These vectors satisfy the standard light-cone relations:
\begin{equation}
n^2 = \bar{n}^2 = 0 \,, \quad n \cdot \bar{n} = 2 \,.
\end{equation}
Throughout this work, we denote $\harpoon{x}_T$ and $\vec{x}$ as the $(d-2)$-dimensional transverse and $(d-1)$-dimensional spatial vectors corresponding to any $d$-dimensional vector $x^\mu$. In particular, we have the identifications:
\begin{equation}
x_\perp^\mu=(0,\harpoon{x}_T,0), \quad x_{-}^\mu=(x_{-}/2,\harpoon{0},-x_{-}/2),\quad x_{+}^\mu=(x_{+}/2,\harpoon{0},x_{+}/2)\,.
\end{equation}
The momenta of the incoming partons and outgoing heavy quarks are given by
\begin{subequations}
  \label{eq:external-momenta}
  \begin{align}
    \label{eq:light-momenta}
    p_{1}^\mu &= \frac{\sqrt{s}}{2}\, n^\mu\,,     \quad \qquad \qquad
    p_{2}^\mu  = \frac{\sqrt{s}}{2}\, \bar{n}^\mu\,,
    \\[0.3em]
    p_i^\mu & = m_Q\, v_i^\mu + k_i^\mu\,,\qquad  \
    v_i^2 = 1\,,\qquad i=3,4\,,
    \label{eq:hq-momenta}
  \end{align}
\end{subequations}
where $k_i^\mu$ is a residual momentum scaling as a soft mode, \ie\ $k_i^\mu/M \sim \lambda = q_T/M \ll 1$.

In the small-$q_T$ ($q_T\equiv |\harpoon{q}_T|$) limit defined in eq.~\eqref{eq:small_qt_limit}, the relevant momentum modes contributing to the factorization of the short-distance cross sections can be classified in the framework of soft-collinear effective theory (SCET)~\cite{Bauer:2000ew, Bauer:2000yr, Bauer:2001yt, Beneke:2002ph, Beneke:2002ni} as follows:
\begin{align*}
  \text{hard:} &\qquad (k_-,k_+,\harpoon{k}_T) \sim M(1,1,\harpoon{1}) \, ,
  \\
  \text{hard-collinear:} &\qquad (k_-,k_+,\harpoon{k}_T) \sim M(\lambda^2,1,\harpoon{\lambda}) \, ,
  \\
  \text{anti-hard-collinear:} &\qquad (k_-,k_+,\harpoon{k}_T) \sim M(1,\lambda^2,\harpoon{\lambda}) \, ,
  \\
  \text{soft:} &\qquad (k_-,k_+,\harpoon{k}_T) \sim M(\lambda,\lambda,\harpoon{\lambda}) \, .
\end{align*}

The small-$q_T$ factorization formula~\cite{Becher:2010tm,Becher:2012yn,Zhu:2012ts,Li:2013mia} for the short-distance cross sections is given by
\begin{eqnarray}
\frac{d\sigma_{\mathrm{X}}(M)}{d^2\harpoon{q}_TdYdM^2d\phi_{Q\bar{Q}}}&=&\frac{1}{S}\int{d\xi_1d\xi_2 \delta\left(\xi_1\xi_2-\frac{M^2}{S}\right)\delta\left(\frac{1}{2}\log{\left(\frac{\xi_1}{\xi_2}\right)}-Y\right)}\nonumber\\
&&\times\int{\frac{d^2\harpoon{b}_{T}}{\left(2\pi\right)^2}e^{i\harpoon{q}_T\cdot \harpoon{b}_T}W_{\mathrm{X}}(\xi_1,\xi_2,M,\harpoon{b}_T)}\,,\label{eq:qtresumgeneralinb}
\end{eqnarray}
where $\mathrm{X}=[C]$ for toponium and $\mathrm{X}=[C,j]$ for charmonium and bottomonium. The one-particle phase-space measure of the heavy quark-antiquark pair is
\begin{equation}
d\phi_{Q\bar{Q}}=\frac{d^4P_{Q\bar{Q}}}{(2\pi)^4}(2\pi)\delta^{(+)}\left(P_{Q\bar{Q}}^2-M^2\right)\left(2\pi\right)^4\delta^{(4)}\left(p_1+p_2-P_{Q\bar{Q}}\right)\,.
\end{equation}
The impact-parameter-dependent function $W_{\mathrm{X}}()$ in eq.~\eqref{eq:qtresumgeneralinb} can be written as
\begin{eqnarray}
W_{\mathrm{X}}(\xi_1,\xi_2,M,\harpoon{b}_T)&=&\left(\frac{b_T^2 M^2}{b_0^2}\right)^{-F_{gg}(b_T^2,\mu)}4B_{g/N_1}^{\alpha\rho}(\xi_1,\harpoon{b}_T,\mu)B_{g/N_2}^{\beta\sigma}(\xi_2,\harpoon{b}_T,\mu)\nonumber\\
&&\times{\rm Tr}_c\left[\bm{H}^{\mathrm{X}}_{gg,\alpha\beta\rho\sigma}(M,\mu)\bm{S}^{\mathrm{X}}_{\perp,gg}(\harpoon{b}_T,\mu)\right]\nonumber\\
&&+\left(\frac{b_T^2 M^2}{b_0^2}\right)^{-F_{qq}(b_T^2,\mu)}\sum_{i=1}^{n_q}{B_{q_i/N_1}(\xi_1,b_T,\mu)B_{\bar{q}_i/N_2}(\xi_2,b_T,\mu)}\nonumber\\
&&\times{\rm Tr}_c\left[\bm{H}^{\mathrm{X}}_{q_i\bar{q}_i}(M,\mu)\bm{S}^{\mathrm{X}}_{\perp,q_i\bar{q}_i}(\harpoon{b}_T,\mu)\right]\nonumber\\
&&+\left(\frac{b_T^2 M^2}{b_0^2}\right)^{-F_{qq}(b_T^2,\mu)}\sum_{i=1}^{n_q}{B_{\bar{q}_i/N_1}(\xi_1,b_T,\mu)B_{q_i/N_2}(\xi_2,b_T,\mu)}\nonumber\\
&&\times{\rm Tr}_c\left[\bm{H}^{\mathrm{X}}_{\bar{q}_iq_i}(M,\mu)\bm{S}^{\mathrm{X}}_{\perp,\bar{q}_iq_i}(\harpoon{b}_T,\mu)\right]\,.\label{eq:Wgeneraldef}
\end{eqnarray}
Here, ${\rm Tr}_c$ denotes the trace over color space, $b_0=2e^{-\gamma_E}$, with $\gamma_E$ the Euler-Mascheroni constant, and $b_T\equiv\left|\harpoon{b}_T\right|$. The index $i$ runs from $1$ to the number of massless quark flavors $n_q$. The exponential factors involving $F_{gg}()$ or $F_{qq}()$ in eq.~\eqref{eq:Wgeneraldef} arise from the collinear anomaly~\cite{Becher:2010tm}. The functions $B_{g/N}()$ and $B_{q/N}()$ are the TMD gluon and quark beam functions, respectively. The hard and soft functions, $\bm{H}^{\mathrm{X}}_{\Ione\Itwo}()$ and $\bm{S}^{\mathrm{X}}_{\perp,\Ione\Itwo}()$, are matrices in color space spanned by a complete basis of the Born-level partonic subprocess, where $\Ione$ and $\Itwo$ are identities of the two initial-state partons. While the hard functions generally depend on both the spin and color of the $Q\bar{Q}$ pair, the soft functions only depend on the color representation $C$, and are insensitive to the spin $j$.

In general, we can assume a complete and orthogonal color basis $|C_l\rangle$, with $l=1,\ldots,n_c$, satisfying 
\begin{equation}
\langle C_{l^\prime}|C_l\rangle=\delta_{l^\prime, l}\langle C_l|C_l\rangle,\quad \bm{1}=\sum_{l=1}^{n_c}{\frac{|C_l\rangle \langle C_l|}{\langle C_l|C_l\rangle}}\,.
\end{equation}
The choice of color basis is not unique and depends on the specific partonic subprocess under consideration. For all subprocesses relevant to this work, we can, for instance, choose the following particular bases:
\begin{itemize}
\item For $q(c_1)\bar{q}(c_2)\to \left(Q\bar{Q}\right)_{1}$, $g(c_1)g(c_2)\to \left(Q\bar{Q}\right)_{1}$, $q(c_1)\bar{q}(c_2)\to \left(Q\bar{Q}\right)[^{(2j+1)}S_j^{[1]}]$, or $g(c_1)g(c_2)\to \left(Q\bar{Q}\right)[^{(2j+1)}S_j^{[1]}]$:
\begin{equation}
n_c=1\,\quad \left|C_1\right\rangle = \delta_{c_1,c_2}\,;\label{eq:color1}
\end{equation}
\item For $q(c_1)\bar{q}(c_2)\to \left(Q\bar{Q}\right)_{8}(c_{Q\bar{Q}})$ or $q(c_1)\bar{q}(c_2)\to \left(Q\bar{Q}\right)[^{(2j+1)}S_j^{[8]}](c_{Q\bar{Q}})$:
\begin{equation}
n_c=1\,,\quad \left|C_1\right\rangle = t^{c_{Q\bar{Q}}}_{c_2c_1}\,;\label{eq:colorqq8}
\end{equation}
\item For $g(c_1)g(c_2)\to \left(Q\bar{Q}\right)_{8}(c_{Q\bar{Q}})$ or $g(c_1)g(c_2)\to \left(Q\bar{Q}\right)[^{(2j+1)}S_j^{[8]}](c_{Q\bar{Q}})$:
\begin{equation}
n_c=2\,,\quad \left|C_1\right\rangle = if^{c_1c_2c_{Q\bar{Q}}}\,,\quad \left|C_2\right\rangle = d^{c_1c_2c_{Q\bar{Q}}}\,.\label{eq:colorgg8}
\end{equation}
\end{itemize}
Some partonic channels, such as $q\bar{q}\to \left(Q\bar{Q}\right)_{1}, \left(Q\bar{Q}\right)[^{(2j+1)}S_j^{[1]}]$, and $gg\to \left(Q\bar{Q}\right)[^{3}S_1^{[8]}]$~\footnote{The Landau-Yang theorem~\cite{Landau:1948kw,Yang:1950rg} applies to this process only at tree level, but not at one-loop level due to color effects, as also discussed in ref.~\cite{Cacciari:2015ela}.}, are loop-induced in QCD. In contrast, the subprocess $q\bar{q}\to \left(Q\bar{Q}\right)[^{1}S_0^{[8]}]$ vanishes in the limit of massless initial-state quarks. We emphasize that our final NNLO soft function is general and does not depend on the particular choice of color basis.

Let us denote the $\ell$-loop ultraviolet (UV)-renormalized amplitude after color and/or spin projection~\footnote{See section 2 in ref.~\cite{AH:2024ueu}.} as $\mathcal{A}^{\mathrm{X},(\ell)}$ for the quark-antiquark channel, and as $\varepsilon_{\lambda_1}^\mu(p_1)\varepsilon_{\lambda_2}^\nu(p_2)\mathcal{A}^{\mathrm{X},(\ell)}_{\mu\nu}$ for the gluon-fusion channel, where $\varepsilon_{\lambda}^\mu(p)$ denotes the external polarization vector of a gluon with momentum $p$ and helicity $\lambda$. The $(l,l^\prime)$ element of the $\ell$-loop hard function in color space is given by
\begin{equation}
\begin{aligned}
\bm{H}^{\mathrm{X},(\ell)}_{q\bar{q},ll^\prime}(M,\mu)=&\frac{1}{\omega(q)\omega(\bar{q})}\frac{1}{2M^2}\sum_{i=0}^{\ell}{\frac{1}{1+\delta_{i,\ell-i}}\frac{\langle C_l|\mathcal{A}^{\mathrm{X},(i)}\rangle \langle \mathcal{A}^{\mathrm{X},(\ell-i)}|C_{l^\prime}\rangle+\langle C_l|\mathcal{A}^{\mathrm{X},(\ell-i)}\rangle \langle \mathcal{A}^{\mathrm{X},(i)}|C_{l^\prime}\rangle}{\langle C_{l^\prime}|C_{l^\prime}\rangle\langle C_{l}|C_{l}\rangle}}\,\label{eq:HF4qq}
\end{aligned}
\end{equation}
for the quark-antiquark channel, and
\begin{equation}
\begin{aligned}
\bm{H}^{\mathrm{X},(\ell)}_{gg,\alpha\beta\rho\sigma,ll^\prime}(M,\mu)=&\frac{1}{\omega(g)\omega(g)}\frac{1}{2M^2}\sum_{i=0}^{\ell}{\frac{1}{1+\delta_{i,\ell-i}}\frac{\langle C_l|\mathcal{A}^{\mathrm{X},(i)}_{\alpha\beta}\rangle \langle \mathcal{A}^{\mathrm{X},(\ell-i)}_{\rho\sigma}|C_{l^\prime}\rangle+\langle C_l|\mathcal{A}^{\mathrm{X},(\ell-i)}_{\alpha\beta}\rangle \langle \mathcal{A}^{\mathrm{X},(i)}_{\rho\sigma}|C_{l^\prime}\rangle}{\langle C_{l^\prime}|C_{l^\prime}\rangle\langle C_{l}|C_{l}\rangle}}\,\label{eq:HF4gg}
\end{aligned}
\end{equation}
for the gluon-fusion channel. Here, $\omega()$ represents the spin and color degrees of freedom of the initial-state particle, given by $\omega(q)=\omega(\bar{q})=2N_c=6$ and $\omega(g)=(N_c^2-1)2(1-\epsilon)=16(1-\epsilon)$ in $d=4-2\epsilon$ dimensions, following conventional dimensional regularization (CDR)~\cite{Collins_1984}. The spin sums over external particles in eqs.\eqref{eq:HF4qq} and \eqref{eq:HF4gg} have been implicitly performed at the amplitude-squared level.

The TMD soft function, whose operator definition will be given in section \ref{sec:softfundef}, can be expanded in powers of $\alpha_s$ as
\begin{equation}
\bm{S}^{\mathrm{X}}_{\perp,\Ione\Itwo}(\harpoon{b}_T,\mu)=\sum_{\ell=0}^{+\infty}{\bm{S}_{\perp,\Ione\Itwo}^{\mathrm{X},(\ell)}(\harpoon{b}_T,\mu)}\,,\label{eq:softfunctionexpandinas}
\end{equation}
where we do not explicitly factor out powers of $\alpha_s$; each term $\bm{S}_{\perp,\Ione\Itwo}^{\mathrm{X},(\ell)}(\harpoon{b}_T,\mu)$ implicitly carries an overall factor of $\alpha_s^\ell$. At LO, the TMD soft function is the identity in color space
\begin{equation}
\bm{S}^{\mathrm{X},(0)}_{\perp,\Ione\Itwo}(\harpoon{b}_T,\mu)=\bm{1}\,,
\end{equation}
and its $(l^\prime, l)$ element is simply a contraction of color basis vectors:
\begin{equation}
\bm{S}^{\mathrm{X},(0)}_{\perp,\Ione\Itwo,l^\prime l}(\harpoon{b}_T,\mu)=\langle C_{l^\prime}|\bm{1}|C_l\rangle=\langle C_{l^\prime}|C_l\rangle=\delta_{l^\prime,l}\langle C_{l}|C_l\rangle\,.\label{eq:SFLO}
\end{equation}
The NLO TMD soft function can be directly extracted from the general results of ref.~\cite{Shao:2025qgv}, using the analytic rapidity regulator~\cite{Becher:2010tm,Becher:2011dz}. After subtraction of infrared (IR) poles in the $\msbar$ scheme, the results are:
\begin{subequations}
  \label{eq:NLOSfun}
  \begin{align}
    \label{eq:NLOSfun1}
    \bm{S}^{[1],(1)}_{\perp,\Ione\Itwo}(\harpoon{b}_T,\mu) &= \bm{0}\,,
    \\[0.3em]
    \label{eq:NLOSfun2}
    \bm{S}^{[8],(1)}_{\perp,q\bar{q}}(\harpoon{b}_T,\mu) &= \frac{\alpha_s}{2\pi}C_AL_\perp\bm{S}^{[8],(0)}_{\perp,q\bar{q}}(\harpoon{b}_T,\mu)\,,
    \\[0.3em]
     \label{eq:NLOSfun3}
    \bm{S}^{[8],(1)}_{\perp,gg}(\harpoon{b}_T,\mu) &= \frac{\alpha_s}{2\pi}C_AL_\perp\bm{S}^{[8],(0)}_{\perp,gg}(\harpoon{b}_T,\mu)\,,   
  \end{align}
\end{subequations}
where
\begin{equation}
L_\perp \equiv \log{\left(\frac{b_T^2\mu^2}{b_0^2}\right)}\,.
\end{equation}
We emphasize that although the matrix elements of the soft function depend on the choice of color basis, the ratios of higher-order soft functions to their LO counterparts are basis-independent in our specific case. Moreover, in the special $2\to 1$ kinematics considered in this paper, the azimuthal-angle-averaged soft function,
\begin{equation}
\bm{S}^{\mathrm{X},(\ell)}_{\perp,\Ione\Itwo}(b_T,\mu)\equiv \int{\frac{d\Omega_b^{(d-2)}}{\Omega_b^{(d-2)}}\bm{S}^{\mathrm{X},(\ell)}_{\perp,\Ione\Itwo}(\harpoon{b}_T,\mu)}\,,\label{eq:Sphiavgatorderl}
\end{equation}
is in fact identical to the full azimuthal-angle-dependent TMD soft function. The azimuthal-angle integration measure and its normalization are given by
\begin{eqnarray}
d\Omega_b^{(d-2)}&=&\prod_{i=1}^{d-3}{d\varphi_i\left(\sin{\varphi_i}\right)^{d-3-i}}\,\\
\Omega_b^{(d-2)}&=&\int{d\Omega_b^{(d-2)}}=\frac{2\pi^{\frac{d-2}{2}}}{\Gamma\left(\frac{d-2}{2}\right)}\,.\label{eq:solidangleint}
\end{eqnarray}

At transverse separation $b_T\ll\Lambda^{-1}_{\mathrm{QCD}}$, the TMD beam function can be matched onto the standard collinear parton distribution function (PDF) via OPE~\cite{Collins:1981uk,Collins:1981uw,Collins:1984kg}:
\begin{eqnarray}
B_{g/N}^{\alpha\beta}(\xi,\harpoon{b}_T,\mu)&=&
\sum_{\ident}{\int_{\xi}^1{\frac{dz}{z}\Bigg[\frac{g^{\alpha\beta}_{\perp}}{d-2}I_{g\ident}\left(z,L_\perp\right)+\left(\frac{g^{\alpha\beta}_{\perp}}{d-2}+\frac{\harpoon{b}_T^\alpha \harpoon{b}_T^\beta}{b_T^2}\right)I_{g\ident}^\prime\left(z,L_\perp\right)\Bigg]f^{(N)}_{\ident}\left(\frac{\xi}{z},\mu^2\right)}}\nonumber\\
&&+\mathcal{O}\left(b_T^2\Lambda_{{\rm QCD}}^2\right)\,,\nonumber\\
B_{q/N}(\xi,b_T,\mu)&=&\sum_{\ident}{\int_{\xi}^{1}{\frac{dz}{z}I_{q\ident}\left(z,L_\perp\right)f^{(N)}_{\ident}\left(\frac{\xi}{z},\mu^2\right)}}+\mathcal{O}\left(b_T^2\Lambda_{{\rm QCD}}^2\right)\,,
\end{eqnarray}
where $f^{(N)}_{\ident}()$ is the $\msbar$-renormalized PDF of parton $\ident$ inside the hadron $N$. The matching coefficients $I_{\ident_1\ident_2}()$ and $I_{g\ident}^\prime()$ encode the perturbative transition from collinear PDFs to the TMD beam functions. At LO, they are given by
\begin{eqnarray}
I_{\ident_1\ident_2}\left(z,L_\perp\right)&=&\delta\left(1-z\right)\delta_{\ident_1,\ident_2}+\mathcal{O}\left(\alpha_s\right),\quad I^\prime_{g\ident}\left(z,L_\perp\right)=\mathcal{O}(\alpha_s).\label{eq:TMDkernelsLO}
\end{eqnarray}
In the gluon TMD beam function, there are both helicity-conserving ($I_{g\ident}()$) and helicity-flip ($I_{g\ident}^\prime()$) components. The azimuthal-angle dependence arises solely from the helicity-flip Lorentz structure multiplying $I_{g\ident}^\prime()$. These matching coefficients have been computed to higher orders in $\alpha_s$ in the literature. The quark kernels $I_{q\ident}()$ are known at NLO~\cite{Becher:2010tm}, NNLO~\cite{Gehrmann:2012ze,Catani:2012qa,Echevarria:2016scs,Luo:2019hmp}, and next-to-NNLO (N$^3$LO)~\cite{Luo:2019szz,Ebert:2020yqt}. The helicity-conserving gluon kernels $I_{g\ident}()$ have been obtained at NLO~\cite{Becher:2012qa,Chiu:2012ir,Becher:2012yn}, NNLO~\cite{Catani:2011kr,Gehrmann:2014yya,Echevarria:2016scs,Luo:2019bmw}, and N$^3$LO~\cite{Ebert:2020yqt,Luo:2020epw}. The helicity-flip kernels $I_{g\ident}^\prime()$, however, are currently only known up to $\mathcal{O}(\alpha_s^2)$~\cite{Catani:2010pd,Becher:2012yn,Gutierrez-Reyes:2019rug,Luo:2019bmw}.

\section{TMD soft function}\label{sec:soft_function}

\subsection{Operator definition\label{sec:softfundef}}

The operator definition of the soft function in impact-parameter space, which enters the factorization formula in eq.\eqref{eq:Wgeneraldef}, reads:
\begin{equation}\label{eq:wsoft}
  \bm{S}_{\perp,\Ione\Itwo}^{[C]}(\harpoon{b}_\perp,\mu) =  
  \langle 0| \bm{\bar{T}} [\bm{O}_{\Ione\Itwo,s}^{[C],\dagger} (\harpoon{b}_\perp)] \bm{T} [\bm{O}^{[C]}_{\Ione\Itwo,s}(\harpoon{0})] |0 \rangle\,,
\end{equation}
where $\bm{T}$ and $\bm{\bar{T}}$ denote time and anti-time ordering operators~\cite{Dyson:1949ha,Schwinger:1960qe,Keldysh:1964ud}, respectively. In eq.\eqref{eq:wsoft}, $\bm{O}^{[C]}_{\Ione\Itwo,s}(\harpoon{b}_T)$ is a Wilson loop operator defined as:
\begin{align}
\bm{O}^{[1]}_{\Ione\Itwo,s}(\harpoon{b}_T) =& \bm{W}_{\Ione}(\harpoon{b}_T)\bm{W}_{\Itwo}(\harpoon{b}_T)\,,\\
\bm{O}^{[8]}_{\Ione\Itwo,s}(\harpoon{b}_T) =& \bm{W}_{\Ione}(\harpoon{b}_T)\bm{W}_{\Itwo}(\harpoon{b}_T)\bm{W}_{(Q\bar{Q})_8}(\harpoon{b}_T)\,,
\end{align}
where the soft Wilson line is given by
\begin{align}
\bm{W}_{\ident}(\harpoon{b}_T) = \mathcal{P}\exp\left[ig_s\int_{-\infty}^0dt\, v\cdot A^a\left(b_\perp+tv\right)\bm{\mathrm{T}}^a(\ident)\right] \,,
\end{align}
with $\mathcal{P}$ denoting the path ordering operator~\cite{Wilson:1974sk}. Here, $b_\perp=(0,\harpoon{b}_T,0)$ and $v$ is a lightlike or timelike four-vector aligned with the momentum of parton $\ident$. The strong coupling is defined as $g_s=\sqrt{4\pi\alpha_s}$. The boldface $\bm{\mathrm{T}}^a(\ident)$ represents the color generator associated with the parton $\ident$:
\begin{align}\label{SUN}
\bm{\mathrm{T}}^a(\ident) = \left\{\begin{array}{cc}t^a & \quad \ident \in \irrep{3}\\
-t^{aT} & \quad \ident \in \irrepbar{3}\\
T^a & \quad \ident \in \irrep{8}\end{array}\right.\,, 
\end{align}
where $t^a$ and $T^a$ are the SU(3) generators in the fundamental and adjoint representations, respectively. We take final-state quarks or initial-state antiquarks to be in the $\irrep{3}$, while initial-state quarks and final-state antiquarks belong to $\irrepbar{3}$. For the adjoint representation, the generators are given by $T^a_{bc}=-if^{abc}$. In the setup considered in this work, we choose $v = n$ for the first initial-state parton and $v=\bar{n}$ for the second one. For the heavy quark pair in the color-octet state, we choose its four-vector $v$ such that $v^2=1$, as the soft Wilson line is invariant under the rescaling $v \to \lambda v$. In the rest frame of the $Q\bar{Q}$ system, a convenient choice is
\begin{equation}
v=\frac{n+\bar{n}}{2}=\left(1,\harpoon{0},0\right)\,.
\end{equation}
By inserting a complete set of final states $X$ into eq.~\eqref{eq:wsoft}, we obtain
\begin{align}
  \bm{S}^{[C]}_{\perp,\Ione\Itwo}(\harpoon{b}_T,\mu) 
  &= 
  \underset{\!\!X}{\int\kern-1.5em\sum} \,
  \langle 0| \bm{\bar T} [\bm{O}_{\Ione\Itwo,s}^{[C],\dagger} (\harpoon{b}_T)] | X \rangle
  \langle X | \bm{T} [\bm{O}_{\Ione\Itwo,s}^{[C]}(\harpoon{0})] |0 \rangle \nonumber \\
  & =
  \underset{\!\!X}{\int\kern-1.5em\sum} e^{-i \harpoon{k}_{X,T} \cdot
  \harpoon{b}_T}\,
  \langle 0| \bm{\bar T} [\bm{O}_{\Ione\Itwo,s}^{[C],\dagger} (\harpoon{0})] | X \rangle
  \langle X | \bm{T} [\bm{O}_{\Ione\Itwo,s}^{[C]}(\harpoon{0})] |0 \rangle \,,
  \label{eq:wsoft2}
\end{align}
where $\harpoon{k}_{X,T}$ is the total transverse momentum carried by the final state~$X$. As in eq.\eqref{eq:Sphiavgatorderl}, the azimuthal-angle-averaged soft function in impact-parameter space is defined as
\begin{equation}
\bm{S}^{[C]}_{\perp,\Ione\Itwo}(b_T,\mu)=\int{\frac{d\Omega_b^{(d-2)}}{\Omega_b^{(d-2)}}\bm{S}^{[C]}_{\perp,\Ione\Itwo}(\harpoon{b}_T,\mu)}\,.
\end{equation}
As pointed out in section \ref{sec:formula}, for our purposes, the azimuthal-angle-dependent soft function is equal to the azimuthal-angle-averaged one to all orders in $\alpha_s$, \ie\
\begin{equation}
\bm{S}^{[C]}_{\perp,\Ione\Itwo}(b_T,\mu)=\bm{S}^{[C]}_{\perp,\Ione\Itwo}(\harpoon{b}_T,\mu)\,.\label{eq:SavgeqSphi}
\end{equation}
The TMD soft function in momentum space can be obtained via Fourier transform:
\begin{equation}
\bm{S}^{[C]}_{\perp,\Ione\Itwo}(\harpoon{q}_T,\mu)=\int{\frac{d^{d-2}\harpoon{b}_T}{(2\pi)^{d-2}}e^{i\harpoon{q}_T\cdot\harpoon{b}_T}\bm{S}^{[C]}_{\perp,\Ione\Itwo}(\harpoon{b}_T,\mu)}\,,\label{eq:SqTinSb}
\end{equation}
and the azimuthal-angle-averaged version is
\begin{equation}
\begin{aligned}
\bm{S}^{[C]}_{\perp,\Ione\Itwo}(q_T,\mu)=&\int{\frac{d\Omega_q^{(d-2)}}{\Omega_{q}^{(d-2)}}\bm{S}^{[C]}_{\perp,\Ione\Itwo}(\harpoon{q}_T,\mu)}\\
=&\int{\frac{d\Omega_q^{(d-2)}}{\Omega_{q}^{(d-2)}}\int{\frac{d^{d-2}\harpoon{b}_T}{(2\pi)^{d-2}}e^{i\harpoon{q}_T\cdot\harpoon{b}_T}\bm{S}^{[C]}_{\perp,\Ione\Itwo}(\harpoon{b}_T,\mu)}}\\
=&~\underset{\!\!X}{\int\kern-1.5em\sum}\int{\frac{d\Omega_q^{(d-2)}}{\Omega_{q}^{(d-2)}}\int{\frac{d^{d-2}\harpoon{b}_T}{(2\pi)^{d-2}} e^{i (\harpoon{q}_T-\harpoon{k}_{X,T}) \cdot
  \harpoon{b}_T}\,
  \langle 0| \bm{\bar T} [\bm{O}_{\Ione\Itwo,s}^{[C],\dagger} (\harpoon{0})] | X \rangle
  \langle X | \bm{T} [\bm{O}_{\Ione\Itwo,s}^{[C]}(\harpoon{0})] |0 \rangle}}\\
  =&~\underset{\!\!X}{\int\kern-1.5em\sum}\int{\frac{d\Omega_q^{(d-2)}}{\Omega_{q}^{(d-2)}}\delta^{(d-2)}\left(\harpoon{q}_T-\harpoon{k}_{X,T}\right)\langle 0| \bm{\bar T} [\bm{O}_{\Ione\Itwo,s}^{[C],\dagger} (\harpoon{0})] | X \rangle
  \langle X | \bm{T} [\bm{O}_{\Ione\Itwo,s}^{[C]}(\harpoon{0})] |0 \rangle}\\
  =&\frac{1}{\Omega_q^{(d-2)} q_T^{d-3}}~\underset{\!\!X}{\int\kern-1.5em\sum}\delta\left(q_T-k_{X,T}\right)\langle 0| \bm{\bar T} [\bm{O}_{\Ione\Itwo,s}^{[C],\dagger} (\harpoon{0})] | X \rangle
  \langle X | \bm{T} [\bm{O}_{\Ione\Itwo,s}^{[C]}(\harpoon{0})] |0 \rangle\,,\label{eq:wsoft3}
\end{aligned}
\end{equation}
where $q_T$ and $k_{X,T}$ denote the magnitudes of the vectors $\harpoon{q}_T$ and $\harpoon{k}_{X,T}$, respectively. The expression for $\Omega_q^{(d-2)}$ is the same as that for $\Omega_b^{(d-2)}$ given in eq.~\eqref{eq:solidangleint}. In our case of interest, since the matrix element does not depend on the azimuthal angle of $\harpoon{q}_T$, we have directly integrated over the azimuthal angle $d\Omega_q^{(d-2)}$ in the last line of eq.~\eqref{eq:wsoft3} using the identity
\begin{equation}
\int{d\Omega_q^{(d-2)}\delta^{(d-2)}(\harpoon{q}_T-\harpoon{k}_{T,X})}=\frac{1}{q_T^{d-3}}\delta(q_T-k_{X,T})\,.
\end{equation}
Since the soft functions in impact-parameter and momentum spaces are equivalent, as shown in eq.~\eqref{eq:SqTinSb}, we are free to work in either representation. In this work, we choose to perform our calculations in momentum space. At LO, the soft functions are given by
\begin{eqnarray}
\bm{S}_{\perp,\Ione\Itwo}^{[C],(0)}(\harpoon{q}_T,\mu)&=&\bm{1}\delta^{(d-2)}(\harpoon{q}_T)\,,\\
\bm{S}_{\perp,\Ione\Itwo}^{[C],(0)}(q_T,\mu)&=&\bm{1}\frac{q_T^{3-d}}{\Omega_q^{(d-2)}}\delta(q_T)\,,
\end{eqnarray}
where $\bm{1}$ denotes the identity in color space.

\subsection{NNLO soft function}
The NNLO soft function receives contributions from double-real (RR), real-virtual (RV), double-virtual (VV), and one-loop $\mathcal{O}(\epsilon)$ radiative corrections. Among these, the double-virtual correction vanishes in dimensional regularization because it involves only scaleless Feynman integrals. The one-loop $\mathcal{O}(\epsilon)$ terms can be directly extracted from the general results provided in ref.~\cite{Shao:2025qgv}. Therefore, in this section, we focus on the computation of the double-real and real-virtual contributions. 

\subsubsection{General structure of the soft amplitude and soft function\label{sec:generalstructure}}

In this section, we demonstrate that the two-loop bare soft function, regardless of its specific form, can be expressed in a general structure. This structure is sufficiently simple to be worked out explicitly.

\begin{figure}[h!]
\subfigure[{\large $\mathrm{R}$}\label{fig:softampR}]{
\begin{tikzpicture}[line width=1 pt, scale=0.9]
  \hspace{0cm}
\draw[wilson] (-2,0.0) -- (0.0,2.0);
\draw[wilson] (-2.0,0.0) -- (0.0,-2.0);
\draw[gluon] (-1.0,-1.0) -- (1.0,-1.0);
\node at (-1.0,-2.0) {\Large $\ident_i, p_i$};
\node at (0.5,-0.5) {\Large $a,k_1$};
\end{tikzpicture}
}
\subfigure[{\large $\mathrm{RR}_1$}\label{fig:softampRR1}]
{
\begin{tikzpicture}[line width=1 pt, scale=0.9]
\hspace{0cm}
\draw[wilson] (-2,0.0) -- (0.0,2.0);
\draw[wilson] (-2,0.0) -- (0.0,-2.0);
\draw[gluon] (-1.0,-1.0) -- (0.7,-1.0);
\draw[fermion] (0.7,-1.0) -- (1.5,-0.2);
\draw[fermionbar] (0.7,-1.0) -- (1.5,-1.8);
\filldraw [red] (0.7,-1.0) circle (3pt);
\node at (-1.0,-2.0) {\Large $\ident_i,p_i$};
\node at (0.0,-0.5) {\Large $c$};
\node at (1.0,-0.0) {\Large $a,k_1$};
\node at (1.0,-2.0) {\Large $b,k_2$};
\end{tikzpicture}
}
\subfigure[{\large $\mathrm{RR}_2$}\label{fig:softampRR2}]{
\begin{tikzpicture}[line width=1 pt, scale=0.9]
\hspace{0cm}
\draw[wilson] (-2,0.0) -- (0.0,2.0);
\draw[wilson] (-2,0.0) -- (0.0,-2.0);
\draw[gluon] (-1.0,-1.0) -- (0.7,-1.0);
\draw[ghost] (0.7,-1.0) -- (1.5,-0.2);
\draw[ghostbar] (0.7,-1.0) -- (1.5,-1.8);
\filldraw [red] (0.7,-1.0) circle (3pt);
\node at (-1.0,-2.0) {\Large $\ident_i,p_i$};
\node at (0.0,-0.5) {\Large $c$};
\node at (1.0,-0.0) {\Large $a,k_1$};
\node at (1.0,-2.0) {\Large $b,k_2$};
\end{tikzpicture}
}
\subfigure[{\large $\mathrm{RR}_3$}\label{fig:softampRR3}]{
\begin{tikzpicture}[line width=1 pt, scale=0.9]
\hspace{0cm}
\draw[wilson] (-2,0.0) -- (0.0,2.0);
\draw[wilson] (-2,0.0) -- (0.0,-2.0);
\draw[gluon] (-1.0,-1.0) -- (0.7,-1.0);
\draw[gluon] (0.7,-1.0) -- (1.5,-0.2);
\draw[gluon] (0.7,-1.0) -- (1.5,-1.8);
\filldraw [red] (0.7,-1.0) circle (3pt);
\node at (-1.0,-2.0) {\Large $\ident_i,p_i$};
\node at (0.0,-0.5) {\Large $c$};
\node at (1.0,-0.0) {\Large $a,k_1$};
\node at (1.0,-2.0) {\Large $b,k_2$};
\end{tikzpicture}
}\\
\subfigure[{\large $\mathrm{RR}_4$}\label{fig:softampRR4}]{
\begin{tikzpicture}[line width=1 pt, scale=0.9]
  \hspace{0cm}
\draw[wilson] (-2,0.0) -- (0.0,2.0);
\draw[wilson] (-2,0.0) -- (0.0,-2.0);
\draw[gluon] (-1.4,-0.6) -- (1.3,-0.6);
\draw[gluon] (-0.7,-1.3) -- (1.3,-1.3);
\node at (-1.0,-2.0) {\Large $\ident_i,p_i$};
\node at (1.0,-2.0) {\Large $a,k_1$};
\node at (1.0, 0.0) {\Large $b,k_2$};
\end{tikzpicture}
}
\subfigure[{\large $\mathrm{RR}_5$}\label{fig:softampRR5}]{
\begin{tikzpicture}[line width=1 pt, scale=0.9]
\hspace{0cm}
\draw[wilson] (-2,0.0) -- (0.0,2.0);
\draw[wilson] (-2,0.0) -- (0.0,-2.0);
\draw[gluon] (-1.4,-0.6) -- (1.3,-0.6);
\draw[gluon] (-0.7,-1.3) -- (1.3,-1.3);
\node at (-1.0,-2.0) {\Large $\ident_i,p_i$};
\node at (1.0, 0.0) {\Large $a,k_1$};
\node at (1.0,-2.0) {\Large $b,k_2$};
\end{tikzpicture}
}
\subfigure[{\large $\mathrm{RR}_6$}\label{fig:softampRR6}]{
\begin{tikzpicture}[line width=1 pt, scale=0.9]
\hspace{0cm}
\draw[wilson] (-2,0.0) -- (0.0,2.0);
\draw[wilson] (-2,0.0) -- (0.0,-2.0);
\draw[gluon] (-1.0, 1.0) -- (1.3, 1.0);
\draw[gluon] (-1.0,-1.0) -- (1.3,-1.0);
\node at (-1.0,-2.0) {\Large $\ident_i,p_i$};
\node at (-1.0, 2.0) {\Large $\ident_j,p_j$};
\node at (1.0, 1.7) {\Large $b,k_2$};
\node at (1.0,-1.7) {\Large $a,k_1$};
\end{tikzpicture}
}
\subfigure[{\large $\mathrm{RV}_1$}\label{fig:softampRV1}]{
\begin{tikzpicture}[line width=1 pt, scale=0.9]
\hspace{0cm}
\draw[wilson] (-2,0.0) -- (0.0,2.0);
\draw[wilson] (-2,0.0) -- (0.0,-2.0);
\draw[gluon] (-1.0,-1.0) -- (1.3,-1.0);
\draw[gluon, segment length=5pt, bend right,looseness = 1.6] (-1.6,-0.4) to (-0.4,-1.6);
\node at (-1.0,-2.0) {\Large $\ident_i,p_i$};
\node at (1.0,-0.5) {\Large $a,k_1$};
\node at (-1.9,-0.7) {\Large $b,$};
\node at (-1.9,-1.2) {\Large $k_2$};
\end{tikzpicture}
}\\
\subfigure[{\large $\mathrm{RV}_2$}\label{fig:softampRV2}]{
\begin{tikzpicture}[line width=1 pt, scale=0.9]
  \hspace{0cm}
\draw[wilson] (-2,0.0) -- (0.0,2.0);
\draw[wilson] (-2,0.0) -- (0.0,-2.0);
\draw[gluon] (-0.4,-0.8) -- (1.3,-0.8);
\draw[gluon, segment length=5pt, bend left,looseness = 1.6] (-1.6,-0.4) to (-0.4,-1.6);
\filldraw [red] (-0.4,-0.8) circle (3pt);
\node at (-1.0,-2.0) {\Large $\ident_i,p_i$};
\node at (1.0,-0.3) {\Large $a,k_1$};
\node at (-0.5, -0.2) {\Large $c$};
\node at (0.3, -1.3) {\Large $b,k_2$};
\end{tikzpicture}
}
\subfigure[{\large $\mathrm{RV}_3$}\label{fig:softampRV3}]{
\begin{tikzpicture}[line width=1 pt, scale=0.9]
\hspace{0cm}
\draw[wilson] (-2,0.0) -- (0.0,2.0);
\draw[wilson] (-2,0.0) -- (0.0,-2.0);
\draw[gluon] (-0.4,-1.6) -- (1.3,-1.6);
\draw[gluon, segment length=5pt, bend left,looseness = 1.6] (-1.6,-0.4) to (-0.6,-1.4);
\node at (-1.0,-2.0) {\Large $\ident_i,p_i$};
\node at (1.0,-1.1) {\Large $a,k_1$};
\node at (-0.5, -0.2) {\Large $b,k_2$};
\end{tikzpicture}
}
\subfigure[{\large $\mathrm{RV}_4$}\label{fig:softampRV4}]{
\begin{tikzpicture}[line width=1 pt, scale=0.9]
\hspace{0cm}
\draw[wilson] (-2,0.0) -- (0.0,2.0);
\draw[wilson] (-2,0.0) -- (0.0,-2.0);
\draw[gluon] (-0.4,-1.6) -- (1.3,-1.6);
\draw[gluon] (-0.6,-1.4) to (-0.6, 1.4);
\node at (-1.0,-2.0) {\Large $\ident_i,p_i$};
\node at (-1.0, 2.0) {\Large $\ident_j,p_j$};
\node at (1.0,-1.1) {\Large $a,k_1$};
\node at (0.2, -0.2) {\Large $b,k_2$};
\end{tikzpicture}
}
\subfigure[{\large $\mathrm{RV}_5$}\label{fig:softampRV5}]{
\begin{tikzpicture}[line width=1 pt, scale=0.9]
\hspace{0cm}
\draw[wilson] (-2,0.0) -- (0.0,2.0);
\draw[wilson] (-2,0.0) -- (0.0,-2.0);
\draw[gluon] (-0.6, 0.0) -- (1.3,0.0);
\draw[gluon] (-0.6,-1.4) to (-0.6, 1.4);
\filldraw [red] (-0.6, 0.0) circle (3pt);
\node at (-1.0,-2.0) {\Large $\ident_i,p_i$};
\node at (-1.0, 2.0) {\Large $\ident_j,p_j$};
\node at (1.0, -0.5) {\Large $a,k_1$};
\node at (0.2, 0.8) {\Large $c,k_2$};
\node at (-0.2, -0.8) {\Large $b$};
\end{tikzpicture}
}\\
\subfigure[{\large $\mathrm{RV}_6$}\label{fig:softampRV6}]{
\begin{tikzpicture}[line width=1 pt, scale=0.9]
  \hspace{0cm}
\draw[wilson] (-2,0.0) -- (0.0,2.0);
\draw[wilson] (-2,0.0) -- (0.0,-2.0);
\draw[gluon] (-1.6,-0.4) -- (1.3,-0.4);
\draw[gluon] (-0.6,-0.2) to (-0.6, 1.4);
\draw[gluon] (-0.6,-1.4) to (-0.6, -0.4);
\node at (-1.0,-2.0) {\Large $\ident_i,p_i$};
\node at (-1.0, 2.0) {\Large $\ident_j,p_j$};
\node at (1.0,-0.8) {\Large $a,k_1$};
\node at (0.2, 0.4) {\Large $b,k_2$};
\end{tikzpicture}
}
\subfigure[{\large $\mathrm{RV}_7$}\label{fig:softampRV7}]{
\begin{tikzpicture}[line width=1 pt, scale=0.9]
\hspace{0cm}
\draw[wilson] (-2,0.0) -- (0.0,2.0);
\draw[wilson] (-2,0.0) -- (-0.9,0.0);
\draw[wilson] (-0.6,0.0) -- (0.0,0.0);
\draw[wilson] (-2,0.0) -- (0.0,-2.0);
\draw[gluon, segment length=6pt, bend right,looseness = 1.2] (-0.4,0.0) to (1.3,-0.5);
\draw[gluon] (-0.7, -1.3) -- (-0.7,1.3);
\node at (-1.0,-2.0) {\Large $\ident_i,p_i$};
\node at (-1.0, 2.0) {\Large $\ident_j,p_j$};
\node at (0.5, 0.3) {\Large $\ident_k,p_k$};
\node at (1.0,-1.1) {\Large $a,k_1$};
\node at (0.2, 0.9) {\Large $b,k_2$};
\end{tikzpicture}
}
\caption{All possible soft configurations at the amplitude level up to NNLO. Thick black lines represent Wilson lines, and soft partons are shown in red.\label{fig:softamplitude}} 
\end{figure}
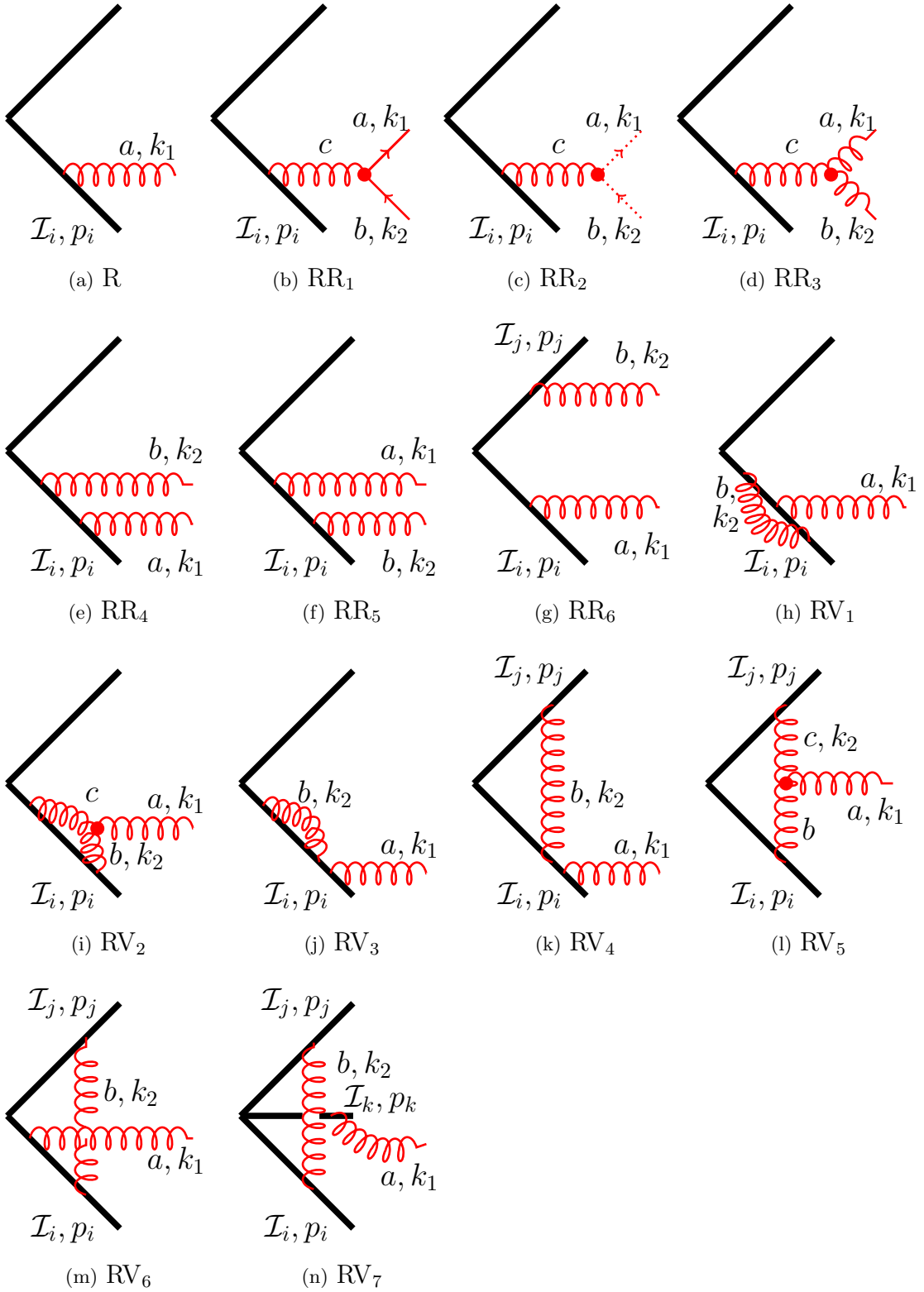

Let us start with the soft amplitudes. The single-real (R) soft amplitude emitted from a Wilson line labeled by $\ident_i$, associated with four-momentum $p_i$ (cf. figure \ref{fig:softampR}), is well known:
\begin{equation}
\bm{A}_{S,i}^{(\mathrm{R})}=g_s\frac{p_i\cdot \varepsilon_{\lambda_1}^*(k_1)}{p_i\cdot k_1}\bm{\mathrm{T}}^a(\ident_i)\,.\label{eq:ARi}
\end{equation}
Summing over all external colored Wilson lines yields the total single-real soft amplitude:
\begin{equation}
\begin{aligned}
\bm{A}_{S}^{(\mathrm{R})}=&\sum_{i}{\bm{A}_{S,i}^{(\mathrm{R})}}\,.\label{eq:ARsum}
\end{aligned}
\end{equation}

Similarly, the double-real (RR) parton emission amplitudes (cf. figures \ref{fig:softampRR1}-\ref{fig:softampRR6}), computed in Feynman gauge, are given by:
\begin{eqnarray}
\bm{A}_{S,i}^{(\mathrm{RR}_1)}&=&g_s^2\frac{\bar{u}_{\lambda_1}(k_1)\slashed{p}_iv_{\lambda_2}(k_2)}{\left(k_1+k_2\right)^2\left(p_i\cdot (k_1+k_2)\right)}t^c_{ab}\bm{\mathrm{T}}^c(\ident_i)\,,\\
\bm{A}_{S,i}^{(\mathrm{RR}_2)}&=&-g_s^2\frac{p_i\cdot k_1}{\left(k_1+k_2\right)^2\left(p_i\cdot (k_1+k_2)\right)}if^{abc}\bm{\mathrm{T}}^c(\ident_i)\,,\\
\bm{A}_{S,i}^{(\mathrm{RR}_3)}&=&g_s^2\frac{1}{\left(k_1+k_2\right)^2\left(p_i\cdot (k_1+k_2)\right)}\Bigg[p_i\cdot\varepsilon^*_{\lambda_1}(k_1)\left(2k_1+k_2\right)\cdot \varepsilon^*_{\lambda_2}(k_2)-p_i\cdot\varepsilon^*_{\lambda_2}(k_2)\left(k_1+2k_2\right)\cdot \varepsilon^*_{\lambda_1}(k_1)\nonumber\\
&&-p_i\cdot\left(k_1-k_2\right)\varepsilon^*_{\lambda_1}(k_1)\cdot \varepsilon^*_{\lambda_2}(k_2)\Bigg]if^{abc}\bm{\mathrm{T}}^c(\ident_i)\,,\\
\bm{A}_{S,i}^{(\mathrm{RR}_4)}&=&g_s^2\frac{p_i\cdot \varepsilon^*_{\lambda_1}(k_1)p_i\cdot \varepsilon^*_{\lambda_2}(k_2)}{\left(p_i\cdot k_1\right)\left(p_i\cdot (k_1+k_2)\right)}\bm{\mathrm{T}}^a(\ident_i)\bm{\mathrm{T}}^b(\ident_i)\,,\\
\bm{A}_{S,i}^{(\mathrm{RR}_5)}&=&\left.\bm{A}_{S,i}^{(\mathrm{RR}_4)}\right|_{(a,k_1,\lambda_1)\leftrightarrow (b,k_2,\lambda_2)}\,,\\
\bm{A}_{S,ij}^{(\mathrm{RR}_6)}&=&g_s^2\frac{p_i\cdot \varepsilon^*_{\lambda_1}(k_1)p_j\cdot \varepsilon^*_{\lambda_2}(k_2)}{(p_i\cdot k_1)(p_j\cdot k_2)}\bm{\mathrm{T}}^b(\ident_j)\bm{\mathrm{T}}^a(\ident_i)\,.
\end{eqnarray}
Here, $\bar{u}_{\lambda_1}(k_1)$ and $v_{\lambda_2}(k_2)$ are Dirac spinors for the outgoing quark and antiquark, respectively. After summing over all possible Wilson lines, the total double-real soft amplitudes consist of contributions from quark–antiquark, ghost, and gluon final states:
\begin{eqnarray}
\bm{A}_{S}^{(\mathrm{RR},q)}&=&\sum_{i}{\bm{A}_{S,i}^{(\mathrm{RR}_1)}}\,,\\
\bm{A}_{S}^{(\mathrm{RR},\mathrm{ghost})}&=&\sum_{i}{\bm{A}_{S,i}^{(\mathrm{RR}_2)}}\,,\\
\bm{A}_{S}^{(\mathrm{RR},g)}&=&\sum_{i}{\left(\bm{A}_{S,i}^{(\mathrm{RR}_3)}+\bm{A}_{S,i}^{(\mathrm{RR}_4)}+\bm{A}_{S,i}^{(\mathrm{RR}_5)}\right)}+\mathop{\sum_{i,j}}_{i\neq j}{\bm{A}_{S,ij}^{(\mathrm{RR}_6)}}\,.
\end{eqnarray}
The real-virtual contributions (cf. figures \ref{fig:softampRV1}-\ref{fig:softampRV7}) are given by:
\begin{eqnarray}
\bm{A}_{S,i}^{(\mathrm{RV}_1)}&=&-ig_s^3\frac{p_i^2p_i\cdot \varepsilon^*_{\lambda_1}(k_1)}{p_i\cdot k_1}\Bigg[\int{[dk_2]\frac{1}{k_2^2\left(k_2\cdot p_i+i\sigma_i 0^+\right)\left(\left(k_2-k_1\right)\cdot p_i+i\sigma_i 0^+\right)}}\Bigg]\nonumber\\
&&\times\underbrace{\bm{\mathrm{T}}^b(\ident_i)\bm{\mathrm{T}}^a(\ident_i)\bm{\mathrm{T}}^b(\ident_i)}_{=(C(\ident_i)-C_A/2)\bm{\mathrm{T}}^a(\ident_i)}\,,\\
\bm{A}_{S,i}^{(\mathrm{RV}_2)}&=&-g_s^3\frac{1}{p_i\cdot k_1}\Bigg[\int{[dk_2]\frac{\left(2k_2\cdot p_i-k_1\cdot p_i\right)p_i\cdot \varepsilon^*_{\lambda_1}(k_1)+p_i^2\left(k_1\cdot \varepsilon^*_{\lambda_1}(k_1)-2k_2\cdot \varepsilon^*_{\lambda_1}(k_1)\right)}{k_2^2\left(k_2\cdot p_i-i\sigma_i0^+\right)\left(k_2-k_1\right)^2}}\Bigg]\nonumber\\
&&\times \underbrace{f^{abc}\bm{\mathrm{T}}^b(\ident_i)\bm{\mathrm{T}}^c(\ident_i)}_{=i\frac{C_A}{2}\bm{\mathrm{T}}^a(\ident_i)}\,,\\
\bm{A}_{S,i}^{(\mathrm{RV}_3)}&=&-ig_s^3\frac{p_i^2p_i\cdot \varepsilon^*_{\lambda_1}(k_1)}{\left(p_i\cdot k_1\right)^2}\Bigg[\int{[dk_2]\frac{1}{k_2^2\left(\left(k_2+k_1\right)\cdot p_i-i\sigma_i0^+\right)}}\Bigg]\underbrace{\bm{\mathrm{T}}^a(\ident_i)\bm{\mathrm{T}}^b(\ident_i)\bm{\mathrm{T}}^b(\ident_i)}_{=C(\ident_i)\bm{\mathrm{T}}^a(\ident_i)}\,,\\
\bm{A}_{S,ij}^{(\mathrm{RV}_4)}&=&ig_s^3\frac{p_i\cdot p_j p_i\cdot \varepsilon^*_{\lambda_1}(k_1)}{p_i\cdot k_1}\Bigg[\int{[dk_2]\frac{1}{k_2^2\left(k_2\cdot p_j+i\sigma_j0^+\right)\left(\left(k_2+k_1\right)\cdot p_i-i\sigma_i0^+\right)}}\Bigg]\nonumber\\
&&\times\bm{\mathrm{T}}^a(\ident_i)\bm{\mathrm{T}}^b(\ident_i)\bm{\mathrm{T}}^b(\ident_j)\,,\\
\bm{A}_{S,ij}^{(\mathrm{RV}_5)}&=&-g_s^3\Bigg[\int{[dk_2]\frac{\left(k_2\cdot p_j-2k_1\cdot p_j\right)p_i\cdot \varepsilon^*_{\lambda_1}(k_1)+p_i\cdot (k_1+k_2)p_j\cdot \varepsilon^*_{\lambda_1}(k_1)+p_i\cdot p_j \left(k_1-2k_2\right)\cdot \varepsilon^*_{\lambda_1}(k_1)}{k_2^2\left(k_2-k_1\right)^2\left(k_2\cdot p_j-i\sigma_j0^+\right)\left(\left(k_2-k_1\right)\cdot p_i+i\sigma_i0^+\right)}}\Bigg]\nonumber\\
&&\times f^{abc}\bm{\mathrm{T}}^b(\ident_i)\bm{\mathrm{T}}^c(\ident_j)\,,\\
\bm{A}_{S,ij}^{(\mathrm{RV}_6)}&=&ig_s^3\Bigg[\int{[dk_2]\frac{p_i\cdot p_j p_i\cdot \varepsilon^*_{\lambda_1}(k_1)}{k_2^2\left(k_2\cdot p_i-i\sigma_i0^+\right)\left(k_2\cdot p_j+i\sigma_j0^+\right)\left((k_2+k_1)\cdot p_i-i\sigma_i0^+\right)}}\Bigg]\nonumber\\
&&\times\bm{\mathrm{T}}^b(\ident_i)\bm{\mathrm{T}}^a(\ident_i)\bm{\mathrm{T}}^b(\ident_j)\,,\\
\bm{A}_{S,ijk}^{(\mathrm{RV}_7)}&=&ig_s^3\frac{p_i\cdot p_jp_k\cdot \varepsilon^*_{\lambda_1}(k_1)}{p_k\cdot k_1}\Bigg[\int{[dk_2]\frac{1}{k_2^2\left(k_2\cdot p_i-i\sigma_i0^+\right)\left(k_2\cdot p_j+i\sigma_j0^+\right)}}\Bigg]\bm{\mathrm{T}}^a(\ident_k)\bm{\mathrm{T}}^b(\ident_i)\bm{\mathrm{T}}^b(\ident_j)=\bm{0}\,.
\end{eqnarray}
We define the loop measure as
\begin{equation}
[dk]\equiv\mu^{4-d}\frac{d^dk}{(2\pi)^d}\,,
\end{equation}
and introduce the sign convention for the infinitesimal imaginary part:
\begin{equation}
\sigma_i=\left\{\begin{array}{cc}+1~& p_i~\mathrm{incoming}\\
-1~& p_i~\mathrm{outgoing}
\end{array}\right.\,.
\end{equation}
The color generator identities used include:
\begin{eqnarray}
\bm{\mathrm{T}}^a(\ident)\bm{\mathrm{T}}^a(\ident)&=&C(\ident)\bm{1}\,,\\
\left[\bm{\mathrm{T}}^a(\ident),\bm{\mathrm{T}}^b(\ident)\right]&=&if^{abc}\bm{\mathrm{T}}^c(\ident)\,,\\
f^{abc}\bm{\mathrm{T}}^b(\ident)\bm{\mathrm{T}}^c(\ident)&=&\frac{1}{2}f^{abc}\left[\bm{\mathrm{T}}^b(\ident),\bm{\mathrm{T}}^c(\ident)\right]=i\frac{C_A}{2}\bm{\mathrm{T}}^a(\ident)\,.
\end{eqnarray}
The QCD Casimir factor $C(\ident)$ depends on the representation of the external particle:
\begin{equation}
\begin{aligned}
C(\ident)&=\left\{\begin{array}{ll}
C_F=(N_c^2-1)/(2N_c)=4/3, &~~{\rm if}~~\ident\in \irrep{3},\irrepbar{3}\\
C_A=N_c=3, &~~{\rm if}~~\ident\in \irrep{8}\\
\end{array}\right..
\end{aligned}
\end{equation}
The full real-virtual soft amplitude is then
\begin{equation}
\begin{aligned}
\bm{A}_{S}^{(\mathrm{RV})}=&\sum_{i}{\left(\bm{A}_{S,i}^{(\mathrm{RV}_1)}+\bm{A}_{S,i}^{(\mathrm{RV}_2)}+\bm{A}_{S,i}^{(\mathrm{RV}_3)}\right)}+\mathop{\sum_{i,j}}_{i\neq j}{\left(\bm{A}_{S,ij}^{(\mathrm{RV}_4)}+\bm{A}_{S,ij}^{(\mathrm{RV}_6)}\right)}\\
&+\mathop{\sum_{i,j}}_{i<j}{\left(\bm{A}_{S,ij}^{(\mathrm{RV}_5)}+\mathop{\sum_{k}}_{k\neq i,j}{\bm{A}_{S,ijk}^{(\mathrm{RV}_7)}}\right)}\,.
\end{aligned}
\end{equation}
Note that the $\bm{A}_{S,ijk}^{(\mathrm{RV}_7)}$ term vanishes, since it corresponds to a scaleless one-loop integral in dimensional regularization. Hence, we omit it in the following discussions. These amplitudes are generated using \feynarts~\cite{Hahn:2000kx}, and cross-checked with \qgraf~\cite{Nogueira:1991ex}.

With the soft amplitudes at hand, the general bare soft function can be expressed in terms of these amplitudes. In particular, the one-loop bare soft function takes the form
\begin{equation}
\bm{S}^{(1)}=\int{[dk_1](2\pi)\delta^{(+)}(k_1^2)K_S(k_1)\bm{A}^{(R)\dagger}_S\bm{A}_S^{(R)}}\,,\label{eq:oneloopsoftfungeneral}
\end{equation}
where the specific form of the kernel function $K_S(k_1)$ depends on the type of soft function considered. For the azimuthal-angle-dependent TMD soft function in impact-parameter space, the kernel is given by
\begin{equation}
K_S(k_1)=e^{-i\harpoon{k}_{1,T}\cdot \harpoon{b}_T}\,,\label{eq:KSfun1}
\end{equation}
while its counterpart in transverse-momentum space ($q_T$ space) is
\begin{equation}
K_S(k_1)=\delta^{(d-2)}\left(\harpoon{q}_T-\harpoon{k}_{1,T}\right)\,.\label{eq:KSfun2}
\end{equation}
For the azimuthal-angle-averaged TMD soft function, the kernel in impact parameter space becomes
\begin{equation}
K_S(k_1)=\int{\frac{d\Omega_b^{(d-2)}}{\Omega_b^{(d-2)}}e^{-i\harpoon{k}_{1,T}\cdot \harpoon{b}_T}}\,\label{eq:KSfun3}
\end{equation}
and in transverse-momentum space, it is
\begin{equation}
K_S(k_1)=\int{\frac{d\Omega_q^{(d-2)}}{\Omega_q^{(d-2)}}\delta^{(d-2)}(\harpoon{q}_T-\harpoon{k}_{1,T})}=\frac{1}{\Omega_q^{(d-2)} q_T^{d-3}}\delta(q_T-k_{1,T})\,.\label{eq:KSfun4}
\end{equation}
In the case of TMD soft functions, a particular type of divergence, known as ``rapidity divergence"~\cite{Collins:2003fm}, appears. These divergences cannot be regularized by dimensional regularization alone. They arise from momentum regions where the invariant mass $k^2$ remains fixed, but the ratio $k_{-}/k_{+}$ or $k_{+}/k_{-}$ diverges. To handle these, a rapidity regulator must be introduced. In this paper, we adopt the analytic rapidity regulator~\cite{Becher:2010tm,Becher:2011dz}, which modifies the kernel function as follows:
\begin{equation}
K_S(k)\quad\to\quad K_S(k)\left(\frac{\bar{n}\cdot k}{n\cdot k}\right)^{\alpha_1}\,,\label{eq:rapidityregulator}
\end{equation}
where $\alpha_1$ is an infinitesimal rapidity regulator. In our specific setup, the final TMD soft function is free of rapidity divergences, although such divergences do appear at intermediate steps of the calculation. The general two-loop bare soft function is given by
\begin{equation}
\begin{aligned}
\bm{S}^{(2)}=&\bm{S}^{(2,\mathrm{RR})}+\bm{S}^{(2,\mathrm{RV})}\\
=&\int{[dk_1][dk_2](2\pi)\delta^{(+)}(k_1^2)(2\pi)\delta^{(+)}(k_2^2)K_S(k_1+k_2)}\\
&\times\Bigg[\frac{1}{2}\bm{A}^{(\mathrm{RR},g)\dagger}_S\bm{A}_S^{(\mathrm{RR},g)}-\bm{A}^{(\mathrm{RR},\mathrm{ghost})\dagger}_S\bm{A}_S^{(\mathrm{RR},\mathrm{ghost})}+n_q\bm{A}^{(\mathrm{RR},q)\dagger}_S\bm{A}_S^{(\mathrm{RR},q)}\Bigg]\\
&+\int{[dk_1](2\pi)\delta^{(+)}(k_1^2)K_S(k_1)\left[\bm{A}_S^{(\mathrm{RV})\dagger}\bm{A}^{(\mathrm{R})}_S+\bm{A}_S^{(\mathrm{R})\dagger}\bm{A}^{(\mathrm{RV})}_S\right]}\,,\label{eq:twoloopsoftfungeneral}
\end{aligned}
\end{equation}
where $n_q$ is the number of massless quark flavors, and the factor $1/2$ in the first square bracket accounts for the final-state symmetry factor. We have implicitly assumed the spin summation over external soft partons in eqs.\eqref{eq:oneloopsoftfungeneral} and \eqref{eq:twoloopsoftfungeneral}. The ghost contribution appears in eq.\eqref{eq:twoloopsoftfungeneral} because the spin sum for external soft gluons is performed in Feynman gauge:
\begin{equation}
\sum_{\lambda}{\varepsilon_{\lambda}^{\mu*}(k)\varepsilon_{\lambda}^{\nu}(k)}=-g^{\mu\nu}\,.
\end{equation}

After performing the spin summation, the explicit expressions for the two-loop double-real and real-virtual soft functions are given by:
\begin{eqnarray}
\bm{S}^{(2,\mathrm{RR})}&=&16\pi^2\alpha_s^2\int{[dk_1][dk_2](2\pi)\delta^{(+)}(k_1^2)(2\pi)\delta^{(+)}(k_2^2)K_S(k_1+k_2)}\nonumber\\
&&\times \Bigg\{\bigg(\sum_{i}{\sum_{n=1}^{3}{\bm{w}_{i}^{(2,D_n)}I_i^{(2,D_n)}}}\bigg)+\mathop{\sum_{i,j}}_{i\neq j}{\bigg[(\sum_{n=7}^{13}{\bm{w}_{ij}^{(2,D_n)}I_{ij}^{(2,D_n)}})}\nonumber\\
&&+(\sum_{n=14}^{15}{\bm{w}_{ij}^{(2,D_{n,1})}I_{ij}^{(2,D_{n,1})}})\bigg]+\mathop{\sum_{i,j,k}}_{i\neq j, i\neq k, j\neq k}{\bigg[\sum_{n=22}^{24}{\bm{w}_{ijk}^{(2,D_n)}I_{ijk}^{(2,D_n)}}\bigg]}\nonumber\\
&&+\mathop{\mathop{\sum_{i,j,k,l}}_{i\neq j, i\neq k, i\neq l}}_{j\neq k, j\neq l, k\neq l}{\bm{w}_{ijkl}^{(2,D_{28})}I_{ijkl}^{(2,D_{28})}}\Bigg\}\,,\label{eq:softfunRR}\\
\bm{S}^{(2,\mathrm{RV})}&=&16\pi^2\alpha_s^2\int{[dk_1](2\pi)\delta^{(+)}(k_1^2)K_S(k_1)}\label{eq:softfunRV}\\
&&\times \Bigg\{\bigg(\sum_{i}{\sum_{n=5}^{6}{2\bm{w}_{i}^{(2,D_n)}\Re{I_i^{(2,D_n)}}}}\bigg)+\mathop{\sum_{i,j}}_{i\neq j}{\bigg[(\sum_{n=14}^{15}{2\bm{w}_{ij}^{(2,D_{n,2})}\Re{I_{ij}^{(2,D_{n,2})}}})}\nonumber\\
&&+(\sum_{n=16}^{21}{2\bm{w}_{ij}^{(2,D_{n})}\Re{I_{ij}^{(2,D_{n})}}})\bigg]+\bigg(\mathop{\sum_{i,j,k}}_{i<j, i\neq k, j\neq k}{2i\bm{w}_{ijk}^{(2,D_{25,2})}\Im{I_{ijk}^{(2,D_{25,2})}}}\bigg)\nonumber\\
&&+\mathop{\sum_{i,j,k}}_{i\neq j, i\neq k, j\neq k}{\bigg[\bm{w}_{ijk}^{(2,D_{26})}(I_{ijk}^{(2,D_{26})}+I_{ijk}^{(2,D_{27})\dagger})+\bm{w}_{ijk}^{(2,D_{23})}(I_{ijk}^{(2,D_{26})\dagger}+I_{ijk}^{(2,D_{27})})\bigg]}\Bigg\}\,.\nonumber
\end{eqnarray}
Each term in the integrands of eqs.\eqref{eq:softfunRR} and \eqref{eq:softfunRV} consists of a product of a color factor $\bm{w}$ and a kinematic function $I$. These terms can be categorized according to whether they involve one, two, three, and four Wilson lines, as shown in figures~\ref{fig:oneWL}, \ref{fig:twoWL}, \ref{fig:threeWL}, and \ref{fig:fourWL}, respectively. In the one-Wilson-line case, the integrands of $D_{4,1}$ and $D_{4,2}$ vanish identically. The non-zero color factors are:
\begin{eqnarray}
\bm{w}_{i}^{(2,D_1)}&=&\bm{\mathrm{T}}^a(\ident_i)\bm{\mathrm{T}}^a(\ident_i)=C(\ident_i)\bm{1}\,,\\
\bm{w}_{i}^{(2,D_2)}&=&\bm{\mathrm{T}}^a(\ident_i)\bm{\mathrm{T}}^b(\ident_i)\bm{\mathrm{T}}^b(\ident_i)\bm{\mathrm{T}}^a(\ident_i)=\left(C(\ident_i)\right)^2\bm{1}\,,\\
\bm{w}_{i}^{(2,D_3)}&=&\bm{\mathrm{T}}^b(\ident_i)\bm{\mathrm{T}}^a(\ident_i)\bm{\mathrm{T}}^b(\ident_i)\bm{\mathrm{T}}^a(\ident_i)=C(\ident_i)\left(C(\ident_i)-\frac{C_A}{2}\right)\bm{1}\,,\\
\bm{w}_{i}^{(2,D_{5})}&=&\bm{\mathrm{T}}^a(\ident_i)\bm{\mathrm{T}}^a(\ident_i)\bm{\mathrm{T}}^b(\ident_i)\bm{\mathrm{T}}^b(\ident_i)=\left(C(\ident_i)\right)^2\bm{1}\,,\\
\bm{w}_{i}^{(2,D_{6})}&=&\bm{w}_{i}^{(2,D_3)}\,.
\end{eqnarray}
The kinematic functions are:
\begin{eqnarray}
I_{i}^{(2,D_1)}&=&\frac{1}{\left(\left(k_1+k_2\right)^2\right)^2\left(\left(k_1+k_2\right)\cdot p_i\right)^2}\Bigg[4T_Fn_q\left(2p_i\cdot k_1 p_i\cdot k_2-p_i^2k_1\cdot k_2\right)\nonumber\\
&&+C_A\bigg(4p_i^2k_1\cdot k_2+\frac{d-7}{2}\left(p_i\cdot k_1\right)^2+\frac{d-7}{2}\left(p_i\cdot k_2\right)^2-(d+3)p_i\cdot k_1 p_i\cdot k_2\bigg)\Bigg]\,,\\
I_{i}^{(2,D_2)}&=&\frac{\left(p_i^2\right)^2}{\left(p_i\cdot k_1\right)^2\left(p_i\cdot (k_1+k_2)\right)^2}\,,\\
I_{i}^{(2,D_3)}&=&\frac{\left(p_i^2\right)^2}{\left(p_i\cdot k_1\right)\left(p_i\cdot k_2\right)\left(p_i\cdot (k_1+k_2)\right)^2}\,,\\
I_{i}^{(2,D_5)}&=&\frac{\left(p_i^2\right)^2}{\left(p_i\cdot k_1\right)^3}\int{[dk_2]\frac{i}{k_2^2\left(p_i\cdot (k_1+k_2)-i\sigma_i0^+\right)}}\,,\\
I_{i}^{(2,D_6)}&=&\frac{\left(p_i^2\right)^2}{\left(p_i\cdot k_1\right)^2}\int{[dk_2]\frac{i}{k_2^2\left(p_i\cdot k_2+i\sigma_i0^+\right)\left(p_i\cdot (k_2-k_1)+i\sigma_i0^+\right)}}\,,
\end{eqnarray}
where $T_F=1/2$. Due to the presence of the ghost, $I_{i}^{(2,D_1)}$ is not symmetric under the exchange of $k_1$ and $k_2$. However, since the integration measures for $k_1$ and $k_2$ are fully symmetric in the double-real soft function, we have symmetrized the ghost amplitude. The same symmetrization is applied to $I_{ij}^{(2,D_7)}$ in the two-Wilson-line case. There are many more non-equivalent soft configurations in the two-Wilson-line case, as shown in figure \ref{fig:twoWL}. The expressions for the color factors are
\begin{eqnarray}
\bm{w}_{ij}^{(2,D_7)}&=&\bm{\mathrm{T}}^a(\ident_j)\bm{\mathrm{T}}^a(\ident_i)\,,\\
\bm{w}_{ij}^{(2,D_8)}&=&\bm{\mathrm{T}}^b(\ident_j)\bm{\mathrm{T}}^a(\ident_j)\bm{\mathrm{T}}^b(\ident_i)\bm{\mathrm{T}}^a(\ident_i)\,,\\
\bm{w}_{ij}^{(2,D_9)}&=&\bm{\mathrm{T}}^a(\ident_j)\bm{\mathrm{T}}^b(\ident_j)\bm{\mathrm{T}}^b(\ident_i)\bm{\mathrm{T}}^a(\ident_i)\,,\\
\bm{w}_{ij}^{(2,D_{10})}&=&\bm{\mathrm{T}}^a(\ident_i)\bm{\mathrm{T}}^b(\ident_j)\bm{\mathrm{T}}^b(\ident_j)\bm{\mathrm{T}}^a(\ident_i)=C(\ident_i)C(\ident_j)\bm{1}\,,\\
\bm{w}_{ij}^{(2,D_{11})}&=&\bm{\mathrm{T}}^a(\ident_j)\bm{\mathrm{T}}^b(\ident_i)\bm{\mathrm{T}}^b(\ident_j)\bm{\mathrm{T}}^a(\ident_i)=\bm{w}_{ij}^{(2,D_{9})}\,,\\
\bm{w}_{ij}^{(2,D_{12})}&=&\bm{\mathrm{T}}^a(\ident_j)\bm{\mathrm{T}}^b(\ident_i)\bm{\mathrm{T}}^b(\ident_i)\bm{\mathrm{T}}^a(\ident_i)=C(\ident_i)\bm{\mathrm{T}}^a(\ident_j)\bm{\mathrm{T}}^a(\ident_i)\,,\\
\bm{w}_{ij}^{(2,D_{13})}&=&\bm{\mathrm{T}}^a(\ident_j)\bm{\mathrm{T}}^b(\ident_i)\bm{\mathrm{T}}^a(\ident_i)\bm{\mathrm{T}}^b(\ident_i)=\left(C(\ident_i)-\frac{C_A}{2}\right)\bm{\mathrm{T}}^a(\ident_j)\bm{\mathrm{T}}^a(\ident_i)\,,\\
\bm{w}_{ij}^{(2,D_{14,1})}&=&if^{abc}\bm{\mathrm{T}}^c(\ident_i)\bm{\mathrm{T}}^b(\ident_i)\bm{\mathrm{T}}^a(\ident_j)=\frac{C_A}{2}\bm{\mathrm{T}}^a(\ident_j)\bm{\mathrm{T}}^a(\ident_i)\,,\\
\bm{w}_{ij}^{(2,D_{14,2})}&=&\bm{w}_{ij}^{(2,D_{15,1})}=\bm{w}_{ij}^{(2,D_{15,2})}=\bm{w}_{ij}^{(2,D_{14,1})}\,,\\
\bm{w}_{ij}^{(2,D_{16})}&=&\bm{\mathrm{T}}^a(\ident_j)\bm{\mathrm{T}}^a(\ident_i)\bm{\mathrm{T}}^b(\ident_i)\bm{\mathrm{T}}^b(\ident_i)=C(\ident_i)\bm{\mathrm{T}}^a(\ident_j)\bm{\mathrm{T}}^a(\ident_i)=\bm{w}_{ij}^{(2,D_{12})}\,,\\
\bm{w}_{ij}^{(2,D_{17})}&=&\bm{w}_{ij}^{(2,D_{20})}=\bm{w}_{ij}^{(2,D_{13})}\,,\\
\bm{w}_{ij}^{(2,D_{18})}&=&\bm{\mathrm{T}}^b(\ident_i)\bm{\mathrm{T}}^a(\ident_j)\bm{\mathrm{T}}^b(\ident_i)\bm{\mathrm{T}}^a(\ident_i)=C(\ident_i)\bm{\mathrm{T}}^a(\ident_j)\bm{\mathrm{T}}^a(\ident_i)=\bm{w}_{ij}^{(2,D_{12})}\,,\\
\bm{w}_{ij}^{(2,D_{19})}&=&\bm{w}_{ij}^{(2,D_{8})}\,,\\
\bm{w}_{ij}^{(2,D_{21})}&=&\bm{w}_{ij}^{(2,D_{9})}\,,
\end{eqnarray}
and for the kinematic functions, they are
\begin{eqnarray}
I_{ij}^{(2,D_7)}&=&\frac{1}{\left(\left(k_1+k_2\right)^2\right)^2\left(\left(k_1+k_2\right)\cdot p_i\right)\left(\left(k_1+k_2\right)\cdot p_j\right)}\nonumber\\
&&\times\Bigg[4T_Fn_q\left(p_i\cdot k_1 p_j\cdot k_2+p_i\cdot k_2 p_j\cdot k_1-p_i\cdot p_j k_1\cdot k_2\right)\nonumber\\
&&+C_A\bigg(4p_i\cdot p_j k_1\cdot k_2+\frac{d-7}{2}p_i\cdot k_1 p_j\cdot k_1+\frac{d-7}{2}p_i\cdot k_2 p_j\cdot k_2\nonumber\\
&&-\frac{d+3}{2}p_i\cdot k_1 p_j\cdot k_2-\frac{d+3}{2}p_i\cdot k_2 p_j\cdot k_1\bigg)\Bigg]\,,\\
I_{ij}^{(2,D_8)}&=&\frac{\left(p_i\cdot p_j\right)^2}{\left(p_i\cdot k_1\right)\left(p_j\cdot k_2\right)\left(p_i\cdot (k_1+k_2)\right)\left(p_j\cdot (k_1+k_2)\right)}\,,\\
I_{ij}^{(2,D_9)}&=&\frac{\left(p_i\cdot p_j\right)^2}{\left(p_i\cdot k_1\right)\left(p_j\cdot k_1\right)\left(p_i\cdot (k_1+k_2)\right)\left(p_j\cdot (k_1+k_2)\right)}\,,\\
I_{ij}^{(2,D_{10})}&=&\frac{\frac{1}{2}p_i^2p_j^2}{\left(p_i\cdot k_1\right)^2\left(p_j\cdot k_2\right)^2}\,,\\
I_{ij}^{(2,D_{11})}&=&\frac{\frac{1}{2}(p_i\cdot p_j)^2}{\left(p_i\cdot k_1\right)\left(p_j\cdot k_1\right)\left(p_i\cdot k_2\right)\left(p_j\cdot k_2\right)}\,,\\
I_{ij}^{(2,D_{12})}&=&\frac{2p_i^2p_i\cdot p_j}{\left(p_i\cdot k_1\right)^2\left(p_j\cdot k_2\right)\left(p_i\cdot (k_1+k_2)\right)}\,,\\
I_{ij}^{(2,D_{13})}&=&\frac{2p_i^2p_i\cdot p_j}{\left(p_i\cdot k_1\right)\left(p_j\cdot k_1\right)\left(p_i\cdot k_2\right)\left(p_i\cdot (k_1+k_2)\right)}\,,\\
I_{ij}^{(2,D_{14,1})}&=&\frac{2\left[p_i^2p_j\cdot \left(2k_1+k_2\right)-p_i\cdot p_j p_i\cdot\left(2k_1+k_2\right)\right]}{\left(k_1+k_2\right)^2\left(p_i\cdot k_1\right)\left(p_j\cdot k_2\right)\left(p_i\cdot(k_1+k_2)\right)}\,,\\
I_{ij}^{(2,D_{14,2})}&=&\frac{1}{p_j\cdot k_1}\int{[dk_2]\frac{-i\left[p_i^2p_j\cdot \left(2k_1-k_2\right)-p_i\cdot p_j p_i\cdot\left(2k_1-k_2\right)\right]}{k_2^2\left(k_2-k_1\right)^2\left(p_j\cdot k_2-i\sigma_j0^+\right)\left(p_i\cdot(k_2-k_1)+i\sigma_i0^+\right)}}\,,\\
I_{ij}^{(2,D_{15,1})}&=&\frac{2\left[p_i\cdot p_j p_i\cdot \left(k_1-k_2\right)-p_i^2 p_j\cdot\left(k_1-k_2\right)\right]}{\left(k_1+k_2\right)^2\left(p_i\cdot k_1\right)\left(p_i\cdot(k_1+k_2)\right)\left(p_j\cdot (k_1+k_2)\right)}\,,\\
I_{ij}^{(2,D_{15,2})}&=&\frac{1}{\left(p_i\cdot k_1\right)\left(p_j\cdot k_1\right)}\int{[dk_2]\frac{-i\left[p_i\cdot p_j p_i\cdot \left(k_1-2k_2\right)-p_i^2 p_j\cdot\left(k_1-2k_2\right)\right]}{k_2^2\left(k_2-k_1\right)^2\left(p_i\cdot k_2-i\sigma_i0^+\right)}}\,,\\
I_{ij}^{(2,D_{16})}&=&\frac{1}{\left(p_i\cdot k_1\right)^2\left(p_j\cdot k_1\right)}\int{[dk_2]\frac{ip_i^2p_i\cdot p_j}{k_2^2\left(p_i\cdot \left(k_1+k_2\right)-i\sigma_i0^+\right)}}\,,\\
I_{ij}^{(2,D_{17})}&=&\frac{1}{\left(p_i\cdot k_1\right)\left(p_j\cdot k_1\right)}\int{[dk_2]\frac{ip_i^2p_i\cdot p_j}{k_2^2\left(p_i\cdot k_2+i\sigma_i0^+\right)\left(p_i\cdot (k_2-k_1)+i\sigma_i0^+\right)}}\,,\\
I_{ij}^{(2,D_{18})}&=&\frac{1}{\left(p_i\cdot k_1\right)^2}\int{[dk_2]\frac{-ip_i^2p_i\cdot p_j}{k_2^2\left(p_j\cdot k_2+i\sigma_j0^+\right)\left(p_i\cdot (k_1+k_2)-i\sigma_i0^+\right)}}\,,\\
I_{ij}^{(2,D_{19})}&=&\frac{1}{\left(p_i\cdot k_1\right)\left(p_j\cdot k_1\right)}\int{[dk_2]\frac{-i\left(p_i\cdot p_j\right)^2}{k_2^2\left(p_j\cdot k_2+i\sigma_j0^+\right)\left(p_i\cdot (k_1+k_2)-i\sigma_i0^+\right)}}\,,\\
I_{ij}^{(2,D_{20})}&=&\frac{1}{\left(p_i\cdot k_1\right)}\int{[dk_2]\frac{-ip_i^2p_i\cdot p_j}{k_2^2\left(p_i\cdot k_2-i\sigma_i0^+\right)\left(p_j\cdot k_2+i\sigma_j0^+\right)\left(p_i\cdot (k_1+k_2)-i\sigma_i0^+\right)}}\,,\\
I_{ij}^{(2,D_{21})}&=&\frac{1}{\left(p_j\cdot k_1\right)}\int{[dk_2]\frac{-i\left(p_i\cdot p_j\right)^2}{k_2^2\left(p_i\cdot k_2-i\sigma_i0^+\right)\left(p_j\cdot k_2+i\sigma_j0^+\right)\left(p_i\cdot (k_1+k_2)-i\sigma_i0^+\right)}}\,.
\end{eqnarray}
In all possible three-Wilson-line configurations outlined in figure \ref{fig:threeWL}, the color factors are
\begin{eqnarray}
\bm{w}_{ijk}^{(2,D_{22})}&=&\bm{\mathrm{T}}^b(\ident_i)\bm{\mathrm{T}}^a(\ident_k)\bm{\mathrm{T}}^a(\ident_j)\bm{\mathrm{T}}^b(\ident_i)=C(\ident_i)\bm{\mathrm{T}}^a(\ident_k)\bm{\mathrm{T}}^a(\ident_j)\,,\\
\bm{w}_{ijk}^{(2,D_{23})}&=&\bm{\mathrm{T}}^a(\ident_k)\bm{\mathrm{T}}^b(\ident_j)\bm{\mathrm{T}}^b(\ident_i)\bm{\mathrm{T}}^a(\ident_i)\,,\\
\bm{w}_{ijk}^{(2,D_{25,1})}&=&-if^{abc}\bm{\mathrm{T}}^c(\ident_k)\bm{\mathrm{T}}^b(\ident_j)\bm{\mathrm{T}}^a(\ident_i)\,,\\
\bm{w}_{ijk}^{(2,D_{26})}&=&\bm{\mathrm{T}}^b(\ident_k)\bm{\mathrm{T}}^a(\ident_j)\bm{\mathrm{T}}^b(\ident_i)\bm{\mathrm{T}}^a(\ident_i)=\bm{w}_{ikj}^{(2,D_{23})}=\left(\bm{w}_{ijk}^{(2,D_{23})}\right)^\dagger\,,\\
\bm{w}_{ijk}^{(2,D_{24})}&=&\bm{w}_{ijk}^{(2,D_{26})}+\bm{w}_{ijk}^{(2,D_{23})}=2\bm{w}_{ijk}^{(2,D_{23})}+\bm{w}_{ijk}^{(2,D_{25,1})}=\left(\bm{w}_{ijk}^{(2,D_{24})}\right)^\dagger\label{eq:wcolorD24}\,,\\
\bm{w}_{ijk}^{(2,D_{25,2})}&=&\bm{w}_{ijk}^{(2,D_{25,1})}\,,\\
\bm{w}_{ijk}^{(2,D_{27})}&=&\bm{w}_{ijk}^{(2,D_{23})}\,.
\end{eqnarray}
The kinematic functions are
\begin{eqnarray}
I_{ijk}^{(2,D_{22})}&=&\frac{p_i^2p_j\cdot p_k}{\left(p_i\cdot k_1\right)^2\left(p_j\cdot k_2\right)\left(p_k\cdot k_2\right)}\,,\\
I_{ijk}^{(2,D_{23})}&=&\frac{p_i\cdot p_j p_i\cdot p_k}{\left(p_i\cdot k_1\right)\left(p_i\cdot k_2\right)\left(p_j\cdot k_2\right)\left(p_k\cdot k_1\right)}\,,\\
I_{ijk}^{(2,D_{24})}&=&\frac{p_i\cdot p_j p_i\cdot p_k}{\left(p_i\cdot k_1\right)\left(p_i\cdot \left(k_1+k_2\right)\right)\left(p_j\cdot k_2\right)\left(p_k\cdot k_1\right)}\,,\\
I_{ijk}^{(2,D_{25,1})}&=&\frac{\frac{1}{2}\left[p_i\cdot p_j p_k\cdot (k_2-k_1)+p_i\cdot p_k p_j\cdot (2k_1+k_2)-p_j\cdot p_kp_i\cdot (k_1+2k_2)\right]}{\left(k_1+k_2\right)^2\left(p_i\cdot k_1\right)\left(p_j\cdot k_2\right)\left(p_k\cdot (k_1+k_2)\right)}\,,\\
I_{ijk}^{(2,D_{25,2})}&=&\frac{1}{\left(p_k\cdot k_1\right)}\int{[dk_2]\frac{i\left[p_i\cdot p_j p_k\cdot(k_1-2k_2)-p_i\cdot p_k p_j\cdot (2k_1-k_2)+p_j\cdot p_k p_i(k_1+k_2)\right]}{k_2^2\left(k_2-k_1\right)^2\left(p_i\cdot (k_2-k_1)+i\sigma_i0^+\right)\left(p_j\cdot k_2-i\sigma_j0^+\right)}}\,,\\
I_{ijk}^{(2,D_{26})}&=&\frac{1}{\left(p_i\cdot k_1\right)\left(p_k\cdot k_1\right)}\int{[dk_2]\frac{-ip_i\cdot p_j p_i\cdot p_k}{k_2^2\left(p_i\cdot (k_1+k_2)-i\sigma_i0^+\right)\left(p_j\cdot k_2+i\sigma_j0^+\right)}}\,,\\
I_{ijk}^{(2,D_{27})}&=&\frac{1}{\left(p_k\cdot k_1\right)}\int{[dk_2]\frac{-ip_i\cdot p_j p_i\cdot p_k}{k_2^2\left(p_i\cdot k_2-i\sigma_i0^+\right)\left(p_i\cdot (k_1+k_2)-i\sigma_i0^+\right)\left(p_j\cdot k_2+i\sigma_j0^+\right)}}\,.
\end{eqnarray}
Note that the configuration $D_{25,1}$ in figure \ref{fig:threeWL} does not contribute to eq.\eqref{eq:softfunRR} since it cancels with its complex conjugate. Additionally, for the diagram $D_{24}$ in figure \ref{fig:threeWL}, we have included its complex conjugate within the color factor in eq.\eqref{eq:wcolorD24}. Finally, there is only one non-isomorphic configuration in the four-Wilson-line case. Its color factor is
\begin{eqnarray}
\bm{w}_{ijkl}^{(2,D_{28})}&=&\bm{\mathrm{T}}^b(\ident_l)\bm{\mathrm{T}}^a(\ident_k)\bm{\mathrm{T}}^b(\ident_j)\bm{\mathrm{T}}^a(\ident_i)\,,
\end{eqnarray}
and the corresponding kinematic function is
\begin{equation}
I_{ijkl}^{(2,D_{28})}=\frac{\frac{1}{2}p_i\cdot p_k p_j\cdot p_l}{\left(p_i\cdot k_1\right)\left(p_k\cdot k_1\right)\left(p_j\cdot k_2\right)\left(p_l\cdot k_2\right)}\,.
\end{equation}
Since we have been unable to find explicit expressions for the two-loop soft function in the generic case anywhere in the literature, we report them here in case someone finds them useful. From these generic results, we again observe that the soft Wilson line is invariant under rescaling of the four-momentum $p_i$ in the form $p_i\to\lambda_ip_i$ with $\lambda_i\neq 0$, so that we have certain freedoms to renormalize these external momenta.

 \begin{figure}[h!]
\begin{tikzpicture}[line width=1 pt, scale=0.9]
  \hspace{0cm}
\draw[wilson] (-2,0.0) -- (0.0,2.0);
\draw[wilson] (-2,0.0) -- (0.0,-2.0);
\draw[wilson] (2.0,2.0) -- (4,0.0);
\draw[wilson] (2.0,-2.0) -- (4,0.0);
\draw[scalarnoarrow] (1.0,2.5) -- (1.0,-2.5);
\draw[scalarnoarrow] (1.0,2.5) -- (1.8,3.0);
\draw[scalarnoarrow] (1.0,-2.5) -- (0.2,-3.0);
\draw[gluon] (-0.6,-1.4) -- (2.6,-1.4);
\filldraw [blue] (1,-1.4) circle (15pt);
\node at (-0.5,-2.0) {\Large $\ident_i$};
\node at (2.5,-2.0) {\Large $\ident_i$};
\node at (1.0,-3.5) {\Large $(D_1)$};
\hspace{8cm}
\draw[wilson] (-2,0.0) -- (0.0,2.0);
\draw[wilson] (-2,0.0) -- (0.0,-2.0);
\draw[wilson] (2.0,2.0) -- (4,0.0);
\draw[wilson] (2.0,-2.0) -- (4,0.0);
\draw[scalarnoarrow] (1.0,2.5) -- (1.0,-2.5);
\draw[scalarnoarrow] (1.0,2.5) -- (1.8,3.0);
\draw[scalarnoarrow] (1.0,-2.5) -- (0.2,-3.0);
\draw[gluon] (-1.0,-1.0) -- (3.0,-1.0);
\draw[gluon] (-0.4,-1.6) -- (2.4,-1.6);
\node at (-0.5,-2.0) {\Large $\ident_i$};
\node at (2.5,-2.0) {\Large $\ident_i$};
\node at (1.0,-3.5) {\Large $(D_2)$};
\end{tikzpicture}\\
\begin{tikzpicture}[line width=1 pt, scale=0.9]
  \hspace{0cm}
\draw[wilson] (-2,0.0) -- (0.0,2.0);
\draw[wilson] (-2,0.0) -- (0.0,-2.0);
\draw[wilson] (2.0,2.0) -- (4,0.0);
\draw[wilson] (2.0,-2.0) -- (4,0.0);
\draw[scalarnoarrow] (1.0,2.5) -- (1.0,-2.5);
\draw[scalarnoarrow] (1.0,2.5) -- (1.8,3.0);
\draw[scalarnoarrow] (1.0,-2.5) -- (0.2,-3.0);
\draw[gluon] (-1.0,-1.0) -- (0.53,-1.27);
\draw[gluon] (1.38,-1.42) -- (2.4,-1.6);
\draw[gluon] (-0.4,-1.6) -- (3.0,-1.0);
\node at (-0.5,-2.0) {\Large $\ident_i$};
\node at (2.5,-2.0) {\Large $\ident_i$};
\node at (1.0,-3.5) {\Large $(D_3)$};
\hspace{8cm}
\draw[wilson] (-2,0.0) -- (0.0,2.0);
\draw[wilson] (-2,0.0) -- (0.0,-2.0);
\draw[wilson] (2.0,2.0) -- (4,0.0);
\draw[wilson] (2.0,-2.0) -- (4,0.0);
\draw[scalarnoarrow] (0.6,2.5) -- (0.6,-2.5);
\draw[scalarnoarrow] (0.6,2.5) -- (1.4,3.0);
\draw[scalarnoarrow] (0.6,-2.5) -- (-0.2,-3.0);
\draw[scalarnoarrow] (1.6,2.5) -- (1.6,-2.5);
\draw[scalarnoarrow] (1.6,2.5) -- (2.4,3.0);
\draw[scalarnoarrow] (1.6,-2.5) -- (0.8,-3.0);
\draw[gluon, segment length=6pt, bend left,looseness = 1.0] (-1.0,-1.0) to (1.2,-1.6);
\filldraw [red] (1.2,-1.6) circle (3pt);
\draw[gluon] (-0.4,-1.6) -- (2.4,-1.6);
\node at (-0.5,-2.0) {\Large $\ident_i$};
\node at (2.5,-2.0) {\Large $\ident_i$};
\node at (-0.2,-3.5) {\Large $(D_{4,1})$};
\node at (1.4,-3.5) {\Large $(D_{4,2})$};
\end{tikzpicture}\\
\begin{tikzpicture}[line width=1 pt, scale=0.9]
  \hspace{0cm}
\draw[wilson] (-2,0.0) -- (0.0,2.0);
\draw[wilson] (-2,0.0) -- (0.0,-2.0);
\draw[wilson] (2.0,2.0) -- (4,0.0);
\draw[wilson] (2.0,-2.0) -- (4,0.0);
\draw[scalarnoarrow] (1.0,2.5) -- (1.0,-2.5);
\draw[scalarnoarrow] (1.0,2.5) -- (1.8,3.0);
\draw[scalarnoarrow] (1.0,-2.5) -- (0.2,-3.0);
\draw[gluon, segment length=5pt, bend left,looseness = 1.6] (-1.6,-0.4) to (-0.6,-1.4);
\draw[gluon] (-0.4,-1.6) -- (2.4,-1.6);
\node at (-0.5,-2.0) {\Large $\ident_i$};
\node at (2.5,-2.0) {\Large $\ident_i$};
\node at (1.0,-3.5) {\Large $(D_5)$};
\hspace{8cm}
\draw[wilson] (-2,0.0) -- (0.0,2.0);
\draw[wilson] (-2,0.0) -- (0.0,-2.0);
\draw[wilson] (2.0,2.0) -- (4,0.0);
\draw[wilson] (2.0,-2.0) -- (4,0.0);
\draw[scalarnoarrow] (1.0,2.5) -- (1.0,-2.5);
\draw[scalarnoarrow] (1.0,2.5) -- (1.8,3.0);
\draw[scalarnoarrow] (1.0,-2.5) -- (0.2,-3.0);
\draw[gluon, segment length=5pt, bend right,looseness = 1.6] (-1.6,-0.4) to (-0.4,-1.6);
\draw[gluon] (-1.0,-1.0) -- (3.0,-1.0);
\node at (-0.5,-2.0) {\Large $\ident_i$};
\node at (2.5,-2.0) {\Large $\ident_i$};
\node at (1.0,-3.5) {\Large $(D_{6})$};
\end{tikzpicture}
\caption{The two-loop one Wilson-line soft function. Each thick black line represents a Wilson line, and soft partons are shown in red. Each dashed line indicates a Cutkosky cut. The blue bubble represents the combined contributions from quark-antiquark, ghosts, gluons, as depicted in figures \ref{fig:softampRR1}, \ref{fig:softampRR2}, and \ref{fig:softampRR3}.\label{fig:oneWL}} 
\end{figure}
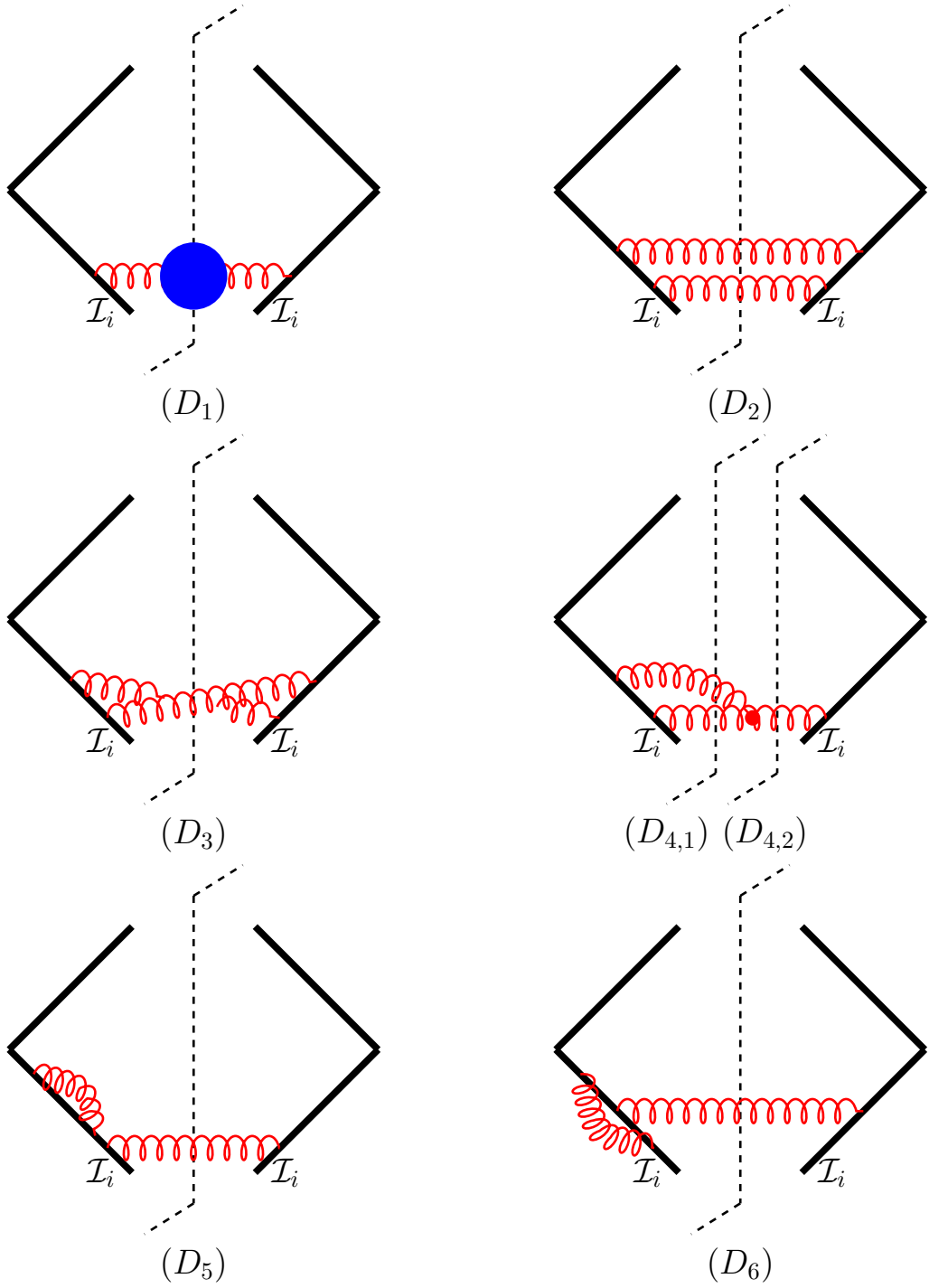

 \begin{figure}[h!]
\begin{tikzpicture}[line width=1 pt, scale=0.7]
  \hspace{0cm}
\draw[wilson] (-2,0.0) -- (0.0,2.0);
\draw[wilson] (-2,0.0) -- (0.0,-2.0);
\draw[wilson] (2.0,2.0) -- (4,0.0);
\draw[wilson] (2.0,-2.0) -- (4,0.0);
\draw[scalarnoarrow] (1.0,2.5) -- (1.0,-2.5);
\draw[scalarnoarrow] (1.0,2.5) -- (1.8,3.0);
\draw[scalarnoarrow] (1.0,-2.5) -- (0.2,-3.0);
\draw[gluon] (-0.6,-1.4) -- (2.6,1.4);
\filldraw [blue] (1,0.0) circle (15pt);
\node at (-0.5,-2.0) {\Large $\ident_i$};
\node at (2.5,2.0) {\Large $\ident_j$};
\node at (1.0,-3.5) {\Large $(D_7)$};
\hspace{5cm}
\draw[wilson] (-2,0.0) -- (0.0,2.0);
\draw[wilson] (-2,0.0) -- (0.0,-2.0);
\draw[wilson] (2.0,2.0) -- (4,0.0);
\draw[wilson] (2.0,-2.0) -- (4,0.0);
\draw[scalarnoarrow] (1.0,2.5) -- (1.0,-2.5);
\draw[scalarnoarrow] (1.0,2.5) -- (1.8,3.0);
\draw[scalarnoarrow] (1.0,-2.5) -- (0.2,-3.0);
\draw[gluon] (-0.8,-1.2) -- (2.4,1.6);
\draw[gluon] (-0.4,-1.6) -- (2.8,1.2);
\node at (-0.5,-2.0) {\Large $\ident_i$};
\node at (2.5,2.0) {\Large $\ident_j$};
\node at (1.0,-3.5) {\Large $(D_8)$};
\hspace{5cm}
\draw[wilson] (-2,0.0) -- (0.0,2.0);
\draw[wilson] (-2,0.0) -- (0.0,-2.0);
\draw[wilson] (2.0,2.0) -- (4,0.0);
\draw[wilson] (2.0,-2.0) -- (4,0.0);
\draw[scalarnoarrow] (1.0,2.5) -- (1.0,-2.5);
\draw[scalarnoarrow] (1.0,2.5) -- (1.8,3.0);
\draw[scalarnoarrow] (1.0,-2.5) -- (0.2,-3.0);
\draw[gluon] (-1.0,-1.0) -- (0.6,-0.2);
\draw[gluon] (1.4,0.2) -- (3.0,1.0);
\draw[gluon] (-0.4,-1.6) -- (2.4,1.6);
\node at (-0.5,-2.0) {\Large $\ident_i$};
\node at (2.5,2.0) {\Large $\ident_j$};
\node at (1.0,-3.5) {\Large $(D_{9})$};
\end{tikzpicture}\\
\begin{tikzpicture}[line width=1 pt, scale=0.7]
  \hspace{0cm}
\draw[wilson] (-2,0.0) -- (0.0,2.0);
\draw[wilson] (-2,0.0) -- (0.0,-2.0);
\draw[wilson] (2.0,2.0) -- (4,0.0);
\draw[wilson] (2.0,-2.0) -- (4,0.0);
\draw[scalarnoarrow] (1.0,2.5) -- (1.0,-2.5);
\draw[scalarnoarrow] (1.0,2.5) -- (1.8,3.0);
\draw[scalarnoarrow] (1.0,-2.5) -- (0.2,-3.0);
\draw[gluon] (-0.8,-1.2) -- (2.8,-1.2);
\draw[gluon] (-0.8, 1.2) -- (2.8,1.2);
\node at (-0.5,-2.0) {\Large $\ident_i$};
\node at (2.5,-2.0) {\Large $\ident_i$};
\node at (2.5,2.0) {\Large $\ident_j$};
\node at (-0.5,2.0) {\Large $\ident_j$};
\node at (1.0,-3.5) {\Large $(D_{10})$};
\hspace{5cm}
\draw[wilson] (-2,0.0) -- (0.0,2.0);
\draw[wilson] (-2,0.0) -- (0.0,-2.0);
\draw[wilson] (2.0,2.0) -- (4,0.0);
\draw[wilson] (2.0,-2.0) -- (4,0.0);
\draw[scalarnoarrow] (1.0,2.5) -- (1.0,-2.5);
\draw[scalarnoarrow] (1.0,2.5) -- (1.8,3.0);
\draw[scalarnoarrow] (1.0,-2.5) -- (0.2,-3.0);
\draw[gluon] (-0.6,-1.4) -- (2.6,1.4);
\draw[gluon] (-0.6, 1.4) -- (0.68,0.28);
\draw[gluon] (1.32, -0.28) -- (2.6,-1.4);
\node at (-0.5,-2.0) {\Large $\ident_i$};
\node at (2.5,-2.0) {\Large $\ident_i$};
\node at (2.5,2.0) {\Large $\ident_j$};
\node at (-0.5,2.0) {\Large $\ident_j$};
\node at (1.0,-3.5) {\Large $(D_{11})$};
\hspace{5cm}
\draw[wilson] (-2,0.0) -- (0.0,2.0);
\draw[wilson] (-2,0.0) -- (0.0,-2.0);
\draw[wilson] (2.0,2.0) -- (4,0.0);
\draw[wilson] (2.0,-2.0) -- (4,0.0);
\draw[scalarnoarrow] (1.0,2.5) -- (1.0,-2.5);
\draw[scalarnoarrow] (1.0,2.5) -- (1.8,3.0);
\draw[scalarnoarrow] (1.0,-2.5) -- (0.5,-3.0);
\draw[gluon] (-1.0,-1.0) -- (3.0,1.0);
\draw[gluon] (-0.6, -1.4) -- (2.6,-1.4);
\node at (-0.5,-2.0) {\Large $\ident_i$};
\node at (2.5,-2.0) {\Large $\ident_i$};
\node at (2.5,2.0) {\Large $\ident_j$};
\node at (1.0,-3.5) {\Large $(D_{12})$};
\end{tikzpicture}\\
\begin{tikzpicture}[line width=1 pt, scale=0.7]
\hspace{0cm}
\draw[wilson] (-2,0.0) -- (0.0,2.0);
\draw[wilson] (-2,0.0) -- (0.0,-2.0);
\draw[wilson] (2.0,2.0) -- (4,0.0);
\draw[wilson] (2.0,-2.0) -- (4,0.0);
\draw[scalarnoarrow] (1.0,2.5) -- (1.0,-2.5);
\draw[scalarnoarrow] (1.0,2.5) -- (1.8,3.0);
\draw[scalarnoarrow] (1.0,-2.5) -- (0.5,-3.0);
\draw[gluon] (-0.6,-1.4) -- (2.6,1.4);
\draw[gluon] (-1.2, -0.8) -- (-0.2,-0.8);
\draw[gluon] ( 0.5, -0.8) -- (3.2,-0.8);
\node at (-0.5,-2.0) {\Large $\ident_i$};
\node at (2.5,-2.0) {\Large $\ident_i$};
\node at (2.5,2.0) {\Large $\ident_j$};
\node at (1.0,-3.5) {\Large $(D_{13})$};
\hspace{5cm}
\draw[wilson] (-2,0.0) -- (0.0,2.0);
\draw[wilson] (-2,0.0) -- (0.0,-2.0);
\draw[wilson] (2.0,2.0) -- (4,0.0);
\draw[wilson] (2.0,-2.0) -- (4,0.0);
\draw[scalarnoarrow] (0.3,2.5) -- (0.3,-2.5);
\draw[scalarnoarrow] (0.3,2.5) -- (1.1,3.0);
\draw[scalarnoarrow] (0.3,-2.5) -- (-0.5,-3.0);
\draw[scalarnoarrow] (1.4,2.5) -- (1.4,-2.5);
\draw[scalarnoarrow] (1.4,2.5) -- (2.2,3.0);
\draw[scalarnoarrow] (1.4,-2.5) -- (0.6,-3.0);
\draw[gluon] (-0.6,-1.4) -- (1.0,0.0);
\draw[gluon] (-0.6, 1.4) -- (2.6,-1.4);
\node at (-0.5,-2.0) {\Large $\ident_i$};
\node at (2.5,-2.0) {\Large $\ident_i$};
\node at (-0.5,2.0) {\Large $\ident_j$};
\filldraw [red] (1.0,0.0) circle (3pt);
\node at (-0.4,-3.5) {\Large $(D_{14,1})$};
\node at (1.6,-3.5) {\Large $(D_{14,2})$};
\hspace{5cm}
\draw[wilson] (-2,0.0) -- (0.0,2.0);
\draw[wilson] (-2,0.0) -- (0.0,-2.0);
\draw[wilson] (2.0,2.0) -- (4,0.0);
\draw[wilson] (2.0,-2.0) -- (4,0.0);
\draw[scalarnoarrow] (0.6,2.5) -- (0.6,-2.5);
\draw[scalarnoarrow] (0.6,2.5) -- (1.4,3.0);
\draw[scalarnoarrow] (0.6,-2.5) -- (-0.2,-3.0);
\draw[scalarnoarrow] (1.6,2.5) -- (1.6,-2.5);
\draw[scalarnoarrow] (1.6,2.5) -- (2.4,3.0);
\draw[scalarnoarrow] (1.6,-2.5) -- (0.8,-3.0);
\draw[gluon, segment length=6pt, bend left,looseness = 1.0] (-0.7,-1.3) to (1.2,0.3);
\filldraw [red] (1.2,0.3) circle (3pt);
\draw[gluon] (-0.4,-1.6) -- (2.4,1.6);
\node at (-0.5,-2.0) {\Large $\ident_i$};
\node at (2.5,2.0) {\Large $\ident_j$};
\node at (-0.4,-3.5) {\Large $(D_{15,1})$};
\node at (1.6,-3.5) {\Large $(D_{15,2})$};
\end{tikzpicture}\\
\begin{tikzpicture}[line width=1 pt, scale=0.7]
\hspace{0cm}
\draw[wilson] (-2,0.0) -- (0.0,2.0);
\draw[wilson] (-2,0.0) -- (0.0,-2.0);
\draw[wilson] (2.0,2.0) -- (4,0.0);
\draw[wilson] (2.0,-2.0) -- (4,0.0);
\draw[scalarnoarrow] (1.0,2.5) -- (1.0,-2.5);
\draw[scalarnoarrow] (1.0,2.5) -- (1.8,3.0);
\draw[scalarnoarrow] (1.0,-2.5) -- (0.2,-3.0);
\draw[gluon, segment length=5pt, bend left,looseness = 1.6] (-1.6,-0.4) to (-0.6,-1.4);
\draw[gluon] (-0.4,-1.6) -- (2.4,1.6);
\node at (-0.5,-2.0) {\Large $\ident_i$};
\node at (2.5, 2.0) {\Large $\ident_j$};
\node at (1.0,-3.5) {\Large $(D_{16})$};
\hspace{5cm}
\draw[wilson] (-2,0.0) -- (0.0,2.0);
\draw[wilson] (-2,0.0) -- (0.0,-2.0);
\draw[wilson] (2.0,2.0) -- (4,0.0);
\draw[wilson] (2.0,-2.0) -- (4,0.0);
\draw[scalarnoarrow] (1.0,2.5) -- (1.0,-2.5);
\draw[scalarnoarrow] (1.0,2.5) -- (1.8,3.0);
\draw[scalarnoarrow] (1.0,-2.5) -- (0.2,-3.0);
\draw[gluon, segment length=5pt, bend right,looseness = 1.6] (-1.6,-0.4) to (-0.4,-1.6);
\draw[gluon] (-1.0,-1.0) -- (3.0,1.0);
\node at (-0.5,-2.0) {\Large $\ident_i$};
\node at (2.5, 2.0) {\Large $\ident_j$};
\node at (1.0,-3.5) {\Large $(D_{17})$};
\hspace{5cm}
\draw[wilson] (-2,0.0) -- (0.0,2.0);
\draw[wilson] (-2,0.0) -- (0.0,-2.0);
\draw[wilson] (2.0,2.0) -- (4,0.0);
\draw[wilson] (2.0,-2.0) -- (4,0.0);
\draw[scalarnoarrow] (1.0,2.5) -- (1.0,-2.5);
\draw[scalarnoarrow] (1.0,2.5) -- (1.8,3.0);
\draw[scalarnoarrow] (1.0,-2.5) -- (0.2,-3.0);
\draw[gluon] (-0.4,-1.6) -- (2.4,-1.6);
\draw[gluon] (-0.7,-1.3) -- (-0.7,1.3);
\node at (-0.5,-2.0) {\Large $\ident_i$};
\node at (-0.5,2.0) {\Large $\ident_j$};
\node at (2.5,-2.0) {\Large $\ident_i$};
\node at (1.0,-3.5) {\Large $(D_{18})$};
\end{tikzpicture}\\
\begin{tikzpicture}[line width=1 pt, scale=0.7]
\hspace{0cm}
\draw[wilson] (-2,0.0) -- (0.0,2.0);
\draw[wilson] (-2,0.0) -- (0.0,-2.0);
\draw[wilson] (2.0,2.0) -- (4,0.0);
\draw[wilson] (2.0,-2.0) -- (4,0.0);
\draw[scalarnoarrow] (1.0,2.5) -- (1.0,-2.5);
\draw[scalarnoarrow] (1.0,2.5) -- (1.8,3.0);
\draw[scalarnoarrow] (1.0,-2.5) -- (0.2,-3.0);
\draw[gluon] (-0.4,-1.6) -- (2.4, 1.6);
\draw[gluon] (-0.7, -1.3) -- (-0.7,1.3);
\node at (-0.5,-2.0) {\Large $\ident_i$};
\node at (-0.5,2.0) {\Large $\ident_j$};
\node at (2.5, 2.0) {\Large $\ident_j$};
\node at (1.0,-3.5) {\Large $(D_{19})$};
\hspace{5cm}
\draw[wilson] (-2,0.0) -- (0.0,2.0);
\draw[wilson] (-2,0.0) -- (0.0,-2.0);
\draw[wilson] (2.0,2.0) -- (4,0.0);
\draw[wilson] (2.0,-2.0) -- (4,0.0);
\draw[scalarnoarrow] (1.0,2.5) -- (1.0,-2.5);
\draw[scalarnoarrow] (1.0,2.5) -- (1.8,3.0);
\draw[scalarnoarrow] (1.0,-2.5) -- (0.2,-3.0);
\draw[gluon] (-1.4,-0.6) -- (-0.4, -0.6);
\draw[gluon] (-0.2,-0.6) -- (3.4, -0.6);
\draw[gluon] (-0.4, -1.6) -- (-0.4,1.6);
\node at (-0.5,-2.0) {\Large $\ident_i$};
\node at (-0.5,2.0) {\Large $\ident_j$};
\node at (2.5, -2.0) {\Large $\ident_i$};
\node at (1.0,-3.5) {\Large $(D_{20})$};
\hspace{5cm}
\draw[wilson] (-2,0.0) -- (0.0,2.0);
\draw[wilson] (-2,0.0) -- (0.0,-2.0);
\draw[wilson] (2.0,2.0) -- (4,0.0);
\draw[wilson] (2.0,-2.0) -- (4,0.0);
\draw[scalarnoarrow] (1.0,2.5) -- (1.0,-2.5);
\draw[scalarnoarrow] (1.0,2.5) -- (1.8,3.0);
\draw[scalarnoarrow] (1.0,-2.5) -- (0.2,-3.0);
\draw[gluon] (-1.0,-1.0) -- (-0.4, -1.0);
\draw[gluon] (-0.2,-0.6) -- (3.0,  1.0);
\draw[gluon] (-0.4, -1.6) -- (-0.4,1.6);
\node at (-0.5,-2.0) {\Large $\ident_i$};
\node at (-0.5,2.0) {\Large $\ident_j$};
\node at (2.5, 2.0) {\Large $\ident_j$};
\node at (1.0,-3.5) {\Large $(D_{21})$};
\end{tikzpicture}
\caption{Same as figure \ref{fig:oneWL}, but for the two-Wilson-line case.\label{fig:twoWL}}
\end{figure}
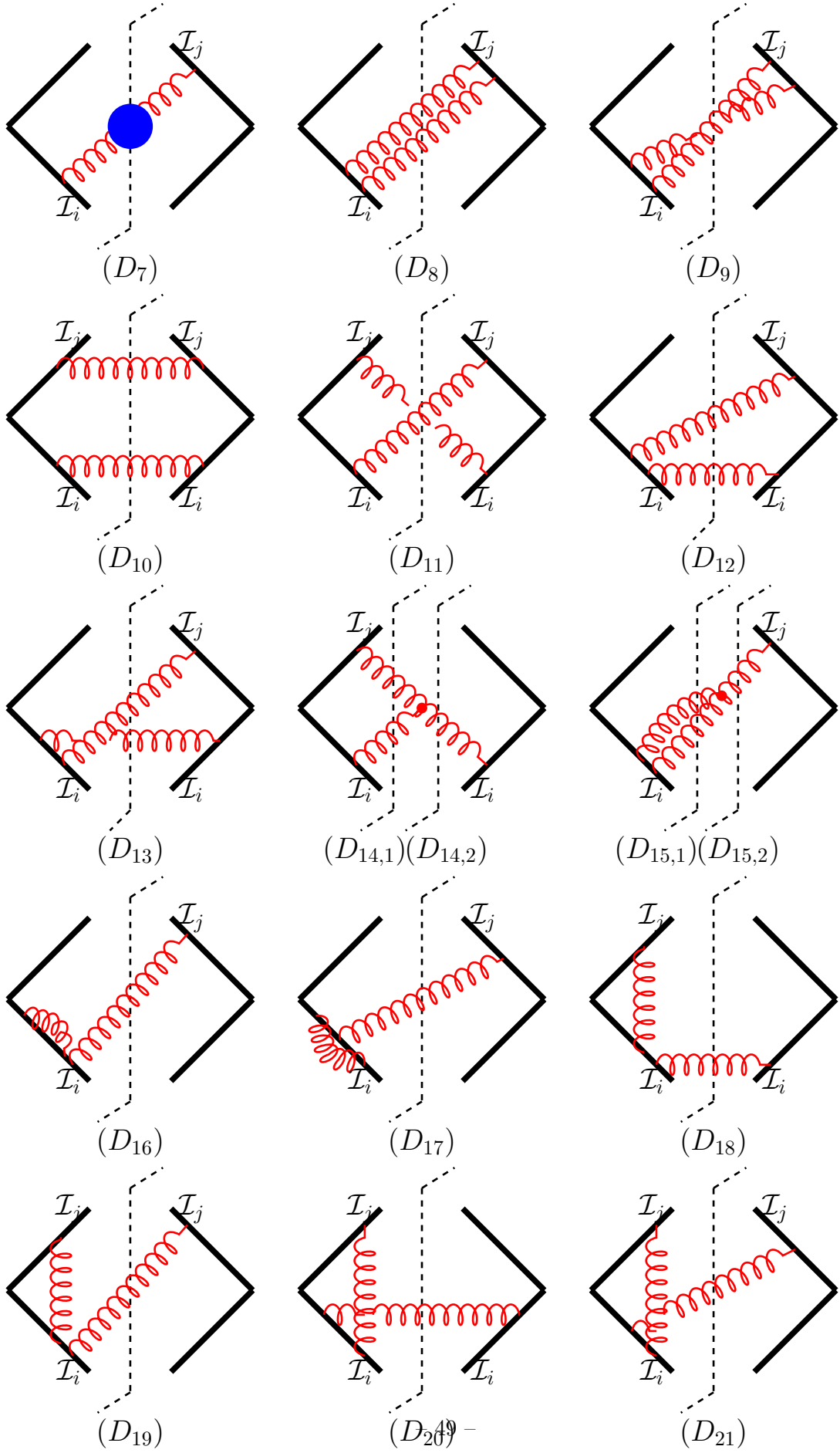

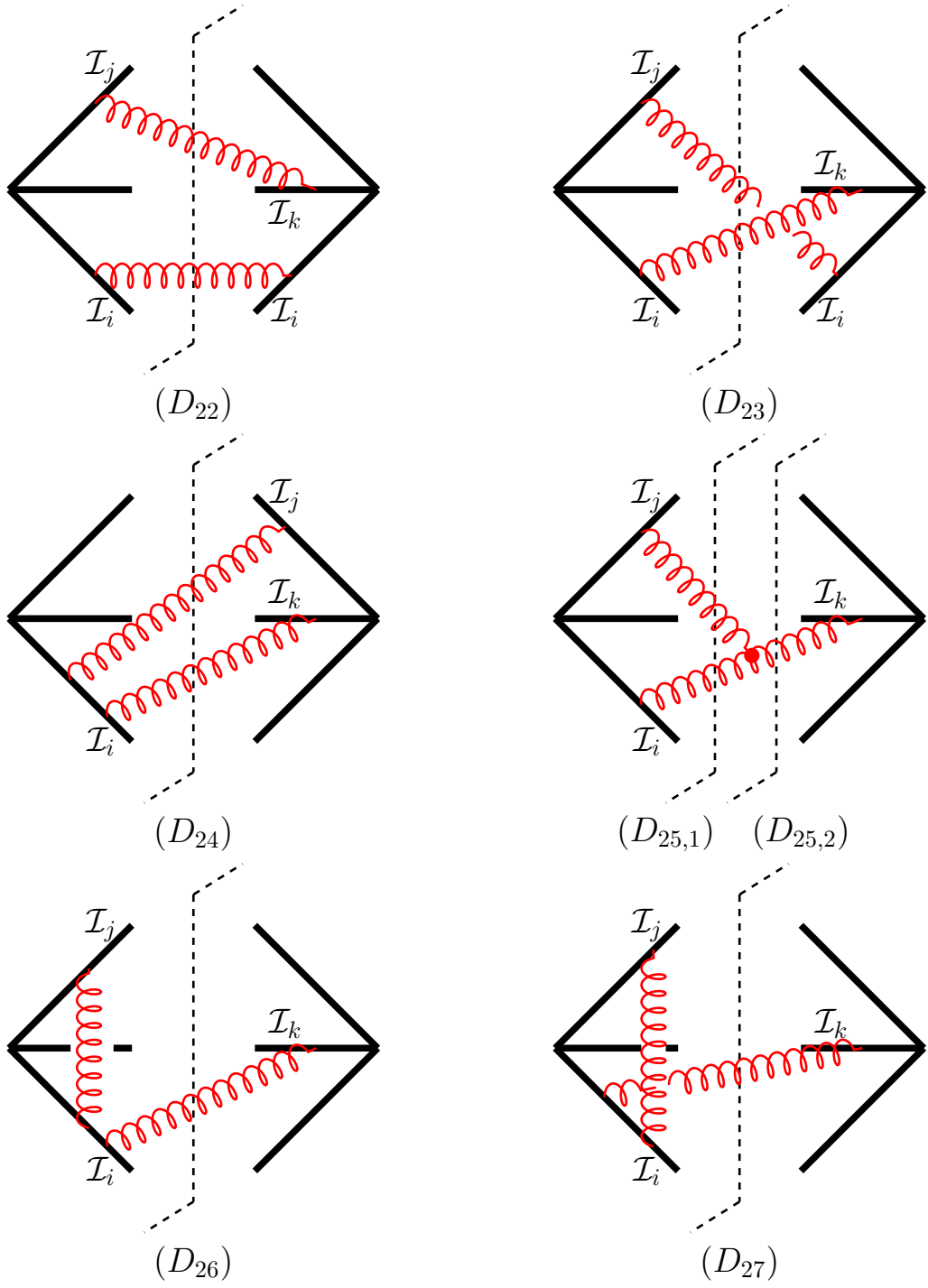
\begin{figure}[h!]
\begin{tikzpicture}[line width=1 pt, scale=0.9]
\hspace{0cm}
\draw[wilson] (-2,0.0) -- (0.0,2.0);
\draw[wilson] (-2,0.0) -- (0.0,0.0);
\draw[wilson] (-2,0.0) -- (0.0,-2.0);
\draw[wilson] (2.0,2.0) -- (4,0.0);
\draw[wilson] (2.0,0.0) -- (4,0.0);
\draw[wilson] (2.0,-2.0) -- (4,0.0);
\draw[scalarnoarrow] (1.0,2.5) -- (1.0,-2.5);
\draw[scalarnoarrow] (1.0,2.5) -- (1.8,3.0);
\draw[scalarnoarrow] (1.0,-2.5) -- (0.2,-3.0);
\draw[gluon] (-0.6,-1.4) -- (2.6,-1.4);
\draw[gluon] (-0.6, 1.4) -- (3.0, 0.0);
\node at (-0.5,-2.0) {\Large $\ident_i$};
\node at (2.5,-2.0) {\Large $\ident_i$};
\node at (2.5,-0.4) {\Large $\ident_k$};
\node at (-0.5,2.0) {\Large $\ident_j$};
\node at (1.0,-3.5) {\Large $(D_{22})$};
\hspace{8cm}
\draw[wilson] (-2,0.0) -- (0.0,2.0);
\draw[wilson] (-2,0.0) -- (0.0,0.0);
\draw[wilson] (-2,0.0) -- (0.0,-2.0);
\draw[wilson] (2.0,2.0) -- (4,0.0);
\draw[wilson] (2.0,0.0) -- (4,0.0);
\draw[wilson] (2.0,-2.0) -- (4,0.0);
\draw[scalarnoarrow] (1.0,2.5) -- (1.0,-2.5);
\draw[scalarnoarrow] (1.0,2.5) -- (1.8,3.0);
\draw[scalarnoarrow] (1.0,-2.5) -- (0.2,-3.0);
\draw[gluon] (-0.6,-1.4) -- (3.0,0.0);
\draw[gluon] (-0.6, 1.4) -- (1.32,-0.28);
\draw[gluon] (1.864, -0.756) -- (2.6,-1.4);
\node at (-0.5,-2.0) {\Large $\ident_i$};
\node at (2.5,-2.0) {\Large $\ident_i$};
\node at (2.5,0.4) {\Large $\ident_k$};
\node at (-0.5,2.0) {\Large $\ident_j$};
\node at (1.0,-3.5) {\Large $(D_{23})$};
\end{tikzpicture}\\
\begin{tikzpicture}[line width=1 pt, scale=0.9]
\hspace{0cm}
\draw[wilson] (-2,0.0) -- (0.0,2.0);
\draw[wilson] (-2,0.0) -- (0.0,0.0);
\draw[wilson] (-2,0.0) -- (0.0,-2.0);
\draw[wilson] (2.0,2.0) -- (4,0.0);
\draw[wilson] (2.0,0.0) -- (4,0.0);
\draw[wilson] (2.0,-2.0) -- (4,0.0);
\draw[scalarnoarrow] (1.0,2.5) -- (1.0,-2.5);
\draw[scalarnoarrow] (1.0,2.5) -- (1.8,3.0);
\draw[scalarnoarrow] (1.0,-2.5) -- (0.2,-3.0);
\draw[gluon] (-1.0,-1.0) -- (2.5,1.5);
\draw[gluon] (-0.4,-1.6) -- (3.0,0.0);
\node at (-0.5,-2.0) {\Large $\ident_i$};
\node at (2.5,2.0) {\Large $\ident_j$};
\node at (2.5,0.4) {\Large $\ident_k$};
\node at (1.0,-3.5) {\Large $(D_{24})$};
\hspace{8cm}
\draw[wilson] (-2,0.0) -- (0.0,2.0);
\draw[wilson] (-2,0.0) -- (0.0,0.0);
\draw[wilson] (-2,0.0) -- (0.0,-2.0);
\draw[wilson] (2.0,2.0) -- (4,0.0);
\draw[wilson] (2.0,0.0) -- (4,0.0);
\draw[wilson] (2.0,-2.0) -- (4,0.0);
\draw[scalarnoarrow] (0.6,2.5) -- (0.6,-2.5);
\draw[scalarnoarrow] (0.6,2.5) -- (1.4,3.0);
\draw[scalarnoarrow] (0.6,-2.5) -- (-0.2,-3.0);
\draw[scalarnoarrow] (1.6,2.5) -- (1.6,-2.5);
\draw[scalarnoarrow] (1.6,2.5) -- (2.4,3.0);
\draw[scalarnoarrow] (1.6,-2.5) -- (0.8,-3.0);
\draw[gluon] (-0.6,-1.4) -- (3.0,0.0);
\draw[gluon] (-0.6, 1.4) -- (1.2,-0.6);
\node at (-0.5,-2.0) {\Large $\ident_i$};
\filldraw [red] (1.2,-0.6) circle (3pt);
\node at (2.5,0.4) {\Large $\ident_k$};
\node at (-0.5,2.0) {\Large $\ident_j$};
\node at (-0.2,-3.5) {\Large $(D_{25,1})$};
\node at (2.0,-3.5) {\Large $(D_{25,2})$};
\end{tikzpicture}\\
\begin{tikzpicture}[line width=1 pt, scale=0.9]
  \hspace{0cm}
\draw[wilson] (-2,0.0) -- (0.0,2.0);
\draw[wilson] (-2,0.0) -- (-1.0,0.0);
\draw[wilson] (-0.3,0.0) -- (0.0,0.0);
\draw[wilson] (-2,0.0) -- (0.0,-2.0);
\draw[wilson] (2.0,2.0) -- (4,0.0);
\draw[wilson] (2.0,0.0) -- (4,0.0);
\draw[wilson] (2.0,-2.0) -- (4,0.0);
\draw[scalarnoarrow] (1.0,2.5) -- (1.0,-2.5);
\draw[scalarnoarrow] (1.0,2.5) -- (1.8,3.0);
\draw[scalarnoarrow] (1.0,-2.5) -- (0.2,-3.0);
\draw[gluon] (-0.4,-1.6) -- (3.0,0.0);
\draw[gluon] (-0.7, -1.3) -- (-0.7,1.3);
\node at (-0.5,-2.0) {\Large $\ident_i$};
\node at (-0.5,2.0) {\Large $\ident_j$};
\node at (2.5,0.4) {\Large $\ident_k$};
\node at (1.0,-3.5) {\Large $(D_{26})$};
\hspace{8cm}
\draw[wilson] (-2,0.0) -- (0.0,2.0);
\draw[wilson] (-2,0.0) -- (-0.6,0.0);
\draw[wilson] (-0.2,0.0) -- (0.0,0.0);
\draw[wilson] (-2,0.0) -- (0.0,-2.0);
\draw[wilson] (2.0,2.0) -- (4,0.0);
\draw[wilson] (2.0,0.0) -- (4,0.0);
\draw[wilson] (2.0,-2.0) -- (4,0.0);
\draw[scalarnoarrow] (1.0,2.5) -- (1.0,-2.5);
\draw[scalarnoarrow] (1.0,2.5) -- (1.8,3.0);
\draw[scalarnoarrow] (1.0,-2.5) -- (0.2,-3.0);
\draw[gluon] (-1.2,-0.8) -- (-0.36,-0.64);
\draw[gluon] (-0.15,-0.6) -- (3.0,0.0);
\draw[gluon] (-0.4, -1.6) -- (-0.4,1.6);
\node at (-0.5,-2.0) {\Large $\ident_i$};
\node at (-0.5,2.0) {\Large $\ident_j$};
\node at (2.5,0.4) {\Large $\ident_k$};
\node at (1.0,-3.5) {\Large $(D_{27})$};
\end{tikzpicture}
\caption{Same as figure \ref{fig:oneWL}, but for the three-Wilson-line case.\label{fig:threeWL}}
\end{figure}

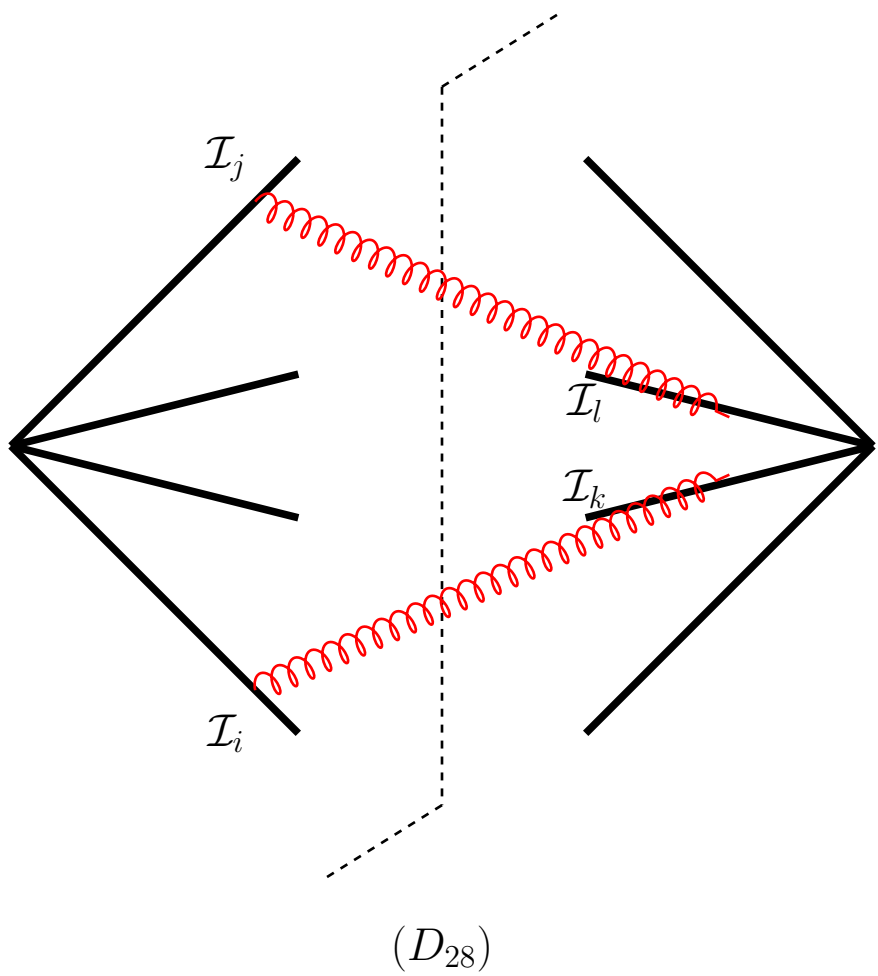
\begin{figure}[h!]
\begin{center}
\begin{tikzpicture}[line width=1 pt, scale=1.9]
\draw[wilson] (-2,0.0) -- (0.0,2.0);
\draw[wilson] (-2,0.0) -- (0.0,0.5);
\draw[wilson] (-2,0.0) -- (0.0,-0.5);
\draw[wilson] (-2,0.0) -- (0.0,-2.0);
\draw[wilson] (2.0,2.0) -- (4,0.0);
\draw[wilson] (2.0,0.5) -- (4,0.0);
\draw[wilson] (2.0,-0.5) -- (4,0.0);
\draw[wilson] (2.0,-2.0) -- (4,0.0);
\draw[scalarnoarrow] (1.0,2.5) -- (1.0,-2.5);
\draw[scalarnoarrow] (1.0,2.5) -- (1.8,3.0);
\draw[scalarnoarrow] (1.0,-2.5) -- (0.2,-3.0);
\draw[gluon] (-0.3,-1.7) -- (3.0,-0.2);
\draw[gluon] (-0.3, 1.7) -- (3.0, 0.2);
\node at (-0.5,-2.0) {\LARGE $\ident_i$};
\node at (2.0, 0.3) {\LARGE $\ident_l$};
\node at (2.0, -0.3) {\LARGE $\ident_k$};
\node at (-0.5,2.0) {\LARGE $\ident_j$};
\node at (1.0,-3.5) {\LARGE $(D_{28})$};
\end{tikzpicture}
\end{center}
\caption{Same as figure \ref{fig:oneWL}, but for the four-Wilson-line case.\label{fig:fourWL}} 
\end{figure}

With the general two-loop soft function at hand, we now discuss how to calculate the two-loop TMD soft function for heavy-quark hadroproduction at threshold. We first consider the azimuthal-angle-averaged soft function in the $q_T$ space, which is equivalent to the azimuthal-angle-dependent case. In this context, the kernel function takes the form given in eq.\eqref{eq:KSfun4}, combined with the rapidity regulator of eq.\eqref{eq:rapidityregulator}. The complex color factors and phase-space integrals in the generic cases simplify considerably, as will become evident in the next two subsections. Without loss of generality, we focus on the color-octet ($C=8$) case, since the integrals in the color-singlet case are a subset of those in the color-octet case. Moreover, the color-singlet TMD soft function--like in any hadroproduction processes involving only colorless final states--consists solely of scaleless integrals in the analytic rapidity regulator scheme and therefore vanishes.

\subsubsection{Calculation of double-real contributions}

In this section, we discuss the calculation of the double-real contributions to the NNLO bare TMD soft function, whose generic expression is given in eq.~\eqref{eq:softfunRR}. Some integrals, such as $I_i^{(2,D_n)}$ for $i=1,2$ and $1\leq n\leq 3$, vanish, and the four Wilson-line diagram $D_{28}$ does not contribute in our specific case. 

In order to evaluate the remaining integrals, similar to the treatment in section 3.4 of ref.~\cite{Catani:2023tby}, we insert the following identity:
\begin{equation}
\mu^{d-4}(2\pi)^d\int{[dq]\delta^{(d)}(k_1+k_2-q)}=1\,.
\end{equation}
This allows the two-body phase space measure in eq.~\eqref{eq:softfunRR} to be rewritten as
\begin{equation}
\begin{aligned}
[dk_1][dk_2](2\pi)\delta^{(+)}(k_1^2)(2\pi)\delta^{(+)}(k_2^2)K_S(k_1+k_2)=&[dq]K_S(q)[dk_2](2\pi)\delta^{(+)}((q-k_2)^2)(2\pi)\delta^{(+)}(k_2^2)\,.
\end{aligned}\label{eq:ps_q}
\end{equation}
Note that the rapidity regulator is applied only to the integration over $q$ via $K_S(q)$ on the right-hand side of eq.~\eqref{eq:ps_q}, which is justified since the integration over $k_2$ does not exhibit any rapidity divergence in our case.
Using the Cutkosky rules~\cite{Cutkosky:1960sp}, the Dirac delta functions involving $k_2$ in the integration measure can be replaced by differences of two Feynman propagators:
\begin{equation}
(2\pi)\delta^{(+)}(k^2-m^2)=\frac{i}{k^2-m^2+i0^+}-\frac{i}{k^2-m^2-i0^+}\,.\label{eq:Cutkoskyeq}
\end{equation}
With this replacement, the phase-space integrals are converted into Feynman loop integrals, which makes it possible to apply integration-by-parts (IBP) identities. These integrals are then reduced to linear combinations of eight distinct master integrals after integrating over $k_2$, but before integrating over $q$. The reductions are performed using the package \kira\ v2.3~\cite{Maierhofer:2017gsa,Klappert:2020nbg}. The master integrals are given by
\begin{equation}
\begin{aligned}
J_m^{(2,\mathrm{RR})}(q)=&\int{[dk_2]\delta^{(+)}((q-k_2)^2)(2\pi)\delta^{(+)}(k_2^2)h_m^{(2,\mathrm{RR})}(q-k_2,k_2)}\,,\quad m\in\left\{1,\ldots,8\right\}\,\label{eq:RRMI1}
\end{aligned}
\end{equation}
with
\begin{equation}
\begin{aligned}
&\left\{h_{1}^{(2,\mathrm{RR})}(q-k_2,k_2),\ldots,h_{8}^{(2,\mathrm{RR})}(q-k_2,k_2)\right\}\\
=&\Bigg\{1,\frac{1}{v\cdot k_2},\frac{1}{n\cdot k_2 \bar{n}\cdot k_2},\frac{1}{n\cdot k_2 \bar{n}\cdot (q-k_2)},\\
&\frac{1}{n\cdot k_2 v\cdot k_2},\frac{1}{\bar{n}\cdot k_2 v\cdot k_2},\frac{1}{n\cdot k_2 v\cdot (q-k_2)},\frac{1}{\bar{n}\cdot k_2 v\cdot (q-k_2)}\Bigg\}\,,\label{eq:RRMI2}
\end{aligned}
\end{equation}
where we have used eq.\eqref{eq:Cutkoskyeq} to revert to Dirac delta functions on the right-hand side. The double-real TMD soft function in the color-octet channel is then
\begin{align}
\bm{S}_{\perp,\Ione\Itwo}^{[8],(2,\mathrm{RR})}(q_T,\mu)&=16\pi^2\alpha_s^2C_A^2\bm{1}\int[dq]{(2\pi)K_S(q)}\nonumber\\
&\times\Bigg\{ \bigg[\frac{2}{q^2}\frac{1}{(v\cdot q)^2-q^2}-\frac{2}{n \cdot q \, \bar{n} \cdot q}\frac{1}{(v \cdot q)^2-q^2 }\nonumber\\
&+ \frac{2 \epsilon ^2+2 \epsilon -9+8\epsilon(1-\epsilon)T_Fn_qC_A^{-1}}{2\epsilon  (3 - 2 \epsilon)} \frac{4}{q^2 \,  n \cdot q \, \bar{n} \cdot q }\nonumber\\
&+ \frac{16 \epsilon ^2-35 \epsilon +17-4(1-\epsilon)T_Fn_qC_A^{-1}}{(3-2 \epsilon )}\frac{1}{q^2 \, (v \cdot q)^2}\bigg] J^{(2, \mathrm{RR})}_1(q)+ \bigg[\frac{2 v \cdot q}{q^2 \, n \cdot q \, \bar{n} \cdot q} \nonumber\\
&+\frac{ v \cdot q}{n \cdot q \, \bar{n} \cdot q }\frac{1}{\, (v \cdot q)^2-q^2}+\left(1+4 \epsilon\right)\frac{1}{q^2 \, v \cdot q}-\frac{v \cdot q}{q^2}\frac{1}{(v \cdot q)^2-q^2 }+\frac{1}{(v \cdot q)^3}\bigg] J^{(2,\mathrm{RR})}_2(q)\nonumber\\
&+ \bigg[\frac{12}{n \cdot q \, \bar{n} \cdot q }-\frac{2}{n \cdot q \, v \cdot q}-4 (1+2 \epsilon) \frac{\bar{n} \cdot q \,}{q^2 \, v \cdot q}-\frac{3}{(v \cdot q)^2}+\frac{4(1+2\epsilon)}{q^2}\bigg]J^{(2,\mathrm{RR})}_3(q) \nonumber \\
&+\bigg(\frac{14}{n \cdot q \, \bar{n} \cdot q}+\frac{2}{q^2}\bigg)J^{(2, \mathrm{RR})}_4(q)+2(1+2\epsilon)\bigg(\frac{\bar{n} \cdot q }{q^2 \, v \cdot q}-\frac{1}{q^2}\bigg)J^{(2,\mathrm{RR})}_5(q) \nonumber \\
&+ \bigg(\frac{2}{n \cdot q \, v \cdot q}-\frac{2}{n \cdot q \, \bar{n} \cdot q }+2(1+2 \epsilon)\frac{ \bar{n} \cdot q}{q^2 \, v \cdot q} - \frac{2(1+2 \epsilon)}{q^2 }\bigg]J^{(2,\mathrm{RR})}_6(q)\nonumber \\
&+\bigg[\frac{1}{q^2}-\frac{n \cdot q }{q^2 \, v \cdot q}-\frac{1}{2n \cdot q \, v \cdot q}-\frac{3}{2(v \cdot q)^2}\bigg]J^{(2,\mathrm{RR})}_7(q)\nonumber \\
&+\bigg[\frac{1}{q^2}-\frac{\bar{n} \cdot q \,}{q^2 \, v \cdot q}-\frac{1}{2\bar{n} \cdot q \, v \cdot q}-\frac{3}{2(v \cdot q)^2}\bigg]J^{(2,\mathrm{RR})}_8(q)-\frac{2}{v \cdot q}\bigg[J^{(2,\mathrm{RR})}_9(q)+J^{(2,\mathrm{RR})}_{10}(q)\bigg]\Bigg\}\nonumber\\
=&16\pi^2\alpha_s^2C_A^2\bm{1}\int[dq]{(2\pi)K_S(q)J^{(2,\mathrm{RR})}_{\mathrm{sum}}(q)}\,.\label{eq:DRSfun1}
\end{align}
In our specific case, the color factor of the soft function reduces to the identity, as a consequence of our special $2\to 1$ kinematics and the presence of only three Wilson lines.
Note that the integrals involving $J_1^{(2,\mathrm{RR})}(q)$ and $J_2^{(2,\mathrm{RR})}(q)$ do not exhibit rapidity divergences; therefore, the rapidity regulator $\alpha_1$ can be set to zero at the integrand level.
Here, $J^{(2,\mathrm{RR})}_9(q)$ and $J^{(2,\mathrm{RR})}_{10}(q)$ can be expressed as linear combinations of the master integrals $J^{(2,\mathrm{RR})}_m(q)$ with $1\leq m\leq 8$, as follows:
\begin{align}
J^{(2,\mathrm{RR})}_9(q) &=\int[dk_2]\delta^{(+)}((q-k_2)^2)(2\pi)\delta^{(+)}(k_2^2)\frac{1}{n\cdot k_2 (v\cdot (q-k_2))^2}  \nonumber \\
&=8(1-2\epsilon)\frac{v\cdot q}{q^2\left(2n\cdot q \, v\cdot q -q^2\right)} J^{(2, \mathrm{RR})}_1(q) + 8\epsilon\frac{\left(v\cdot q\right)^2 - q^2}{q^2\left(2n\cdot q \, v\cdot q -q^2\right)}J^{(2,\mathrm{RR})}_2(q) \nonumber \\
&+ 2(1+2\epsilon)\frac{n\cdot q}{2n\cdot q \, v\cdot q -q^2}J^{(2,\mathrm{RR})}_7(q) \,,\nonumber \\
J^{(2,\mathrm{RR})}_{10}(q) &= J^{(2,\mathrm{RR})}_9(q) \bigg|_{n\to \bar{n}} \,.
\end{align}
However, because the new denominators $(2n\cdot q \, v\cdot q -q^2)$ and $(2\bar{n}\cdot q \, v\cdot q -q^2)$ introduced in the IBP reductions of $J^{(2,\mathrm{RR})}_9(q)$ and $J^{(2,\mathrm{RR})}_{10}(q)$ complicate the subsequent integration over $q$, we choose to retain $J^{(2,\mathrm{RR})}_9(q)$ and $J^{(2, \mathrm{RR})}_{10}(q)$ in their original form rather than rewriting them in terms of the other $J_m^{(2,\mathrm{RR})}(q)$ functions. 

To calculate the master integrals in eqs.\eqref{eq:RRMI1} and \eqref{eq:RRMI2}, we begin with the following expression:
\begin{equation}
\begin{aligned}
J_m^{(2,\mathrm{RR})}(q)=&\int{[dk_2](2\pi)\delta^{(+)}(k_2^2)\delta^{(+)}((q-k_2)^2)h_m^{(2,\mathrm{RR})}(q-k_2,k_2)}\\
=&\mu^{4-d}\int{\frac{d^{d-1}\vec{k}_2}{(2\pi)^{d-1}2|\vec{k}_2|}\delta^{(+)}((q-k_2)^2)h_m^{(2,\mathrm{RR})}(q-k_2,k_2)}\\
=&\mu^{4-d}\int{d\Omega^{(d-1)}\int_0^{+\infty}{\frac{d|\vec{k}_2|}{2(2\pi)^{d-1}}|\vec{k}_2|^{d-3}\delta^{(+)}((q-k_2)^2)h_m^{(2,\mathrm{RR})}(q-k_2,k_2)}}\,.
\end{aligned}
\end{equation}
Since the integral $J_m^{(2,\mathrm{RR})}(q)$ is Lorentz invariant, it is convenient to evaluate it in the rest frame of $q$, where $q^\mu=(\sqrt{q^2},\vec{0})=(Q,\vec{0})$. In this frame, the delta function becomes
\begin{equation}
\delta^{(+)}((q-k_2)^2)=\delta(Q(Q-2|\vec{k}_2|))\Theta(Q-|\vec{k}_2|)\,,
\end{equation}
which allows us to perform the radial integration straightforwardly and obtain
\begin{equation}
\begin{aligned}
J_m^{(2,\mathrm{RR})}(q)=&\frac{1}{(4\pi)^{d-1}}\left(\frac{\mu^2}{q^2}\right)^{\frac{4-d}{2}}\int{d\Omega^{(d-1)}h_m^{(2,\mathrm{RR})}(q-k_2,k_2)}\,,
\end{aligned}
\end{equation}
where $|\vec{k}_2|=Q/2$. For the case $m=1$, the angular integral is straightforward, and we find
\begin{equation}
\begin{aligned}
J_1^{(2,\mathrm{RR})}(q)=&\frac{\Omega^{(d-1)}}{(4\pi)^{d-1}}\left(\frac{\mu^2}{q^2}\right)^{\frac{4-d}{2}}=\frac{\pi^{1-\epsilon}}{(2\pi)^{3-2\epsilon}}\frac{\Gamma(2-\epsilon)}{\Gamma(3-2\epsilon)}\left(\frac{\mu^2}{q^2}\right)^{\epsilon}\,.\label{eq:J1RRexp}
\end{aligned}
\end{equation}
For the remaining integrals, we evaluate the angular part using the techniques outlined in ref.~\cite{Somogyi:2011ir}. In general, these integrals can be expressed in terms of Mellin–Barnes representations (cf. eq.(14) in ref.~\cite{Somogyi:2011ir}):
\begin{eqnarray}
J_2^{(2,\mathrm{RR})}(q)&=&\left(\frac{\mu^2}{q^2}\right)^\epsilon\frac{1}{v\cdot q}\frac{1}{(4\pi)^{3-2\epsilon}}\frac{2^{2-2\epsilon}\pi^{1-\epsilon}}{\Gamma(1-2\epsilon)}\label{eq:J2RRexp}\\
&&\times\int_{-i\infty}^{+i\infty}{\frac{dz_1}{2\pi i}\left(\frac{q^2}{4(v\cdot q)^2}\right)^{z_1}\Gamma(-z_1)\Gamma(1+2z_1)\Gamma(-\epsilon-z_1)}\nonumber\\
&=&\left(\frac{\mu^2}{q^2}\right)^\epsilon\frac{1}{v\cdot q}\frac{\pi^{1-\epsilon}}{(2\pi)^{3-2\epsilon}}\frac{\Gamma(1-\epsilon)}{\Gamma(2-2\epsilon)}{}_2\hspace{-0.7mm}F_1\left(\frac{1}{2},1; \frac{3}{2}-\epsilon; 1-\frac{q^2}{(v\cdot q)^2}\right)\nonumber\,,\\
J_3^{(2,\mathrm{RR})}(q)&=&\left(\frac{\mu^2}{q^2}\right)^\epsilon\frac{1}{n\cdot q\bar{n}\cdot q}\frac{1}{(4\pi)^{3-2\epsilon}}\frac{2^{2-2\epsilon}\pi^{1-\epsilon}}{\Gamma(-2\epsilon)}\label{eq:J3RRexp}\\
&&\times\int_{-i\infty}^{+i\infty}{\frac{dz_1}{2\pi i}\left(\frac{q^2}{(n\cdot q)(\bar{n}\cdot q)}\right)^{z_1}\Gamma(-z_1)\Gamma^2(1+z_1)\Gamma(-1-\epsilon-z_1)}\nonumber\,,\\
J_4^{(2,\mathrm{RR})}(q)&=&\left(\frac{\mu^2}{q^2}\right)^\epsilon\frac{1}{n\cdot q\bar{n}\cdot q}\frac{1}{(4\pi)^{3-2\epsilon}}\frac{2^{2-2\epsilon}\pi^{1-\epsilon}}{\Gamma(-2\epsilon)}\label{eq:J4RRexp}\\
&&\times\int_{-i\infty}^{+i\infty}{\frac{dz_1}{2\pi i}\left(1-\frac{q^2}{(n\cdot q)(\bar{n}\cdot q)}\right)^{z_1}\Gamma(-z_1)\Gamma^2(1+z_1)\Gamma(-1-\epsilon-z_1)}\nonumber\\
&=&\left(\frac{\mu^2}{q^2}\right)^\epsilon\frac{1}{n\cdot q\bar{n}\cdot q}\frac{1}{(4\pi)^{3-2\epsilon}}\frac{2^{2-2\epsilon}\pi^{1-\epsilon}\Gamma^2(-\epsilon)}{\Gamma(-2\epsilon)\Gamma(1-\epsilon)}{}_2\hspace{-0.7mm}F_1\left(1,1; 1-\epsilon; \frac{q^2}{(n\cdot q)(\bar{n}\cdot q)}\right)\,,\nonumber\\
J_5^{(2,\mathrm{RR})}(q)&=&\left(\frac{\mu^2}{q^2}\right)^\epsilon\frac{1}{n\cdot q v\cdot q}\frac{1}{(4\pi)^{3-2\epsilon}}\frac{2^{2-2\epsilon}\pi^{1-\epsilon}}{\Gamma(-2\epsilon)}\label{eq:J5RRexp}\\
&&\times\int_{-i\infty}^{+i\infty}{\frac{dz_1}{2\pi i}\frac{dz_2}{2\pi i}\left(\frac{q^2}{4(v\cdot q)^2}\right)^{z_1}\left(\frac{q^2}{2(n\cdot q)(v\cdot q)}\right)^{z_2}}\nonumber\\
&&\times\Gamma(-z_1)\Gamma(-z_2)\Gamma(1+z_2)\Gamma(1+2z_1+z_2)\Gamma(-1-\epsilon-z_1-z_2)\,,\nonumber\\
J_6^{(2,\mathrm{RR})}(q)&=&\left.J_5^{(2,\mathrm{RR})}(q)\right|_{n\to\bar{n}}\,,\label{eq:J6RRexp}\\
J_7^{(2,\mathrm{RR})}(q)&=&\left(\frac{\mu^2}{q^2}\right)^\epsilon\frac{1}{n\cdot q v\cdot q}\frac{1}{(4\pi)^{3-2\epsilon}}\frac{2^{2-2\epsilon}\pi^{1-\epsilon}}{\Gamma(-2\epsilon)}\label{eq:J7RRexp}\\
&&\times\int_{-i\infty}^{+i\infty}{\frac{dz_1}{2\pi i}\frac{dz_2}{2\pi i}\left(\frac{q^2}{4(v\cdot q)^2}\right)^{z_1}\left(1-\frac{q^2}{2(n\cdot q)(v\cdot q)}\right)^{z_2}}\nonumber\\
&&\times\Gamma(-z_1)\Gamma(-z_2)\Gamma(1+z_2)\Gamma(1+2z_1+z_2)\Gamma(-1-\epsilon-z_1-z_2)\,,\nonumber\\
J_8^{(2,\mathrm{RR})}(q)&=&\left.J_7^{(2,\mathrm{RR})}(q)\right|_{n\to\bar{n}}\,,\label{eq:J8RRexp}\\
J_9^{(2,\mathrm{RR})}(q)&=&\left(\frac{\mu^2}{q^2}\right)^\epsilon\frac{1}{n\cdot q \, \left(v\cdot q\right)^2}\frac{1}{\left(4\pi\right)^{3-2\epsilon}}\frac{2^{2-2\epsilon}\pi^{1-\epsilon}}{\Gamma\left(-1-2\epsilon\right)}\label{eq:J9RRexp}\\
&&\times\int_{-i\infty}^{+i\infty}\frac{dz_1}{2\pi i}\frac{dz_2}{2\pi i}\left(\frac{q^2}{4\left(v\cdot q\right)^2}\right)^{z_1}\left(1-\frac{q^2}{2(n\cdot q)(v\cdot q)}\right)^{z_2} \nonumber \\ 
&&\times\Gamma\left(-z_1\right)\Gamma\left(-z_2\right)\Gamma\left(1+z_2\right)\Gamma\left(2+2z_1+z_2\right)\Gamma\left(-2-\epsilon-z_1-z_2\right)\,, \nonumber\\
J_{10}^{(2,\mathrm{RR})}(q)&=&\left.J_9^{(2,\mathrm{RR})}(q)\right|_{n \to \bar{n}}\,.\label{eq:J10RRexp}
\end{eqnarray}  
In the integrals involving denominators with $q-k_2$, we have used the fact that in the rest frame of $q$
\begin{equation}
(q-k_2)^\mu=(|\vec{k}_2|,-\vec{k}_2)\,,
\end{equation}
so that
\begin{equation}
p\cdot (q-k_2)=\bar{p}\cdot k_2\,,
\end{equation}
where $\bar{p}^\mu=(p^0,-\vec{p})$ for any four-momentum $p^\mu=(p^0,\vec{p})$.~\footnote{Since this may be confusing, we explicitly emphasize that the bar operation on momentum is not Lorentz invariant and therefore frame dependent. Although $n^\mu$ and $\bar{n}^\mu$ are conjugate in the partonic center-of-mass frame, their three-momenta are not generally opposite in the rest frame of $q$. In other words, in this frame, $\bar{\bar{n}}^\mu\neq n^\mu$, where $\bar{n}^\mu$ refers to $p_2^\mu/p_2^0$, with $p_2^\mu$ being the four-momentum of the second initial-state particle.} In this case, we also have
\begin{eqnarray}
\bar{p}\cdot q&=&p\cdot q,\quad \frac{q^2 p_i\cdot \bar{p}_j}{2p_i\cdot q \bar{p}_j\cdot q}=\frac{2p_i^0 p_j^0-p_i\cdot p_j}{2p_i^0 p_j^0}=1-\frac{q^2p_i\cdot p_j}{2p_i\cdot q p_j\cdot q}\,.
\end{eqnarray}
Although the above master integrals can, in principle, be expressed in terms of Gaussian hypergeometric functions, ${}_2\hspace{-0.7mm}F_1()$, or the Appell hypergeometric function of the first kind, the Mellin-Barnes representation is more convenient for performing the integration over $q$.

To evaluate the integration over $q$, we first introduce the identity
\begin{equation}
1=\int{dQ^2\delta^{(+)}(q^2-Q^2)}\,.
\end{equation}
In light-cone coordinates in the partonic rest frame, we have
\begin{equation}
\int{[dq](2\pi)\delta^{(+)}(q^2-Q^2)}=\frac{\mu^{4-d}}{2(2\pi)^{d-1}}\int_0^{+\infty}{dq_{+}dq_{-}\int_{-\infty}^{+\infty}{d^{d-2}\harpoon{q}_{T}}\delta(q_{+}q_{-}-q_{T}^2-Q^2)}\,.
\end{equation}
We then perform the change of variables
\begin{equation}
x\equiv \frac{Q^2}{q_T^2}\,,\quad y\equiv \frac{q_+^2}{q_T^2}\,,
\end{equation}
and integrate out $q_{-}$ using the Dirac delta function. Equation \eqref{eq:DRSfun1} becomes
\begin{equation}
\begin{aligned}
\bm{S}_{\perp,\Ione\Itwo}^{[8],(2,\mathrm{RR})}(q_T,\mu)=&
16\pi^2\alpha_s^2C_A^2\bm{1}\int_0^{+\infty}{dQ^2\int{[dq](2\pi)\delta^{(+)}(q^2-Q^2)}K_S(q)J^{(2,\mathrm{RR})}_{\mathrm{sum}}(q)}\\
=&16\pi^2\alpha_s^2C_A^2\bm{1}\frac{\mu^{4-d}}{4(2\pi)^{d-1}}\\
&\times\int_0^{+\infty}{dx \int_0^{+\infty}{\frac{dy}{y}\int_{-\infty}^{+\infty}{\frac{d^{d-2}\harpoon{q}_{T}}{\Omega_q^{(d-2)}}q_T^{5-d}\delta(|\harpoon{q}_T|-q_T)\left(\frac{q_+}{q_-}\right)^{\alpha_1}J_{\mathrm{sum}}^{(2,\mathrm{RR})}(q)}}}\\
=&16\pi^2\alpha_s^2C_A^2\bm{1}\frac{\mu^{4-d}q_T^2}{4(2\pi)^{d-1}}\int_0^{+\infty}{dx dy (1+x)^{-\alpha_1}y^{-1+\alpha_1} J_{\mathrm{sum}}^{(2,\mathrm{RR})}(q)}\,,\label{eq:DRSfun2}
\end{aligned}
\end{equation}
where we have used the fact that none of the scalar products $q^2$, $n\cdot q$, $\bar{n}\cdot q$, or $v\cdot q$ depend on the angular direction of $\harpoon{q}_T$, so the integration over $\harpoon{q}_T$ can be performed trivially. We are now in a position to explain how the integration over $x$ and $y$ is evaluated. In the following, we will take a few representative examples to illustrate the techniques we apply. 

For integrals involving both $1/((v\cdot q)^2-q^2)$ and $J_1^{(2,\mathrm{RR})}(q)$, we first apply the Mellin-Barnes representation, which can be generally written as
\begin{equation}
\frac{1}{\left(\sum_{j=1}^{n}{A_j}\right)^\lambda}=\frac{2\pi i}{\Gamma(\lambda)}\int_{-i\infty}^{+i\infty}{\left(\prod_{j=1}^{n}{\frac{dz_j}{2\pi i}}\right)\delta(\lambda+\sum_{j=1}^{n}{z_j})\left(\prod_{j=1}^{n}{\Gamma(-z_j)A_j^{z_j}}\right)}\,,\label{eq:MellinBarnesrelation}
\end{equation}
as a standard technique. In our case, we express
\begin{equation}
\frac{1}{(v\cdot q)^2-q^2}=\int_{-i\infty}^{+i\infty}{\frac{dz_1}{2\pi i} (v\cdot q)^{2z_1} (-q^2)^{-1-z_1}\Gamma(-z_1) \Gamma(1+z_1)}\,.\label{eq:vqmq2MBint1}
\end{equation}
We then consider the following integral (setting $\alpha_1=0$, since there is no rapidity divergence in this case):
\begin{equation}
\begin{aligned}
&\int_0^{+\infty}{dxdy y^{-1} \frac{1}{q^2}\frac{1}{(v\cdot q)^2-q^2}J_1^{(2,\mathrm{RR})}(q)}\\
=&\mu^{2\epsilon}\frac{\pi^{1-\epsilon}}{(2\pi)^{3-2\epsilon}}\frac{\Gamma(2-\epsilon)}{\Gamma(3-2\epsilon)}\int_{-i\infty}^{+i\infty}{\frac{dz_1}{2\pi i}(-1)^{-1-z_1}\Gamma(-z_1)\Gamma(1+z_1)\int_0^{+\infty}{dx dy y^{-1}(v\cdot q)^{2z_1}(q^2)^{-2-z_1-\epsilon}}}\\
=&\left(\frac{\mu^2}{q_T^2}\right)^{\epsilon}\frac{1}{q_T^4}\frac{\pi^{1-\epsilon}}{(2\pi)^{3-2\epsilon}}\frac{\Gamma(2-\epsilon)\Gamma(1+\epsilon)}{\Gamma(3-2\epsilon)}\int_{-i\infty}^{+i\infty}{\frac{dz_1}{2\pi i}(-1)^{-1-z_1}2^{-2z_1}\frac{\Gamma^2(-z_1)\Gamma(1+z_1)\Gamma(-1-\epsilon-z_1)}{\Gamma(-2z_1)}}\\
=&\left(\frac{\mu^2}{q_T^2}\right)^{\epsilon}\frac{1}{q_T^4}\frac{\pi^{1-\epsilon}}{(2\pi)^{3-2\epsilon}}\frac{4\Gamma(-\epsilon)\Gamma(2-\epsilon)\Gamma(1+\epsilon)}{(1+2\epsilon)\Gamma(3-2\epsilon)}\,,
\end{aligned}
\end{equation}
where in the last step we have evaluated the Mellin–Barnes integral using Cauchy's residue theorem. Using the same method, we also obtain:
\begin{equation}
\begin{aligned}
&\int_0^{+\infty}{dxdy y^{-1} \frac{1}{n\cdot q \bar{n}\cdot q}\frac{1}{(v\cdot q)^2-q^2}J_1^{(2,\mathrm{RR})}(q)}\\
=&\left(\frac{\mu^2}{q_T^2}\right)^{\epsilon}\frac{1}{q_T^4}\frac{\pi^{1-\epsilon}}{(2\pi)^{3-2\epsilon}}\frac{2\Gamma(-\epsilon)\Gamma(2-\epsilon)\Gamma(1+\epsilon)}{\Gamma(3-2\epsilon)}\Bigg(\psi(\epsilon)-2\psi(2\epsilon)-\gamma_E\Bigg)\,,
\end{aligned}
\end{equation}
where $\psi(x)=d\log{\Gamma(x)}/dx$ is the digamma function.

To evaluate integrals involving both $1/((v\cdot q)^2-q^2)$ and $J_2^{(2,\mathrm{RR})}(q)$, we use the series representation of the hypergeometric function given in eq.\eqref{eq:J2RRexp}, which takes the form
\begin{eqnarray}
{}_2\hspace{-0.7mm}F_1(a,b;c;x)&=&\sum_{i=0}^{+\infty}{\frac{(a)_i(b)_i}{(c)_i}\frac{x^i}{i!}}\,,\label{eq:series42F1}
\end{eqnarray}
where $(a)_i\equiv\Gamma(a+i)/\Gamma(a)$ denotes the Pochhammer symbol. Using this expansion, we can express
\begin{equation}
\begin{aligned}
&\frac{1}{(v\cdot q)^2-q^2}{}_2\hspace{-0.7mm}F_1\left(\frac{1}{2},1; \frac{3}{2}-\epsilon; 1-\frac{q^2}{(v\cdot q)^2}\right)\\
=&\frac{1}{(v\cdot q)^2}\sum_{i=0}^{+\infty}{\frac{\left(\frac{1}{2}\right)_i(1)_i}{\left(\frac{3}{2}-\epsilon\right)_i}\frac{1}{i!}\left(1-\frac{q^2}{(v\cdot q)^2}\right)^{i-1}}\\
=&\frac{1}{(v\cdot q)^2-q^2}+\frac{1}{(v\cdot q)^2}\sum_{n=1}^{+\infty}{\frac{\left(\frac{1}{2}\right)_i(1)_i}{\left(\frac{3}{2}-\epsilon\right)_i}\frac{1}{i!}\left(1-\frac{q^2}{(v\cdot q)^2}\right)^{i-1}}\\
=&\frac{1}{(v\cdot q)^2-q^2}+\frac{1}{(v\cdot q)^2}\frac{1}{3-2\epsilon}{}_2\hspace{-0.7mm}F_1\left(1,\frac{3}{2}; \frac{5}{2}-\epsilon; 1-\frac{q^2}{(v\cdot q)^2}\right)\,,\label{eq:vqmq22F1trick}
\end{aligned}
\end{equation}
where in the last step we used a known identity to re-sum the remaining series.
This decomposition allows us to avoid introducing two-fold Mellin-Barnes representations. Instead, each term in eq.\eqref{eq:vqmq22F1trick} can be written as a one-fold Mellin-Barnes integral using eq.\eqref{eq:vqmq2MBint1} and the Barnes integral representation of the hypergeometric function (see eq.(6) in ref.~\cite{marcovecchio2019classicalfunctionalgeneralizationbarnes}):
\begin{equation}
\begin{aligned}
{}_2\hspace{-0.7mm}F_1(a,b;c;x)=&\frac{\Gamma(c)}{\Gamma(a)\Gamma(b)\Gamma(c-a)\Gamma(c-b)}\\
&\times\int_{-i\infty}^{+i\infty}{\frac{dz_1}{2\pi i}\Gamma(a+z_1)\Gamma(b+z_1)\Gamma(c-a-b-z_1)\Gamma(-z_1)(1-x)^{z_1}}\,.
\end{aligned}
\end{equation}
Using this representation, the integrations over $x$ and $y$ can be performed straightforwardly using properties of Euler’s gamma function. The final integration over $z_1$ then be evaluated using the residue theorem. As a result, we obtain:
\begin{eqnarray}
&&\int_0^{+\infty}{dx dy y^{-1}\frac{v\cdot q}{n\cdot q\bar{n}\cdot q}\frac{1}{(v\cdot q)^2-q^2}J_2^{(2,\mathrm{RR})}(q)}\nonumber\\
&=&\left(\frac{\mu^2}{q_T^2}\right)^\epsilon\frac{1}{q_T^4}\frac{\pi^{1-\epsilon}}{(2\pi)^{3-2\epsilon}}\frac{2\Gamma^2(1-\epsilon)\Gamma(\epsilon)}{\Gamma(2-2\epsilon)}\Bigg(\Gamma(1-2\epsilon)\Gamma(2\epsilon)+\psi(2\epsilon)+\gamma_E\Bigg)\,,\\
&&\int_0^{+\infty}{dx dy y^{-1}\frac{v\cdot q}{q^2}\frac{1}{(v\cdot q)^2-q^2}J_2^{(2,\mathrm{RR})}(q)}\nonumber\\
&=&\left(\frac{\mu^2}{q_T^2}\right)^\epsilon\frac{1}{q_T^4}\frac{\pi^{1-\epsilon}}{(2\pi)^{3-2\epsilon}}\frac{\Gamma^2(-\epsilon)\Gamma(\epsilon)\Gamma(-1-2\epsilon)}{2\Gamma^2(-2\epsilon)}\,.
\end{eqnarray}

The integrals involving $J_4^{(2,\mathrm{RR})}(q)$ vanish because they are scaleless. This becomes evident by examining the hypergeometric function representation in eq.\eqref{eq:J4RRexp} and the following Barnes integral representation of the hypergeometric function:
\begin{equation}
\begin{aligned}
{}_2\hspace{-0.7mm}F_1(a,b;c;x)=&\frac{\Gamma(c)}{\Gamma(a)\Gamma(b)}\int_{-i\infty}^{+i\infty}{\frac{dz_1}{2\pi i}\frac{\Gamma(a+z_1)\Gamma(b+z_1)}{\Gamma(c+z_1)}\Gamma(-z_1)(-x)^{z_1}}\,.
\end{aligned}
\end{equation}

For the remaining integrals that involve at least two-fold Mellin-Barnes representations, we employ the public {\sc\small Mathematica} packages {\tt MB.m}~\cite{Czakon:2005rk} and {\tt MBresolve.m}~\cite{Smirnov:2009up} for analytic continuation, {\tt barnesroutines.m}~\cite{Kosower:barnes} for applying the first and second Barnes lemmas, and {\tt MBsums}~\cite{Ochman:2015fho} for transforming the integrals into infinite series by closing contours and taking
residues. All of these open-source toolkits are available as part of the {\tt MBTools} ensemble~\cite{Belitsky:2022gba}. To evaluate these integrals, we first perform a Laurent expansion in the two infinitesimal regulators, $\alpha_1$ and $\epsilon$, prior to carrying out the Mellin–Barnes integrations. Since the limits $\alpha_1\to 0$ and $\epsilon\to 0$ do not generally commute, we adopt the prescription that the $\alpha_1\to 0$ limit is always taken before the $\epsilon\to 0$ expansion. At NNLO,  it suffices to retain terms up to $\mathcal{O}(\alpha_1^0)$ and $\mathcal{O}(\epsilon)$ in the respective expansions. However, as not all infinite series obtained from {\tt MBsums} can be evaluated using the built-in {\sc\small Mathematica} function {\tt Sum}, we also resort to numerical Mellin–Barnes integration via {\tt NIntegrate}, followed by application of the {\tt PSLQ} algorithm~\cite{Ferguson:1992,Ferguson:1999} to reconstruct analytic results. The analytic basis consists of Riemann zeta values $\zeta_n\equiv \zeta(n)$ and $\log{(2)}$, up to transcendental weight four.

A few comments are in order regarding the evaluation of the integrals involving $J_m^{(2,\mathrm{RR})}(q)$ for $7\leq m\leq 10$. When performing the integrations over the variables $x$ and $y$ in evaluating these integrals, we make use of eq.\eqref{eq:MellinBarnesrelation} to first rewrite expressions such as
\begin{equation}
\left[(1+x)^2+y\right]^{z_2}=\int_{-i\infty}^{+i\infty}{\frac{dz_3}{2\pi i}(1+x)^{2z_3}y^{z_2-z_3}\frac{\Gamma(-z_3)\Gamma(z_3-z_2)}{\Gamma(-z_2)}}\,,
\end{equation}
or
\begin{equation}
\left(1+x\right)^{-\alpha_1}=\int_{-i\infty}^{+i\infty}{\frac{dz_3}{2\pi i}x^{z_3}\frac{\Gamma(-z_3)\Gamma(z_3+\alpha_1)}{\Gamma(\alpha_1)}}\,.
\end{equation}
Additionally, we use the fact that the integrals involving $J_7^{(2,\mathrm{RR})}(q)$ (respectively $J_9^{(2,\mathrm{RR})}(q)$) are related to those involving $J_8^{(2,\mathrm{RR})}(q)$ (respectively $J_{10}^{(2,\mathrm{RR})}(q)$) by the replacement $n\leftrightarrow\bar{n}$, which is equivalent to taking $\alpha_1\to-\alpha_1$. All results of the integrals appearing in eq.\eqref{eq:DRSfun1} are collected in table \ref{tab:DRints} in appendix \ref{sec:tabofints}.

After combining all terms, we finally obtain
\begin{equation}
\begin{aligned}
\bm{S}_{\perp,\Ione\Itwo}^{[8],(2,\mathrm{RR})}(q_T,\mu)=&\left(\frac{\alpha_s}{2\pi}\right)^2\bm{1}\frac{\mu^{4\epsilon}q_T^{-2-2\epsilon}}{\pi^{1-\epsilon}}(4\pi)^{2\epsilon}e^{-\epsilon\gamma_E}\Bigg\{C_A^2\bigg[-\frac{1}{\epsilon^2}-\frac{1}{\epsilon}\left(\frac{11}{6}+\zeta_2\right)\\
&+\frac{7}{2}\zeta_2-\frac{49}{18}+\epsilon\left(-\frac{29}{4}\zeta_4+\frac{71}{6}\zeta_3+\zeta_2(\frac{95}{12}-6\log{(2)})-\frac{256}{27}\right)\bigg]\\
&+C_AT_Fn_q\bigg[\frac{2}{3\epsilon}+\frac{10}{9}+\epsilon\left(-\frac{\zeta_2}{3}+\frac{56}{27}\right)\bigg]+\mathcal{O}(\epsilon^2)\Bigg\}+\mathcal{O}(\alpha_1)\,,\label{eq:DRSfun3}
\end{aligned}
\end{equation}
where we have used the {\sc\small Mathematica} package {\tt HypExp}~\cite{Huber:2005yg,Huber:2007dx} to perform the series expansions of two generalized hypergeometric functions in $\epsilon$. Although the individual integrals in eq.\eqref{eq:DRSfun1} may contain rapidity divergences of the form $1/\alpha_1$, the combined result in eq.\eqref{eq:DRSfun3} is free of such divergences.

\subsubsection{Calculation of real-virtual contributions}

In this section, we turn to the calculation of the real-virtual contributions to the NNLO bare TMD soft function, whose general expression is given in eq.~\eqref{eq:softfunRV}. Some integrands, such as $I_i^{(2,D_n)}$ for $i=1,2$ and $5\leq n\leq 6$, vanish, while $I_{ijk}^{(2,D_{25,2})}$ does not contribute in our specific case, as its color structure does not contribute to color-singlet or color-octet heavy-quark pair production at threshold.~\footnote{The three-Wilson-line correlator $I_{ijk}^{(2,D_{25,2})}$ does, however, contribute to the imaginary part of the TMD soft function in generic cases, such as heavy-quark pair hadroproduction~\cite{Angeles-Martinez:2018mqh}.} 

The remaining one-loop integrals over the loop momentum $k_2$ are reduced to linear combinations of seven master integrals by applying the IBP identities generated with \kira. These master integrals are defined as
\begin{equation}
\begin{aligned}
J_m^{(2,\mathrm{RV})}(q)=&i\int{[dk_2]\frac{1}{k_2^2 + i0^+}h_m^{(2,\mathrm{RV})}(k_1,k_2)}\,,\quad m\in\left\{1,\ldots,7\right\}\,,\label{eq:RVMI1}
\end{aligned}
\end{equation}
where
\begin{equation}
\begin{aligned}
&\left\{h_{1}^{(2,\mathrm{RV})}(k_1,k_2),\ldots,h_{7}^{(2,\mathrm{RV})}(k_1,k_2)\right\}\\
=&\Bigg\{\frac{1}{v\cdot(k_1+k_2)+i0^+},\frac{1}{n\cdot(k_1+k_2)-i0^+}\frac{1}{v\cdot k_2-i0^+},\frac{1}{\bar{n}\cdot(k_1+k_2)-i0^+}\frac{1}{v\cdot k_2-i0^+},\\
&\frac{k_2^2+i0^+}{(k_1+k_2)^2+i0^+}\frac{1}{v\cdot k_2-i0^+},\frac{1}{(k_1+k_2)^2+i0^+}\frac{1}{n\cdot k_2+i0^+}\frac{1}{\bar{n}\cdot(k_1+k_2)-i0^+},\\
&\frac{1}{(k_1+k_2)^2+i0^+}\frac{1}{n\cdot k_2+i0^+}\frac{1}{v\cdot(k_1+k_2)+i0^+},\\
&\frac{1}{(k_1+k_2)^2+i0^+}\frac{1}{\bar{n}\cdot k_2+i0^+}\frac{1}{v\cdot(k_1+k_2)+i0^+}\Bigg\}\,.\label{eq:RVMI2}
\end{aligned}
\end{equation}
As a result, the real-virtual contribution to the TMD soft function for the color-octet channel is given by
\begin{align}
\bm{S}_{\perp,\Ione\Itwo}^{[8],(2,\mathrm{RV})}(q_T,\mu)&=16\pi^2\alpha_s^2C_A^2\bm{1}\int[dk_1]{(2\pi)\delta^{(+)}(k_1^2)K_S(k_1)}\nonumber\\
&\times\Bigg\{ \frac{1+2\epsilon}{8 \epsilon } \left[-\frac{2}{\left(n \cdot k_1\right)^2 \bar{n} \cdot k_1}+\frac{1}{\left(n \cdot k_1\right)^2 v \cdot k_1}+\frac{3-2 \epsilon}{1+2 \epsilon}\frac{1}{ \left(v \cdot k_1\right){}^3}\right]J^{(2, \mathrm{RV})}_1(k_1)\nonumber\\
&-\frac{1}{4}\frac{1}{ n \cdot k_1 v \cdot k_1}J^{(2, \mathrm{RV})}_2(k_1)-\frac{1}{4}\frac{1}{\bar{n} \cdot k_1 v \cdot k_1}J^{(2, \mathrm{RV})}_3(k_1)\nonumber \\
&+\frac{1+2\epsilon}{8 \epsilon } \left[-\frac{2}{\left(n \cdot k_1\right)^2 \bar{n} \cdot k_1}+\frac{1}{\left(n \cdot k_1\right)^2 v \cdot k_1}+\frac{1}{ \left(v \cdot k_1\right){}^3}\right]J^{(2, \mathrm{RV})}_4(k_1) \nonumber \\
&+\left(1-\frac{C(\Ione)+C(\Itwo)}{C_A}\right)J^{(2, \mathrm{RV})}_5(k_1)+\frac{(n-\bar{n})\cdot k_1}{4 v\cdot k_1}\left[J^{(2, \mathrm{RV})}_6(k_1)-J^{(2, \mathrm{RV})}_7(k_1)\right]\Bigg\}+\mathrm{c.c.}\nonumber\\
=&16\pi^2\alpha_s^2C_A^2\bm{1}\int[dk_1]{(2\pi)\delta^{(+)}(k_1^2)K_S(k_1)J^{(2,\mathrm{RV})}_{\mathrm{sum}}(k_1)}\,,\label{eq:RVSfun1}
\end{align}
where `c.c.' denotes the complex conjugate, which should be taken on the right-hand side of the first expression.

The master integrals defined in eqs.\eqref{eq:RVMI1} and \eqref{eq:RVMI2} can be evaluated using Feynman parameterization, the Cheng-Wu theorem~\cite{Cheng:1987ga}, and Mellin-Barnes representations. Their explicit expressions are given by
\begin{align}
J^{(2, \mathrm{RV})}_1(k_1)&= \frac{2e^{2\epsilon \pi i}\mu^{2\epsilon} }{\left(4\pi\right) ^{2-\epsilon}}  \Gamma (1-\epsilon ) \Gamma (2 \epsilon -1) \left(2v \cdot k_1\right)^{1-2 \epsilon }\,,  \\
J^{(2, \mathrm{RV})}_2(k_1)&= \frac{2\mu^{2\epsilon}}{\left(4\pi\right) ^{2-\epsilon}}\Gamma (1-2 \epsilon ) \Gamma (\epsilon ) \Gamma (2 \epsilon ) \left(n \cdot k_1\right){}^{-2 \epsilon }\,,  \\
J^{(2, \mathrm{RV})}_3(k_1)&=\left.J^{(2, \mathrm{RV})}_2(k_1)\right|_{n\to \bar{n}} \,, \\
J^{(2, \mathrm{RV})}_4(k_1)&= -\frac{2e^{2 \epsilon \pi i}\mu^{2\epsilon} }{\left(4\pi\right) ^{2-\epsilon}} \Gamma (1-\epsilon ) \Gamma (2 \epsilon -1) \left(2v \cdot k_1\right){}^{1-2 \epsilon } \,,\\
J^{(2, \mathrm{RV})}_5(k_1)&= \frac{e^{\epsilon \pi i}\mu^{2\epsilon} }{\left(4\pi\right) ^{2-\epsilon}}\frac{ \Gamma^3(-\epsilon )\Gamma^2(1+\epsilon)}{\Gamma (-2 \epsilon )}\left(n \cdot k_1\right)^{-\epsilon -1} \left(\bar{n} \cdot k_1\right)^{-\epsilon -1}\,,\\
J^{(2, \mathrm{RV})}_6(k_1)&= \frac{2e^{\epsilon \pi i}\mu^{2\epsilon} }{\left(4\pi\right) ^{2-\epsilon}\Gamma (-2 \epsilon )}\int_{-i\infty}^{+i\infty}\frac{dz_1}{2\pi i}\left(n \cdot k_1\right)^{z_1-\epsilon -1} \left(2v \cdot k_1\right)^{-z_1-\epsilon -1} \nonumber \\
&\times  e^{z_1\pi i} \Gamma \left(-z_1\right) \Gamma \left(-\epsilon -z_1\right) \Gamma \left(\epsilon -z_1+1\right) \Gamma^2\left(z_1-\epsilon \right) \Gamma \left(\epsilon +z_1+1\right) \nonumber \\
&=\frac{2e^{\epsilon \pi i}\mu^{2\epsilon} }{\left(4\pi\right) ^{2-\epsilon}\Gamma (-2 \epsilon )}\bigg[\Gamma^3(-\epsilon) \Gamma^2(1+\epsilon) \left(n \cdot k_1\right){}^{-\epsilon -1} \left(2v \cdot k_1\right)^{-\epsilon -1} \left(1-\frac{n \cdot k_1}{2 v \cdot k_1}\right)^{\epsilon } \nonumber \\
&+e^{(1+\epsilon)\pi i} \Gamma (-2 \epsilon -1) \Gamma (-\epsilon -1) \Gamma(2+2\epsilon) \left(2v \cdot k_1\right)^{-2 \epsilon -2} \, _2\hspace{-0.7mm}F_1\left(1,1;2+\epsilon;\frac{n \cdot k_1}{2 v \cdot k_1}\right) \nonumber  \\
&+\frac{e^{-\epsilon \pi i} }{2 v \cdot k_1}\Gamma^2(-2\epsilon) \Gamma (\epsilon) \Gamma(1+2\epsilon) \left(n \cdot k_1\right){}^{-2 \epsilon -1} \, _2\hspace{-0.7mm}F_1\left(1,-2 \epsilon ;1-\epsilon ;\frac{n \cdot k_1}{2 v \cdot k_1}\right)\bigg] \,, \label{eq:J6RVexp}\\
J^{(2, \mathrm{RV})}_7(k_1)&=\left.J^{(2, \mathrm{RV})}_6(k_1)\right|_{n\to \bar{n}} \,.
\label{eq:J7RVexp}
\end{align}

Analogous to the double-real contribution, we introduce the integration variable
\begin{equation}
y_1 \equiv \frac{k_{1,+}^2}{k_{1,T}^2}\,,
\end{equation}
such that the real-virtual soft function can be expressed as a one-fold integral over $y_1$:
\begin{equation}
\begin{aligned}
\bm{S}_{\perp,\Ione\Itwo}^{[8],(2,\mathrm{RV})}(q_T,\mu)=&
16\pi^2\alpha_s^2C_A^2\bm{1}\int{[dk_1](2\pi)\delta^{(+)}(k_1^2)}K_S(k_1)J^{(2,\mathrm{RV})}_{\mathrm{sum}}(k_1)\\
=&16\pi^2\alpha_s^2C_A^2\bm{1}\frac{\mu^{4-d}}{4(2\pi)^{d-1}}\\
&\times {\int_0^{+\infty}{\frac{dy_1}{y_1}\int_{-\infty}^{+\infty}{\frac{d^{d-2}\harpoon{k}_{1,T}}{\Omega_q^{(d-2)}}q_T^{3-d}\delta(|\harpoon{k}_{1,T}|-q_T)\left(\frac{k_{1,+}}{k_{1,-}}\right)^{\alpha_1}J_{\mathrm{sum}}^{(2,\mathrm{RV})}(k_1)}}}\\
=&16\pi^2\alpha_s^2C_A^2\bm{1}\frac{\mu^{4-d}}{4(2\pi)^{d-1}}\int_0^{+\infty}{ dy_1 y_1^{-1+\alpha_1} J_{\mathrm{sum}}^{(2,\mathrm{RV})}(k_1)}\,.\label{eq:RVSfun2}
\end{aligned}
\end{equation}
The integration over $y_1$ in eq.\eqref{eq:RVSfun2} is straightforward for the terms involving $J_m^{(2,\mathrm{RV})}(k_1)$ with $1\leq m\leq 4$, while the contribution from $J_5^{(2,\mathrm{RV})}(k_1)$ vanishes due to its scalelessness. All terms that do not involve $J_6^{(2,\mathrm{RV})}(k_1)$ or $J_7^{(2,\mathrm{RV})}(k_1)$ are free of rapidity divergences, and the regulator $\alpha_1$ can thus be safely set to zero prior to integration. To evaluate the integrals involving $J_6^{(2,\mathrm{RV})}(k_1)$ or $J_7^{(2,\mathrm{RV})}(k_1)$--which are related by the replacement $n\leftrightarrow\bar{n}$ (or equivalently $\alpha_1\to -\alpha_1$)--we employ the Mellin-Barnes representation instead of the hypergeometric form given in eq.\eqref{eq:J6RVexp}. The integration procedure closely parallels that used in the double-real case: we first perform the $y_1$ integration using the Mellin-Barnes representation and then apply Cauchy's residue theorem to obtain an analytic expression valid to all orders in $\alpha_1$ and $\epsilon$:
\begin{equation}
\begin{aligned}
&\int_0^{+\infty}{dy_1 y_1^{-1+\alpha_1}\frac{(n-\bar{n})\cdot k_1}{v\cdot k_1}J_6^{(2,\mathrm{RV})}(k_1)}\\
=&\left(\frac{\mu^2}{q_T^2}\right)^\epsilon\frac{1}{q_T^2}\frac{\pi^{1-\epsilon}}{(2\pi)^{3-2\epsilon}}\Bigg\{-\frac{2e^{\epsilon \pi i}(1+2\alpha_1+2\epsilon)\Gamma(-\alpha_1)\Gamma^2(1-\epsilon)\Gamma(-\epsilon)\Gamma^2(\epsilon)\Gamma(1+\alpha_1+2\epsilon)}{\Gamma(-2\epsilon)\Gamma(2+2\epsilon)}\\
&-\frac{2\Gamma(1-2\epsilon)\Gamma(1-\alpha_1-\epsilon)\Gamma(2\epsilon)\Gamma(\alpha_1+\epsilon)}{(1+2\epsilon)}\bigg[2\Gamma(1+\epsilon)+\frac{(1+2\alpha_1+2\epsilon)\Gamma(1-\epsilon)\Gamma(\epsilon)\Gamma(1+\alpha_1+2\epsilon)}{\Gamma(1+\alpha_1)\Gamma(1+\epsilon)}\bigg]\\
&-\frac{2e^{2\epsilon \pi i}\Gamma(1-2\epsilon)\Gamma(1-\epsilon)\Gamma(\epsilon)\Gamma(2\epsilon)\Gamma(1-\alpha_1+\epsilon)\Gamma(1+\alpha_1+\epsilon)}{\Gamma(-2\epsilon)}\\
&\times\bigg[\frac{1-\alpha_1+\epsilon}{\Gamma(3+\epsilon)\Gamma(4+2\epsilon)}{}_3F_2(2,2,2-\alpha_1+\epsilon;3+\epsilon,4+2\epsilon;1)\\
&-\frac{2\alpha_1}{\Gamma(2+\epsilon)\Gamma(3+2\epsilon)}{}_3F_2(1,1,1-\alpha_1+\epsilon;2+\epsilon,3+2\epsilon;1)\bigg]\Bigg\}\,.\label{eq:int4J2RV6}
\end{aligned}
\end{equation}
The two generalized hypergeometric functions ${}_3F_2()$ in eq.\eqref{eq:int4J2RV6} depend on both regulators $\alpha_1$ and $\epsilon$, and must be expanded in a Laurent series. For our purposes, it suffices to set $\alpha_1\to 0$ directly within the hypergeometric functions. The expansion in $\epsilon$ then be performed using the package {\tt HypExp}. Expressions for the remaining integrals in eq.\eqref{eq:RVSfun1} are listed in table \ref{tab:RVints} in appendix \ref{sec:tabofints}. The final real-virtual contribution to the NNLO TMD soft function in the color-octet channel is
\begin{equation}
\begin{aligned}
\bm{S}_{\perp,\Ione\Itwo}^{[8],(2,\mathrm{RV})}(q_T,\mu)=&\left(\frac{\alpha_s}{2\pi}\right)^2\bm{1}\frac{\mu^{4\epsilon}q_T^{-2-2\epsilon}}{\pi^{1-\epsilon}}(4\pi)^{2\epsilon}e^{-\epsilon\gamma_E}\Bigg\{C_A^2\bigg[\frac{1}{\epsilon^2}+\frac{1}{\epsilon}\left(\zeta_2-1\right)-\zeta_3-\frac{5}{2}\zeta_2\\
&+\epsilon\left(-\frac{33}{4}\zeta_4-\frac{13}{3}\zeta_3-\frac{11}{2}\zeta_2+4\right)\bigg]+\mathcal{O}(\epsilon^2)\Bigg\}+\mathcal{O}(\alpha_1)\,.\label{eq:RVSfun3}
\end{aligned}
\end{equation}
Once again, we observe that the rapidity divergence pole $1/\alpha_1$ cancels between the contributions from $J_6^{(2,\mathrm{RV})}(k_1)$ and $J_7^{(2,\mathrm{RV})}(k_1)$. Moreover, the double poles in $\epsilon$, \ie\ $1/\epsilon^2$, cancel out upon summing the double-real contribution in eq.\eqref{eq:DRSfun3} and the real-virtual part in eq.\eqref{eq:RVSfun3}. Finally, we find that the soft function is independent of the color representation of the initial-state partons $\Ione$ and $\Itwo$. In other words, the quark-antiquark and gluon-gluon initiated soft functions yield the same expression, up to differences in the color basis. This universality arises due to the simplicity of the specific process under consideration and is not expected to generalize to more complex cases.

Since we require the soft function in impact-parameter space (cf. eq.\eqref{eq:Wgeneraldef}), we now proceed to transform the $q_T$-dependent two-loop soft function into its $b_T$-dependent counterpart. This amounts to performing the Fourier transform of $q_T^{-2-2\epsilon}$, namely,
\begin{equation}
\int{d^{d-2}\harpoon{q}_T e^{-i\harpoon{q}_T\cdot \harpoon{b}_T} q_T^{-2-2\epsilon}}=\left(\frac{b_T^2}{4}\right)^{2\epsilon}\pi^{1-\epsilon}\frac{\Gamma(-2\epsilon)}{\Gamma(1+\epsilon)}\,.
\end{equation}
Applying this result to the sum of eqs.~\eqref{eq:DRSfun3} and \eqref{eq:RVSfun3}, we obtain the combined double-real and real-virtual contribution to the NNLO soft function in impact-parameter space:
\begin{equation}
\begin{aligned}
&\bm{S}_{\perp,\Ione\Itwo}^{[8],(2,\mathrm{RR}+\mathrm{RV})}(b_T,\mu)=\bm{S}_{\perp,\Ione\Itwo}^{[8],(2,\mathrm{RR}+\mathrm{RV})}(\harpoon{b}_T,\mu)\\
=&\left(\frac{\alpha_s}{2\pi}\right)^2\bm{1}e^{2\epsilon L_\perp}S_\epsilon^2\Bigg\{C_A^2\bigg[\frac{17}{12\epsilon^2}+\frac{1}{\epsilon}\left(\frac{\zeta_3}{2}-\frac{\zeta_2}{2}+\frac{49}{36}\right)+\frac{31}{4}\zeta_4-\frac{15}{4}\zeta_3\\
&+\zeta_2\left(3\log{(2)}+\frac{11}{12}\right)+\frac{74}{27}\bigg]+C_AT_Fn_q\bigg[-\frac{1}{3\epsilon^2}-\frac{5}{9\epsilon}-\frac{\zeta_2}{3}-\frac{28}{27}\bigg]+\mathcal{O}(\epsilon)\Bigg\}+\mathcal{O}(\alpha_1)\,,\label{eq:DRRVSfun1}
\end{aligned}
\end{equation}
where we have introduced the shorthand $S_\epsilon\equiv\left(4\pi e^{-\gamma_E}\right)^\epsilon$.

\subsubsection{UV renormalization and IR subtraction}

To perform the UV renormalization, we require the one-loop TMD soft function up to $\mathcal{O}(\epsilon)$, which can be readily extracted from the general results available in ref.~\cite{Shao:2025qgv}. The expression for the color-octet channel in impact-parameter space reads
\begin{equation}
\begin{aligned}
\bm{S}_{\perp,\Ione\Itwo}^{[8],(1,\mathrm{R}_0)}(\harpoon{b}_T,\mu)=&\frac{\alpha_{s,0}}{2\pi}\bm{1}S_\epsilon C_A\Bigg\{\bigg[\frac{1}{\epsilon}+L_\perp+\epsilon\left(\frac{L_\perp^2}{2}+\frac{\zeta_2}{2}\right)\bigg]+\mathcal{O}(\epsilon^2)\Bigg\}+\mathcal{O}(\alpha_1)\,,\label{eq:RSfun1ep1}
\end{aligned}
\end{equation}
where $\alpha_{s,0}$ denotes the bare strong coupling constant. In order to carry out the UV renormalization, we define the strong coupling in $d=4-2\epsilon$ dimensions as
\begin{equation}
\alpha_s^{(4-2\epsilon)}\equiv S_\epsilon \alpha_s\,,
\end{equation}
where $\alpha_s$ is the renormalized coupling in four dimensions. The bare coupling can then be expressed in terms of $\alpha_s^{(4-2\epsilon)}$ as~\footnote{We identify the dimensional regularization scale $\mu$ with the renormalization scale $\mu_R$.}
\begin{equation}
S_\epsilon \alpha_{s,0}=Z_{\alpha_s} \alpha_s^{(4-2\epsilon)}\,,
\end{equation}
where the renromalization constant $Z_{\alpha_s}$ is given by
\begin{equation}
Z_{\alpha_s}=1+\frac{\alpha_s^{(4-2\epsilon)}}{2\pi}\left(-\frac{\beta_0}{\epsilon}\right)+\mathcal{O}(\alpha_s^2)\,,
\end{equation}
and the LO QCD beta-function coefficient is
\begin{equation}
\beta_0=\frac{11C_A-4T_Fn_q}{6}\,.
\end{equation}
From this, we obtain the UV renormalization contribution to the two-loop TMD soft function:
\begin{equation}
\begin{aligned}
\bm{S}_{\perp,\Ione\Itwo}^{[8],(2,\mathrm{UV})}(b_T,\mu)=&\bm{S}_{\perp,\Ione\Itwo}^{[8],(2,\mathrm{UV})}(\harpoon{b}_T,\mu)=\left.\bm{S}_{\perp,\Ione\Itwo}^{[8],(1,\mathrm{R}_0)}(\harpoon{b}_T,\mu)\right|_{\mathcal{O}(\alpha_s^2)}\\
=&\left(\frac{\alpha_s}{2\pi}\right)^2\bm{1}S_\epsilon^2 C_A\beta_0\Bigg\{\bigg[-\frac{1}{\epsilon^2}-\frac{L_\perp}{\epsilon}-\left(\frac{L_\perp^2}{2}+\frac{\zeta_2}{2}\right)\bigg]+\mathcal{O}(\epsilon)\Bigg\}+\mathcal{O}(\alpha_1)\,.\label{eq:UVSfun1}
\end{aligned}
\end{equation}
The UV-renormalized two-loop TMD soft function is then obtained by combining eqs.~\eqref{eq:DRRVSfun1} and \eqref{eq:UVSfun1}:
\begin{equation}
\begin{aligned}
&\bm{S}_{\perp,\Ione\Itwo}^{[8],(2,\mathrm{RR}+\mathrm{RV}+\mathrm{UV})}(b_T,\mu)=\bm{S}_{\perp,\Ione\Itwo}^{[8],(2,\mathrm{RR}+\mathrm{RV}+\mathrm{UV})}(\harpoon{b}_T,\mu)\\
=&\left(\frac{\alpha_s}{2\pi}\right)^2\bm{1}S_\epsilon^2\Bigg\{C_A^2\bigg[-\frac{5}{12\epsilon^2}+\frac{1}{\epsilon}\left(L_\perp+\frac{\zeta_3}{2}-\frac{\zeta_2}{2}+\frac{49}{36}\right)+\frac{23}{12}L_\perp^2+L_\perp\left(\zeta_3-\zeta_2+\frac{49}{18}\right)\\
&+\frac{31}{4}\zeta_4-\frac{15}{4}\zeta_3+3\log{(2)}\zeta_2+\frac{74}{27}\bigg]+C_AT_Fn_q\bigg[\frac{1}{3\epsilon^2}-\frac{5}{9\epsilon}-\frac{L_\perp^2}{3}-\frac{10}{9}L_\perp-\frac{28}{27}\bigg]+\mathcal{O}(\epsilon)\Bigg\}+\mathcal{O}(\alpha_1)\,.\label{eq:DRRVUVSfun1}
\end{aligned}
\end{equation}
Similarly, the UV-renormalized one-loop TMD soft function is
\begin{equation}
\begin{aligned}
&\bm{S}_{\perp,\Ione\Itwo}^{[8],(2,\mathrm{R})}(b_T,\mu)=\bm{S}_{\perp,\Ione\Itwo}^{[8],(2,\mathrm{R})}(\harpoon{b}_T,\mu)=\left.\bm{S}_{\perp,\Ione\Itwo}^{[8],(1,\mathrm{R}_0)}(\harpoon{b}_T,\mu)\right|_{\mathcal{O}(\alpha_s)}\\
=&\frac{\alpha_{s}}{2\pi}\bm{1}S_\epsilon C_A\Bigg\{\bigg[\frac{1}{\epsilon}+L_\perp+\epsilon\left(\frac{L_\perp^2}{2}+\frac{\zeta_2}{2}\right)\bigg]+\mathcal{O}(\epsilon^2)\Bigg\}+\mathcal{O}(\alpha_1)\,.\label{eq:RUVSfun1ep1}
\end{aligned}
\end{equation}

We now perform the IR pole subtraction of the soft function in the $\msbar$ scheme. This is implemented via the transformation
\begin{equation}
\bm{Z}_S^{\dagger}\bm{S}_{\perp,\Ione\Itwo}^{[8]}(\harpoon{b}_T,\mu)\bm{Z}_S\,,
\end{equation}
where the IR renormalization factor $\bm{Z}_S(\mu)$ is related to the soft anomalous dimension $\bm{\Gamma}_S(\alpha_s)$ by the integral
\begin{equation}
\log{\left(\bm{Z}_S\right)}=\int_{0}^{\alpha_s^{(4-2\epsilon)}}{\frac{d\alpha_s^\prime}{\alpha_s^\prime}\frac{1}{2(\epsilon-\beta(\alpha_s^\prime)/\alpha_s^\prime)}\bm{\Gamma}_S(\alpha_s^\prime)}\,.\label{eq:logZsgeneral}
\end{equation}
The QCD beta function and the soft anomalous dimension admit perturbative expansions in $\alpha_s$ as follows:
\begin{eqnarray}
\beta(\alpha_s)&=&-\alpha_s\sum_{i=0}^{+\infty}{\left(\frac{\alpha_s}{2\pi}\right)^{i+1}\beta_{i}}\,,\\
\bm{\Gamma}_S(\alpha_s)&=&\sum_{i=0}^{+\infty}{\left(\frac{\alpha_s}{2\pi}\right)^{i+1}\bm{\Gamma}_{S,i}}\,.
\end{eqnarray}
Consequently, eq.\eqref{eq:logZsgeneral} yields the expansion
\begin{equation}
\log{\left(\bm{Z}_S\right)}=\frac{\alpha_s^{(4-2\epsilon)}}{2\pi}\frac{\bm{\Gamma}_{S,0}}{2\epsilon}+\left(\frac{\alpha_s^{(4-2\epsilon)}}{2\pi}\right)^2\left(-\frac{\beta_0 \bm{\Gamma}_{S,0}}{4\epsilon^2}+\frac{\bm{\Gamma}_{S,1}}{4\epsilon}\right)+\mathcal{O}(\alpha_s^3)\,.
\end{equation}
From this, the IR renormalization factor $\bm{Z}_S$ can be written explicitly as
\begin{equation}
\begin{aligned}
\bm{Z}_S=&\underbrace{\bm{1}}_{\equiv \bm{Z}_S^{(0)}}+\underbrace{\frac{\alpha_s^{(4-2\epsilon)}}{2\pi}\frac{\bm{\Gamma}_{S,0}}{2\epsilon}}_{\equiv \bm{Z}_S^{(1)}}+\underbrace{\left(\frac{\alpha_s^{(4-2\epsilon)}}{2\pi}\right)^2\left(\frac{\bm{\Gamma}_{S,0}^2-2\beta_0 \bm{\Gamma}_{S,0}}{8\epsilon^2}+\frac{\bm{\Gamma}_{S,1}}{4\epsilon}\right)}_{\equiv \bm{Z}_S^{(2)}}+\mathcal{O}(\alpha_s^3)\,.
\end{aligned}
\end{equation}
For the color-octet final state, the soft anomalous dimensions are given by~\cite{Becher:2009kw}
\begin{eqnarray}
\bm{\Gamma}_{S,0}&=&-\bm{1}C_A\left(1-i\pi\right)\,,\\
\bm{\Gamma}_{S,1}&=&\bm{1}\Bigg[C_A^2\left(-\zeta_3+\zeta_2(1-i\pi)-\frac{49}{18}+\frac{67}{18}i\pi\right)+C_AT_Fn_q\frac{10}{9}(1-i\pi)\Bigg]\,.
\end{eqnarray}
The IR-subtracted NLO TMD soft function is obtained as
\begin{equation}
\begin{aligned}
\bm{S}_{\perp,\Ione\Itwo}^{[8],(1)}(\harpoon{b}_T,\mu)=&\lim_{\epsilon\to 0}\left[\bm{S}_{\perp,\Ione\Itwo}^{[8],(1,\mathrm{R})}(\harpoon{b}_T,\mu)+\bm{Z}_S^{(1)}+\left(\bm{Z}_S^{(1)}\right)^\dagger\right]\\
=&\bm{1}\frac{\alpha_s}{2\pi}C_A L_\perp\,,
\end{aligned}
\end{equation}
which is in agreement with eqs.\eqref{eq:NLOSfun2} and \eqref{eq:NLOSfun3}. The NNLO soft function after IR subtraction reads
\begin{equation}
\begin{aligned}
\bm{S}_{\perp,\Ione\Itwo}^{[8],(2)}(\harpoon{b}_T,\mu)=&\lim_{\epsilon\to 0}\Bigg[\bm{S}_{\perp,\Ione\Itwo}^{[8],(2,\mathrm{RR}+\mathrm{RV}+\mathrm{UV})}(\harpoon{b}_T,\mu)+\left(\bm{Z}_S^{(1)}\right)^\dagger \bm{Z}_S^{(1)}+\left(\bm{Z}_S^{(1)}\right)^\dagger \bm{S}_{\perp,\Ione\Itwo}^{[8],(1,\mathrm{R})}(\harpoon{b}_T,\mu)\\
&+\bm{S}_{\perp,\Ione\Itwo}^{[8],(1,\mathrm{R})}(\harpoon{b}_T,\mu)\bm{Z}_S^{(1)}+\bm{Z}_S^{(2)}+\left(\bm{Z}_S^{(2)}\right)^\dagger\Bigg]\\
=&\bm{1}\left(\frac{\alpha_s}{2\pi}\right)^2\Bigg[C_A^2 \bigg(\frac{17}{12}L_\perp^2+L_\perp\left(\zeta_3-\zeta_2+\frac{49}{18}\right)+\frac{31}{4}\zeta_4-\frac{15}{4}\zeta_3+\zeta_2\left(3\log{(2)}-\frac{1}{2}\right)\\
&+\frac{74}{27}\bigg)+C_AT_Fn_q\bigg(-\frac{L_\perp^2}{3}-\frac{10}{9}L_\perp-\frac{28}{27}\bigg)\Bigg]=\bm{S}_{\perp,\Ione\Itwo}^{[8],(2)}(b_T,\mu)\,.\label{eq:NNLOSfun}
\end{aligned}
\end{equation}
Equation \eqref{eq:NNLOSfun} represents the main result of this work. Despite the complexity of the intermediate steps, the final expression is remarkably simple. Furthermore, we have verified that the soft function satisfies the renormalization group equation
\begin{equation}
\frac{d}{d\log{\mu}}\bm{S}_{\perp,\Ione\Itwo}^{[8]}(\harpoon{b}_T,\mu)=-\bm{\Gamma}_S^\dagger(\alpha_s)\bm{S}_{\perp,\Ione\Itwo}^{[8]}(\harpoon{b}_T,\mu)-\bm{S}_{\perp,\Ione\Itwo}^{[8]}(\harpoon{b}_T,\mu)\bm{\Gamma}_S(\alpha_s)\,
\end{equation}
up to $\mathcal{O}(\alpha_s^2)$. 

We have also numerically compared our final result, eq.~\eqref{eq:NNLOSfun} with $n_q=0$, to $S_{33}$ at $\beta=0.1$ in figure 17 (bottom-right panel) of ref.~\cite{Angeles-Martinez:2018mqh}, and found a relative difference below $1\%$ after accounting for different normalization conventions. For the $n_q$-dependent term, we compared our bare soft function in eq.~\eqref{eq:DRRVSfun1} to $S_{33}$ at $\beta=0.1$ in figure 15 (bottom-right panel) of ref.~\cite{Angeles-Martinez:2018mqh}, again finding a relative difference less than $1\%$.

\section{Summary}\label{sec:conclusion}

We have presented an analytic calculation of the NNLO TMD soft function for heavy-quark pair production in the hadronic collision process, in the non-relativistic limit. While the soft function for a color-singlet $Q\bar{Q}$ pair vanishes due to scalelessness in both analytic rapidity regularization and dimensional regularization, the soft function for a color-octet configuration is non-zero. Our main result--the NNLO TMD soft function in the color-octet channel--is given in eq.\eqref{eq:NNLOSfun}. Owing to its single-scale nature, the result takes a particularly simple form and is identical for both quark-antiquark and gluon-gluon initial states (up to differences in the color basis). As a byproduct of this work, we have also outlined in section \ref{sec:generalstructure} the general structure of the soft amplitude and soft function at the two-loop level, which is applicable to arbitrary processes in the SM. Additionally, we have detailed our computational approach and provided extensive intermediate results to facilitate the reproduction of our calculation by interested readers. 

Our result provides an important ingredient for NNLO QCD calculations of non-relativistic heavy-quark hadroproduction, including $S$-wave quarkonium states, within the $q_T$-slicing framework. It is also relevant for the resummation of small-$q_T$ logarithmically enhanced terms in the cross section at next-to-next-to-next-to-leading-logarithmic (N$^3$LL) accuracy. In these contexts, the NNLO hard function is already known~\cite{Abreu:2022vei,Abreu:2022cco}, and the four-loop cusp anomalous dimension, along with the three-loop collinear and soft anomalous dimensions, have been computed in the literature~\cite{Becher:2009qa,Becher:2009kw,Grozin:2014hna,Grozin:2015kna,Henn:2019swt,Bruser:2019yjk,vonManteuffel:2020vjv}.~\footnote{For massless (anti)quarks and gluons, the four-loop collinear anomalous dimensions have also been obtained~\cite{Das:2019btv,Das:2020adl,Chakraborty:2022yan}.}

\acknowledgments
 G.W. thanks Wan-Li Ju and Li Lin Yang for helpful discussions. This work is supported by the ERC (grant 101041109 ``BOSON") and the French ANR (grant
ANR-20-CE31-0015, ``PrecisOnium"). Views and opinions expressed are however those of the authors only and do not necessarily reflect those of the European Union or the European Research Council Executive Agency. Neither the European Union nor the granting authority can be held responsible for them.

\appendix

\section{Tables of double-real and real-virtual integrals\label{sec:tabofints}}

In this appendix, we collect the results of the double-real and real-virtual integrals, which we hope will be useful to others, particularly those wishing to reproduce our calculations. Table \ref{tab:DRints} summarizes all integrals appearing in the double-real contribution to the soft function in eq.\eqref{eq:DRSfun1}. The left column shows the integrands to be integrated over $x$ and $y$ (cf. eq.~\eqref{eq:DRSfun2}), while the right column lists the results of these integrals after factoring out the common prefactor
\begin{equation}
C_{\mathrm{RR}}(\mu,q_T)\equiv 4\left(\frac{\mu^2}{q_T^2}\right)^\epsilon\frac{1}{q_T^4}\frac{\pi^{1-\epsilon}}{(2\pi)^{3-2\epsilon}}\,.\label{eq:CRR}
\end{equation}
Similarly, the results of all integrals in the real-virtual contribution in eq.\eqref{eq:RVSfun1} are given in table \ref{tab:RVints}, where the integrands are integrated over $y_1$ (cf. eq.~\eqref{eq:RVSfun2}), and the results are normalized by the common prefactor
\begin{equation}
C_{\mathrm{RV}}(\mu,q_T)\equiv\left(\frac{\mu^2}{q_T^2}\right)^\epsilon\frac{1}{q_T^2}\frac{\pi^{1-\epsilon}}{(2\pi)^{3-2\epsilon}}\,.\label{eq:CRV}
\end{equation}

\setlength\LTleft{-10mm}
\begin{longtable}[H]{| p{.38\textwidth} | p{.75\textwidth} |}
\hline
Integrand & Result$/C_{\mathrm{RR}}(\mu,q_T)$ \\
\hline\hline
$y^{-1}\frac{1}{q^2}\frac{1}{(v\cdot q)^2-q^2}J_1^{(2,\mathrm{RR})}(q)$ &  $\frac{\Gamma(-\epsilon)\Gamma(2-\epsilon)\Gamma(1+\epsilon)}{(1+2\epsilon)\Gamma(3-2\epsilon)}$ \\\hline
$y^{-1}\frac{1}{n\cdot q \bar{n}\cdot q}\frac{1}{(v\cdot q)^2-q^2}J_1^{(2,\mathrm{RR})}(q)$ &  $\frac{\Gamma(-\epsilon)\Gamma(2-\epsilon)\Gamma(1+\epsilon)}{2\Gamma(3-2\epsilon)}\left(\psi(\epsilon)-2\psi(2\epsilon)-\gamma_E\right)$ \\\hline
$y^{-1}\frac{1}{q^2 n\cdot q \bar{n}\cdot q}J_1^{(2,\mathrm{RR})}(q)$ &  $0$\\\hline
$y^{-1}\frac{1}{q^2 (v\cdot q)^2}J_1^{(2,\mathrm{RR})}(q)$ &  $\frac{\Gamma(-\epsilon)\Gamma(2-\epsilon)\Gamma(1+\epsilon)}{\Gamma(3-2\epsilon)}$\\\hline
$y^{-1}\frac{v\cdot q}{q^2 n\cdot q\bar{n}\cdot q}J_2^{(2,\mathrm{RR})}(q)$ &  $\frac{\Gamma^2(-\epsilon)\Gamma(\epsilon)}{8\Gamma(-2\epsilon)}\left(\Gamma(1+2\epsilon)\Gamma(-2\epsilon)-\psi(1+\epsilon)+\psi(1+2\epsilon)\right)$\\\hline
$y^{-1}\frac{v\cdot q}{n\cdot q\bar{n}\cdot q}\frac{1}{(v\cdot q)^2-q^2}J_2^{(2,\mathrm{RR})}(q)$ &  $\frac{\Gamma^2(1-\epsilon)\Gamma(\epsilon)}{2\Gamma(2-2\epsilon)}\left(\Gamma(1-2\epsilon)\Gamma(2\epsilon)+\psi(2\epsilon)+\gamma_E\right)$\\\hline
$y^{-1}\frac{1}{q^2 v\cdot q}J_2^{(2,\mathrm{RR})}(q)$ &  $\frac{\Gamma(\epsilon)\Gamma(-2\epsilon)}{2}\bigg[\frac{2^{4\epsilon}\Gamma^2(1+\epsilon)}{1+2\epsilon}-\frac{\Gamma^2(-\epsilon)}{2\Gamma^2(-2\epsilon)}{}_3F_2(\frac{1}{2},1,-\epsilon;\frac{3}{2},1+\epsilon;1)\bigg]$\\\hline
$y^{-1}\frac{v\cdot q}{q^2}\frac{1}{(v\cdot q)^2-q^2}J_2^{(2,\mathrm{RR})}(q)$ &  $\frac{\Gamma^2(-\epsilon)\Gamma(\epsilon)\Gamma(-1-2\epsilon)}{8\Gamma^2(-2\epsilon)}$\\\hline
$y^{-1}\frac{1}{(v\cdot q)^3}J_2^{(2,\mathrm{RR})}(q)$ &  $\Gamma(1+\epsilon)\Gamma(-2\epsilon)\bigg[-\frac{2^{4\epsilon}\Gamma^2(1+\epsilon)}{1+2\epsilon}+\frac{\Gamma^2(-\epsilon)}{6\Gamma^2(-2\epsilon)}{}_3F_2(\frac{1}{2},1,1-\epsilon;\frac{5}{2},1+\epsilon;1)\bigg]$\\\hline
$(1+x)^{-\alpha_1}y^{-1+\alpha_1}\frac{1}{n\cdot q \bar{n}\cdot q}J_3^{(2,\mathrm{RR})}(q)$ &  $0$\\
\hline
$(1+x)^{-\alpha_1}y^{-1+\alpha_1}\frac{1}{n\cdot q v\cdot q}J_3^{(2,\mathrm{RR})}(q)$ &  $-\frac{2^{-1+2\epsilon}\sqrt{\pi}\Gamma(1-\alpha_1)\Gamma(\alpha_1)\Gamma(1-\epsilon)\Gamma(\epsilon)}{\Gamma(\frac{1}{2}-\epsilon)}\bigg[\Gamma(1-2\epsilon)\Gamma(2\epsilon)+\psi(1+\epsilon)-\psi(1+2\epsilon)\bigg]$\\\hline
$(1+x)^{-\alpha_1}y^{-1+\alpha_1}\frac{\bar{n}\cdot q}{q^2 v\cdot q}J_3^{(2,\mathrm{RR})}(q)$ &  $\frac{2^{2\epsilon} \sqrt{\pi}\Gamma(1-\alpha_1)\Gamma(\alpha_1)\Gamma(1-\epsilon)\Gamma(\epsilon)}{(1+2\epsilon)^2\Gamma(-\frac{1}{2}-\epsilon)}$\\\hline
$y^{-1}\frac{1}{(v\cdot q)^2}J_3^{(2,\mathrm{RR})}(q)$ &  $\frac{2^{2\epsilon} \sqrt{\pi}\Gamma(1-\epsilon)\Gamma(\epsilon)}{\Gamma(\frac{1}{2}-\epsilon)}\bigg[\Gamma(1-2\epsilon)\Gamma(2\epsilon)+\psi(1+\epsilon)-\psi(1+2\epsilon)\bigg]$\\\hline
$(1+x)^{-\alpha_1}y^{-1+\alpha_1}\frac{1}{q^2}J_3^{(2,\mathrm{RR})}(q)$ &  $0$\\\hline
$(1+x)^{-\alpha_1}y^{-1+\alpha_1}\frac{1}{n\cdot q \bar{n}\cdot q}J_4^{(2,\mathrm{RR})}(q)$ &  $0$\\\hline
$(1+x)^{-\alpha_1}y^{-1+\alpha_1}\frac{1}{q^2}J_4^{(2,\mathrm{RR})}(q)$ &  $0$\\\hline
\multirow{2}{*}{$(1+x)^{-\alpha_1}y^{-1+\alpha_1}\frac{\bar{n}\cdot q}{q^2 v\cdot q}J_5^{(2,\mathrm{RR})}(q)$} &  $\Gamma(1+\epsilon)\bigg[-\frac{1}{\alpha_1\epsilon}+\frac{2}{\alpha_1}+\frac{\epsilon}{\alpha_1}\left(\zeta_2-4\right)+\frac{1}{4\epsilon^2}-\frac{1}{\epsilon}$\\
& $+\frac{\zeta_2}{4}+3+\epsilon(\zeta_3+3\zeta_2-8)+\mathcal{O}(\alpha_1,\epsilon^2)\bigg]$\\\hline
$(1+x)^{-\alpha_1}y^{-1+\alpha_1}\frac{1}{q^2}J_5^{(2,\mathrm{RR})}(q)$ &  $\Gamma(1+\epsilon)\bigg[-\frac{1}{2\alpha_1\epsilon}+\frac{1}{\alpha_1}+\frac{\epsilon}{\alpha_1}(\frac{\zeta_2}{2}-2)+\mathcal{O}(\alpha_1,\epsilon^2)\bigg]$\\\hline
$(1+x)^{-\alpha_1}y^{-1+\alpha_1}\frac{1}{n\cdot qv\cdot q}J_6^{(2,\mathrm{RR})}(q)$ &  $\Gamma(1+\epsilon)\bigg[-\frac{1}{4\epsilon^3}-\frac{3\zeta_2}{4\epsilon}-\frac{3\zeta_3}{2}-\epsilon\frac{43\zeta_4}{16}+\mathcal{O}(\alpha_1,\epsilon^2)\bigg]$\\\hline
$(1+x)^{-\alpha_1}y^{-1+\alpha_1}\frac{1}{n\cdot q\bar{n}\cdot q}J_6^{(2,\mathrm{RR})}(q)$ &  $\Gamma(1+\epsilon)\bigg[\frac{1}{4\alpha_1\epsilon^2}+\frac{\zeta_2}{4\alpha_1}+\frac{\epsilon}{\alpha_1}\zeta_3+\mathcal{O}(\alpha_1,\epsilon^2)\bigg]$\\\hline
$(1+x)^{-\alpha_1}y^{-1+\alpha_1}\frac{\bar{n}\cdot q}{q^2v\cdot q}J_6^{(2,\mathrm{RR})}(q)$ &  $\Gamma(1+\epsilon)\bigg[-\frac{1}{4\epsilon^2}+\frac{1}{\epsilon}-3-\frac{\zeta_2}{4}-\epsilon(\zeta_3+3\zeta_2-8)+\mathcal{O}(\alpha_1,\epsilon^2)\bigg]$\\\hline
$(1+x)^{-\alpha_1}y^{-1+\alpha_1}\frac{1}{q^2}J_6^{(2,\mathrm{RR})}(q)$ &  $\Gamma(1+\epsilon)\bigg[\frac{1}{2\alpha_1\epsilon}-\frac{1}{\alpha_1}+\frac{\epsilon}{\alpha_1}(-\frac{\zeta_2}{2}+2)+\mathcal{O}(\alpha_1,\epsilon^2)\bigg]$\\\hline
\multirow{2}{*}{$(1+x)^{-\alpha_1}y^{-1+\alpha_1}\frac{1}{q^2}J_7^{(2,\mathrm{RR})}(q)$} &  $\Gamma(1+\epsilon)\bigg[-\frac{1}{2\alpha_1^2\epsilon}+\frac{\epsilon}{\alpha_1^2}\frac{\zeta_2}{2}-\frac{1}{4\alpha_1\epsilon^2}+\frac{5\zeta_2}{4\alpha_1}+\frac{\epsilon}{\alpha_1}\frac{5\zeta_3}{2}$\\
 & $-\frac{1}{8\epsilon^3}-\frac{3\zeta_2}{8\epsilon}+\frac{\zeta_3}{4}-\epsilon\frac{63\zeta_4}{32}+\mathcal{O}(\alpha_1,\epsilon^2)\bigg]$ \\\hline
 $(1+x)^{-\alpha_1}y^{-1+\alpha_1}\frac{n\cdot q}{q^2v\cdot q}J_7^{(2,\mathrm{RR})}(q)$ &  $\Gamma(1+\epsilon)\bigg[\frac{3}{4\epsilon^2}-\frac{1}{\epsilon}+1-\frac{9\zeta_2}{4}+\epsilon(-4\zeta_3+\zeta_2)+\mathcal{O}(\alpha_1,\epsilon^2)\bigg]$\\\hline
 \multirow{2}{*}{$(1+x)^{-\alpha_1}y^{-1+\alpha_1}\frac{1}{n\cdot q v\cdot q}J_7^{(2,\mathrm{RR})}(q)$} &  $\Gamma(1+\epsilon)\bigg[-\frac{1}{\alpha_1^2\epsilon}+\frac{\epsilon}{\alpha_1^2}\zeta_2+\frac{1}{2\alpha_1\epsilon^2}+\frac{\zeta_2}{2\alpha_1}+\frac{\epsilon}{\alpha_1}2\zeta_3$\\
 & $-\frac{1}{2\epsilon^3}-\frac{\zeta_2}{\epsilon}-\frac{3\zeta_3}{2}-\epsilon\frac{13\zeta_4}{8}+\mathcal{O}(\alpha_1,\epsilon^2)\bigg]$\\\hline
  $(1+x)^{-\alpha_1}y^{-1+\alpha_1}\frac{1}{(v\cdot q)^2}J_7^{(2,\mathrm{RR})}(q)$ &  $\Gamma(1+\epsilon)\bigg[-\frac{1}{\epsilon}+2+\epsilon(7\zeta_2-4)+\mathcal{O}(\alpha_1,\epsilon^2)\bigg]$\\\hline
\multirow{2}{*}{$(1+x)^{-\alpha_1}y^{-1+\alpha_1}\frac{1}{q^2}J_8^{(2,\mathrm{RR})}(q)$} &  $\Gamma(1+\epsilon)\bigg[-\frac{1}{2\alpha_1^2\epsilon}+\frac{\epsilon}{\alpha_1^2}\frac{\zeta_2}{2}+\frac{1}{4\alpha_1\epsilon^2}-\frac{5\zeta_2}{4\alpha_1}-\frac{\epsilon}{\alpha_1}\frac{5\zeta_3}{2}$\\
& $-\frac{1}{8\epsilon^3}-\frac{3\zeta_2}{8\epsilon}+\frac{\zeta_3}{4}-\epsilon\frac{63\zeta_4}{32}+\mathcal{O}(\alpha_1,\epsilon^2)\bigg]$\\\hline
$(1+x)^{-\alpha_1}y^{-1+\alpha_1}\frac{\bar{n}\cdot q}{q^2 v\cdot q}J_8^{(2,\mathrm{RR})}(q)$ &  $\Gamma(1+\epsilon)\bigg[\frac{3}{4\epsilon^2}-\frac{1}{\epsilon}+1-\frac{9\zeta_2}{4}+\epsilon(-4\zeta_3+\zeta_2)+\mathcal{O}(\alpha_1,\epsilon^2)\bigg]$\\\hline
\multirow{2}{*}{$(1+x)^{-\alpha_1}y^{-1+\alpha_1}\frac{1}{\bar{n}\cdot q v\cdot q}J_8^{(2,\mathrm{RR})}(q)$} &  $\Gamma(1+\epsilon)\bigg[-\frac{1}{\alpha_1^2\epsilon}+\frac{\epsilon}{\alpha_1^2}\zeta_2-\frac{1}{2\alpha_1\epsilon^2}-\frac{\zeta_2}{2\alpha_1}-\frac{\epsilon}{\alpha_1}2\zeta_3$\\
 & $-\frac{1}{2\epsilon^3}-\frac{\zeta_2}{\epsilon}-\frac{3\zeta_3}{2}-\epsilon\frac{13\zeta_4}{8}+\mathcal{O}(\alpha_1,\epsilon^2)\bigg]$\\\hline
$(1+x)^{-\alpha_1}y^{-1+\alpha_1}\frac{1}{(v\cdot q)^2}J_8^{(2,\mathrm{RR})}(q)$ &  $\Gamma(1+\epsilon)\bigg[-\frac{1}{\epsilon}+2+\epsilon(7\zeta_2-4)+\mathcal{O}(\alpha_1,\epsilon^2)\bigg]$\\\hline
$(1+x)^{-\alpha_1}y^{-1+\alpha_1}\frac{1}{v\cdot q}J_9^{(2,\mathrm{RR})}(q)$ &  $\Gamma(1+\epsilon)\bigg[\frac{2}{\alpha_1\epsilon}-\frac{\epsilon}{\alpha_1}2\zeta_2-\frac{1}{2\epsilon^2}+\frac{1}{\epsilon}-2-\frac{\zeta_2}{2}+\epsilon(-2\zeta_3-7\zeta_2+4)+\mathcal{O}(\alpha_1,\epsilon^2)\bigg]$\\\hline
$(1+x)^{-\alpha_1}y^{-1+\alpha_1}\frac{1}{v\cdot q}J_{10}^{(2,\mathrm{RR})}(q)$ &  $\Gamma(1+\epsilon)\bigg[-\frac{2}{\alpha_1\epsilon}+\frac{\epsilon}{\alpha_1}2\zeta_2-\frac{1}{2\epsilon^2}+\frac{1}{\epsilon}-2-\frac{\zeta_2}{2}+\epsilon(-2\zeta_3-7\zeta_2+4)+\mathcal{O}(\alpha_1,\epsilon^2)\bigg]$\\
\hline
\caption{Results of the double-real integrals appearing in eq.~\eqref{eq:DRSfun1}. A common prefactor $C_{\mathrm{RR}}(\mu,q_T)$ defined in eq.~\eqref{eq:CRR} has been factored out in the ``Result" column.\label{tab:DRints}}
\end{longtable}

\setlength\LTleft{-10mm}
\begin{longtable}[H]{| p{.38\textwidth} | p{.75\textwidth} |}
\hline
Integrand & Result$/C_{\mathrm{RV}}(\mu,q_T)$ \\
\hline\hline
$y_1^{-1}\frac{1}{(n\cdot k_1)^2\bar{n}\cdot k_1}J_1^{(2,\mathrm{RV})}(k_1)$ &  $e^{2\epsilon\pi i}\Gamma(1-\epsilon)\Gamma(\epsilon-1)\Gamma(\epsilon)$ \\\hline
$y_1^{-1}\frac{1}{(n\cdot k_1)^2 v\cdot k_1}J_1^{(2,\mathrm{RV})}(k_1)$ &  $\frac{2e^{2\epsilon \pi i}\Gamma(1-\epsilon)\Gamma(\epsilon-1)\Gamma(1+\epsilon)\Gamma(2\epsilon-1)}{\Gamma(2\epsilon)}$ \\\hline
$y_1^{-1}\frac{1}{(v\cdot k_1)^3}J_1^{(2,\mathrm{RV})}(k_1)$ &  $\frac{8e^{2\epsilon \pi i}\Gamma(1-\epsilon)\Gamma^2(1+\epsilon)\Gamma(2\epsilon-1)}{\Gamma(2+2\epsilon)}$ \\\hline
$y_1^{-1}\frac{1}{n\cdot k_1 v\cdot k_1}J_2^{(2,\mathrm{RV})}(k_1)$ &  $-2\Gamma(1-2\epsilon)\Gamma(1-\epsilon)\Gamma^2(\epsilon)\Gamma(2\epsilon)$ \\\hline
$y_1^{-1}\frac{1}{\bar{n}\cdot k_1 v\cdot k_1}J_3^{(2,\mathrm{RV})}(k_1)$ &  $-2\Gamma(1-2\epsilon)\Gamma(1-\epsilon)\Gamma^2(\epsilon)\Gamma(2\epsilon)$ \\\hline
$y_1^{-1}\frac{1}{(n\cdot k_1)^2 \bar{n}\cdot k_1}J_4^{(2,\mathrm{RV})}(k_1)$ &  $-e^{2\epsilon \pi i}\Gamma(1-\epsilon)\Gamma(\epsilon-1)\Gamma(\epsilon)$\\\hline
$y_1^{-1}\frac{1}{(n\cdot k_1)^2 v\cdot k_1}J_4^{(2,\mathrm{RV})}(k_1)$ &  $-\frac{2e^{2\epsilon \pi i}\Gamma(1-\epsilon)\Gamma(\epsilon-1)\Gamma(1+\epsilon)\Gamma(2\epsilon-1)}{\Gamma(2\epsilon)}$\\\hline
$y_1^{-1}\frac{1}{(v\cdot k_1)^3}J_4^{(2,\mathrm{RV})}(k_1)$ &  $-\frac{8e^{2\epsilon \pi i}\Gamma(1-\epsilon)\Gamma^2(1+\epsilon)\Gamma(2\epsilon-1)}{\Gamma(2+2\epsilon)}$\\\hline
$y_1^{-1+\alpha_1}J_5^{(2,\mathrm{RV})}(k_1)$ &  $0$\\\hline
$y_1^{-1+\alpha_1}\frac{n\cdot k_1-\bar{n}\cdot k_1}{v\cdot k_1}J_6^{(2,\mathrm{RV})}(k_1)$ &  eq.\eqref{eq:int4J2RV6}$/C_{\mathrm{RV}}(\mu,q_T)$\\\hline
$y_1^{-1+\alpha_1}\frac{n\cdot k_1-\bar{n}\cdot k_1}{v\cdot k_1}J_7^{(2,\mathrm{RV})}(k_1)$ &  $\left.[-\mathrm{eq.}\eqref{eq:int4J2RV6}/C_{\mathrm{RV}}(\mu,q_T)\right]_{\alpha_1\to -\alpha_1}$\\
\hline
\caption{Results of the real-virtual integrals appearing in eq.~\eqref{eq:RVSfun1}. A common prefactor $C_{\mathrm{RV}}(\mu,q_T)$ defined in eq.~\eqref{eq:CRV} has been factored out in the ``Result" column.\label{tab:RVints}}
\end{longtable}

\bibliographystyle{JHEP}
\bibliography{references_inspire.bib}

\end{document}